%% file: Extended-Lee-Bounds.tex
\documentclass[11pt,reqno]{amsart}
\usepackage{graphics,amsmath,amssymb,epsfig,stmaryrd,color}
\usepackage{subfigure,amsfonts,geometry,array,cancel}
\usepackage{graphicx}
\usepackage{pdflscape}
\usepackage{amsthm}
\usepackage{flafter}
\usepackage{tikz}
\usetikzlibrary{positioning, arrows.meta}
\usepackage{cancel}
\usepackage{natbib}
\usepackage[normalem]{ulem}
\usepackage{caption}
\usepackage{float}

\newtheorem{theorem}{Theorem}
\theoremstyle{plain}

\newtheorem{assumption}{Assumption}

\newtheorem{lemma}{Lemma}

\newtheorem{remark}{Remark}

\numberwithin{equation}{section}
\newcommand{\given}{\ \vert\ }

\usepackage{booktabs,dcolumn}
\usepackage{multirow}
\usepackage{longtable}
\usepackage[hidelinks]{hyperref}
\usepackage{cleveref}
\crefname{assumption}{Assumption}{Assumptions}
\newcolumntype{M}[1]{>{\centering\arraybackslash}m{#1}}
\newcolumntype{L}[1]{>{\raggedright\arraybackslash}m{#1}}
\newcolumntype{R}[1]{>{\raggedleft\arraybackslash}m{#1}}
\newcolumntype{d}[1]{D{.}{.}{#1}}
\newcolumntype{N}{@{}m{0pt}@{}}

\begin{document}


\title[bounds with  Multilayered sample selection]{Horowitz-Manski-Lee bounds with multilayered sample selection}
\author{Kory Kroft$^\ast$, Ismael Mourifi\'e $^\dag$, and Atom Vayalinkal$^\ddag$}
\date{This present version of the paper is \today. {\footnotesize 
We thank David Lee and seminar audiences at Brown University, Carnegie Mellon University, National University of Singapore, New York University, Princeton University, University of Toronto, as well as the Rochester Labor Economics and Southern Economic Association conferences. We thank Stephen Claassen for providing outstanding research assistance.\\
$^\ast$ Department of Economics, University of Toronto, \& NBER. 150 St. George Street, Toronto ON M5S 3G7, Canada, kory.kroft@utoronto.ca.\\
$^\dag$ Department of Economics, Washington University in St. Louis \& NBER, ismaelm@wustl.edu}\\
$^\ddag$ Department of Economics, University of Toronto, 150 St. George Street, Toronto ON M5S 3G7, Canada, atom.vayalinkal@mail.utoronto.ca.\\}


\begin{abstract}

This paper investigates the causal effect of job training on wage rates in the presence of firm heterogeneity. When training affects the sorting of workers to firms, sample selection is no longer binary but is ``multilayered". This paper extends the canonical Heckman (1979) sample selection model -- which assumes selection is binary -- to a setting where it is multilayered. In this setting Lee bounds set identifies a total effect that combines a weighted-average of the causal effect of job training on wage rates across firms with a weighted-average of the contrast in wages between different firms for a fixed level of training. Thus, Lee bounds set identifies a policy-relevant estimand only when firms pay homogeneous wages and/or when job training does not affect worker sorting across firms. We derive analytic expressions for sharp  
bounds for the causal effect of job training on wage rates at each firm that leverage  information on firm-specific wages. We illustrate our partial identification approach with two empirical applications to job training experiments. 
Our estimates demonstrate that even when conventional Lee bounds are strictly positive, our within-firm bounds can be tight around 0, showing that the 
canonical 
Lee bounds may capture only a pure sorting effect of job training.

\end{abstract}
\maketitle

\section{Introduction}\label{sec:intro}

Governments allocate substantial funds to job training programs that are designed to improve worker skills. The United States (U.S.) federal government spends roughly $19$ billion U.S. dollars annually on employment and training programs.\footnote{See the Council of Economic Advisers 2019 report: ``Government Employment and Training Programs: Assessing the Evidence on their Performance" (\citeauthor{the_council_of_economic_advisers_government_2019} \citeyear{the_council_of_economic_advisers_government_2019}).} Federal agencies administer roughly 40 employment and training programs to assist job seekers in gaining employment. Against this backdrop, there is an ongoing debate about the appropriate level of spending on such programs. Advocates argue that they help close the ``skills gap'' and address worker shortages, while critics argue that they are ineffective and socially wasteful. 

At the center of the debate is the long-standing question of whether job training has a causal effect on the labor market outcomes of participants. This question has garnered significant interest from both academics and policymakers and knowing the answer is essential for determining whether to continue spending on training programs. Substantial progress has been made to answer this question as a result of randomized evaluations of job training programs. These evaluations have been the subject of several comprehensive meta-analyzes (see \citeauthor{heckman_economics_1999} \citeyear{heckman_economics_1999} and \citeauthor{card_active_2010} \citeyear{card_active_2010}, \citeyear{card_what_2018}).

To date, most training program evaluations have focused on total earnings as the main outcome of analysis. While the earnings impact is surely important for answering some questions, for others it is important to narrow the focus. Earnings naturally reflect both labor supply decisions (employment and hours margin) and wage rates. To better understand whether job training increases worker skills and welfare, standard economic models show that it is important to focus on the latter.\footnote{Labor supply decisions could also be impacted by job training via an increase in human capital. In particular, workers with higher skills are more likely to be offered -- and accept -- better paying jobs.}\textsuperscript{,}\footnote{\citet{hendren_unified_2020} measure the willingness to pay for job training using the treatment effect on total earnings. This assumes that all increases in earnings come from returns to human capital (higher wage rate), not from higher levels of labor supply.} However, identifying the causal effect of job training on wage rates is empirically challenging due to the well-known sample selection problem (\citeauthor{heckman_sample_1979} \citeyear{heckman_sample_1979}). This arises since a researcher only observes wages of the employed, and the likelihood of employment can itself be impacted by job training. 


In a seminal contribution, \cite{lee_training_2009} showed that the causal effect of job training on wages for the always-employed can partially identify under the \cite{imbens_identification_1994} (``IA'') monotonicity assumption. Lee's key insight was to reduce the partial identification problem in this framework to the one considered by \cite{horowitz_identification_1995} – the problem of finding sharp bounds for the mean of an unobserved potential outcome that is a component of an observed mixing distribution with set-identified mixing probabilities. IA’s monotonicity condition delivers point identification of both the mixing weight (i.e., the share of always-employed) and the mean of the ``untrained'' potential outcome for the always-employed. Therefore, to obtain sharp bounds on the causal effect of job training on wages, one needs only to find sharp bounds on the mean of the ``trained'' potential outcome for the always-employed. Lee's bounding approach has become influential in empirical research.\footnote{In \Cref{sec:app_d_litreview}, we describe results from an informal survey we conducted. We counted 56 papers published in `top 5' general interest economic journals -- the American Economic Review, Econometrica, the Journal of Political Economy, the Quarterly Journal of Economics and the Review of Economic Studies -- that cited \cite{lee_training_2009} or earlier working paper versions. The purpose of our survey was to determine whether the nature of sample selection in these papers was binary or multilayered. Of these, 42 empirically implemented Lee bounds to address sample selection, and 7 of them featured multilayered selection. To apply Lee bounds in these settings, the researchers collapsed the sample selection problem to a single dimension.}

In developing his approach, Lee focused on the potential for job training to affect labor supply along the extensive margin (work vs. no work). Although it is important to know whether training increases employment, a fundamental question for policymakers is whether job training improves labor market outcomes by increasing job quality. If this is the case, training may increase both the probability individuals are employed, as well as the probability individuals are employed at ``better jobs”, both of which can increase earnings. For example, this is a key feature of President Biden's workforce training initiative, the ``American Rescue Plan’s Good Jobs Challenge'', which prioritizes job quality and is designed to ensure that workers can access good jobs.

Despite the emphasis on job quality by policymakers, the academic literature on job training programs has mostly ignored firms. In the standard competitive model considered by Lee and most of the training literature, the firm for which an individual works does not matter for wages. However, there is growing empirical evidence that demonstrates the importance of firms for wage determination.\footnote{See, for example, \cite{abowd_high_1999}, \cite{card_workplace_2013}, \cite{song_firming_2019}, \cite{bonhomme_distributional_2019} and \cite{bonhomme_how_2023}.} This research highlights the importance of having a ``good job'' which can be interpreted as working at a ``good firm'' that offers a higher wage for all its employees. Thus, an open question is whether job training raises earnings by moving participants to higher-paying firms.


The presence of firm heterogeneity and the potential for worker sorting raises several new questions of interest. First, what estimand do ``Lee bounds'' partially identify when there is firm heterogeneity in wages? This question cannot be addressed with the canonical \cite{heckman_sample_1979} sample selection model since this assumes that sample selection is binary, i.e., job training can increase employment but has no effect on sorting to firms. The first contribution of this paper is to extend the standard sample selection model to a setting where selection is multilayered, and to show that the conventional Lee bounds set identifies (for the always-employed population) a total effect that combines a weighted average of the causal effect of job training on wage rates across firms (we label this the “within-firm effect'') with a weighted average of the contrast in wages between different firms for a fixed level of training (we label this the “sorting effect”).

A natural question is whether it is possible to separate the within-firm wage effect of job training from the sorting effect in the presence of heterogeneous firms.
There are several reasons why one would want to separately identify these effects. First, some features of job training programs affect sorting (job search assistance), whereas others affect skill acquisition (classroom and vocational training). Thus, the decomposition could potentially highlight which investments – job search assistance or classroom training – are effective for raising wages and thus improve targeting. Second, the within-firm wage effect is arguably better able to shed light on the causal effect of \emph{employer-sponsored} job training. Third, for a welfare analysis of job training programs, it is important to focus on the direct wage effects of job training, since labor supply effects have second-order effects on utility (via the envelope theorem) (see \citeauthor{hendren_unified_2020} \citeyear{hendren_unified_2020}).\footnote{This logic requires that the government is increasing spending on job training by a sufficiently small amount.} 
Thus, we argue that the within-firm wage effect is the more relevant causal effect for evaluating welfare effects.



The second contribution of this paper is to derive sharp bounds on the within-firm wage effect. Our bounding approach proceeds in two steps. In the first step, we derive sharp closed-form bounds on the \emph{response type} probabilities.\footnote{The \emph{response type} represents the pair of firms that an individual would choose to work at if she were externally assigned to the control group or the treatment group, respectively.} In deriving these bounds, we exploit a unique feature of our setting, which is that (unlike in the traditional instrumental variables framework), the exclusion restriction does not hold, since job training can have a direct causal effect on the outcome (wages). We show that this feature implies that the distribution of response types does not depend on the outcome (wage) distribution and allows us to derive closed-form bounds on the distribution of response types.\footnote{While we derive closed-form bounds on the distribution of response types, we show that one can obtain them equivalently using a linear programming approach.}
The second step provides closed-form bounds on the treatment effects, as a function of the sharp bounds on the response types derived in the first step. This step involves extending the \cite{horowitz_identification_1995} approach (which involves a single-equation mixture model with two components) to our setting which involves two mixture model equations with unknown weights that are interdependent across the equations. 
Importantly, we show that while this two-step approach provides an easy and tractable way to construct closed-form bounds, it does not entail any loss of information and provides sharp bounds. 
We also consider a set of additional restrictions on response types and show that they naturally lead to tighter bounds on the treatment effects of interest. Using a numerical illustration, we demonstrate that our bounds can be informative enough to potentially separate cases with a positive within-firm effect from those with no within-firm effect, even when both cases lead to strictly positive Lee bounds.

 We consider two empirical applications. Our first application is based on the randomized evaluation of Job Corps following \cite{lee_training_2009}. We classify firms into observable firm types taking advantage of the fact that in the publicly available survey data there are direct measures of firm amenities, such as the availability of health insurance, paid vacation and retirement, or pension benefits. We show that, on average, firms that offer amenities pay higher wages than firms that do not. Moreover, the wage distribution for amenity-offering firms stochastically dominates the wage distribution for non-amenity-offering, within both the treatment and control groups. We then go on to show that being randomly assigned to Job Corps led individuals to work at firms with better job amenities compared to the control group. This combined evidence suggests that selection is multilayered and motivates our implementation of sharp bounds to evaluate Job Corps. We begin by replicating the findings of \cite{lee_training_2009}. 
 Our estimates reveal that even in the cases where 
conventional Lee bounds are strictly positive 
($[0.047,0.048]$, 90 weeks after training),
our multilayered bounds for the within-firm wage effect (which hold the sorting effect constant) include 0. 
 This suggests that Lee bounds may capture a pure sorting effect of job training rather than a direct human capital effect. Indeed, under additional plausible assumptions on the distribution of response-types, our bounds on the within-firm wage effect become tight but continue to contain 0. 

Our second empirical application is based on the WorkAdvance experiment recently examined in \cite{katz_why_2022}. This is a sectoral employment program that targets high-quality jobs in specific industries with strong labor demand. We classify firms based on whether a firm operates in a WorkAdvance targeted sector. Similar to Job Corps, wages are higher at target sector firms and random assignment to job training increases sorting to these firms. We again document a case where Lee bounds are strictly positive and tight, but our within-firm wage effect includes 0.

The remainder of the paper is organized as follows. Section \ref{sec:AF} considers the multilayered sample selection problem in both a parametric model along the lines of \cite{heckman_sample_1979} and a more general treatment effects framework; it defines the key causal estimands of interest. Section \ref{Sec: Causal} considers the causal interpretation of Lee bounds in the presence of multilayered sample selection and presents the general decomposition. Section 4 derives the sharp bounds of a large class of parameters of interest in the multilayered sample selection model.  Section \ref{sec:emp_jobcorps} presents empirical applications that implement the sharp bounds for Job Corps and WorkAdvance. The last Section concludes. The Appendices contain simulations of the model assuming that there are two types of firms, and presents all the proofs for the paper.

\subsection*{Related Literature}
Our paper builds on and contributes to the following literature. First, there is a large literature on active labor market programs that is reviewed in \cite{heckman_economics_1999} and \citeauthor{card_active_2010} (\citeyear{card_active_2010}, \citeyear{card_what_2018}). Our contribution to this literature is to examine whether, and to what extent, worker sorting to firms affects the wage impacts of job training. To our knowledge, there are only a few empirical analyses that have examined the impact of training on worker sorting to firms. \cite{andersson_does_2022} find suggestive evidence of a positive impact of training on firm characteristics, as well as effects on industry of employment. Another related study is \cite{katz_why_2022}, who evaluate sector-based training programs. Examining evidence from randomized evaluations of programs that combine upfront screening, occupational and soft skills training, wrap-around services, and targeted low-wage workers, \cite{katz_why_2022} find substantial and persistent earnings gains after training. With regard to mechanisms, 
they 
interpret the earnings gain as driven in part by the sorting of workers to higher-paying industries and occupations. They do not, however, provide a framework for isolating the sorting effect as a causal mechanism. Finally, \cite{schochet_does_2008} evaluate the impact of Job Corps on the sorting of workers to jobs with different amenities, such as the availability of health insurance, and retirement or pension benefits, and report positive impacts. However, 
they do not disentangle the effects on these characteristics for those who would be employed in any case from a selection effect that comes from the impact of Job Corps on employment.

Second, our paper relates to the literature that has documented firm heterogeneity in wages. Firms have been shown to be important for wage inequality (\citeauthor{abowd_high_1999} \citeyear{abowd_high_1999}), the cyclicality of wages and early career progression (\citeauthor{card_workplace_2013} \citeyear{card_workplace_2013}), the earnings losses of displaced workers (\citeauthor{lachowska_sources_2020} \citeyear{lachowska_sources_2020}; \citeauthor{schmieder_costs_2023} \citeyear{schmieder_costs_2023}), and gender (\citeauthor{card_bargaining_2016} \citeyear{card_bargaining_2016}) and racial wage gaps (\citeauthor{gerard_assortative_2021} \citeyear{gerard_assortative_2021}). Our contribution to this literature is to examine the role of firms in understanding the wage effect of job training.

Third, our paper relates to econometric approaches that address the sample selection problem. The \cite{heckman_sample_1979} sample selection model has been extended in various dimensions. First, a series of papers, including \cite{gallant_semi-nonparametric_1987}, \cite{newey_semiparametric_1990}, and \cite{ahn_semiparametric_1993}, propose estimation and inference methods that relax the normality assumption imposed by \cite{heckman_sample_1979}; see \citeauthor{li_nonparametric_2007} (\citeyear{li_nonparametric_2007}, Chapter 10) for a review of such extensions. Second, \cite{lee_training_2009} extends \cite{heckman_sample_1979} by relaxing the exclusion restriction of instrumental variables and derives bounds on the parameters of interest. \cite{honore_selection_2020} study a semiparametric version of Lee's model. Additionally, \cite{semenova_generalized_2020} and \cite{olma_nonparametric_2021} propose various approaches for inference on Lee's bounds conditional on (potentially continuous) covariates. To our knowledge, this paper represents the first attempt to extend the seminal \cite{heckman_sample_1979} sample selection model to multilayered settings. Although the focus of our paper is primarily on firms, which we consider to be the main layer of interest, our analysis can be extended in various directions. For example, one could consider occupation as a layer and examine the returns to occupation while controlling for sorting, similar to the approach taken by \cite{gottschalk_taking_2014}.


Finally, one can view the firm as a ``mediator'' in the context of the literature on mediation analysis (see, for example, \cite{robins_identifiability_1992} and \cite{pearl_direct_2001}).
Traditionally, most of this literature ignores sample selection where the outcome is not observed at some mediator values. For example, recently \cite{kwon2024testing} developed a method to test for the presence of a mediator but abstracts from sample selection. 
A rare exception is \cite{zuo_mediation_2022}, who consider identification of direct and indirect effects within a mediation analysis framework, when both the outcome and the mediator are missing. They focus on point identification under various assumptions, including the abstract and non-falsifiable assumption of completeness; however, many of their assumptions do not apply to our setting.\footnote{For an in-depth review of completeness, see \cite{dhaultfoeuille_completeness_2011}, and also \cite{canay_testability_2013}.} For instance, in our setting, whether the wage is observed can depend on wage rate offered by firms (the mediator), and this situation is ruled out by their assumptions. Our paper therefore complements \cite{zuo_mediation_2022} by establishing partial identification of the direct and indirect effects without imposing completeness. 
Our maintained assumptions are transparent and directly apply to the primitives of our model. Moreover, our approach accommodates an endogenous mediator and allows the outcome to be missing non-randomly, even conditional on covariates.

\section{Analytical Framework}\label{sec:AF}
 \subsection{Multilayered Sample Selection: A parametric model}
Since Heckman's seminal work in 1979, the sample selection model has been formalized as follows:

\begin{eqnarray}
Y&=& 
\begin{cases}
\alpha Z+ \beta X^{'} + U, \;\;\; \text{ if } D=1,\\
\\ \\
\text{unobserved}, \;\;\;  \;\;\;\;\;\; \text{ if } D=0,\\ 
\end{cases} \label{eq: Heck1}\\
D&=&1\{\eta Z + \theta X^{'} + V>0\},\label{eq: Heck2}
\end{eqnarray}
where  $D$ captures the binary sample selection model (i.e., $D$ is equal to $1$ if employed and $0$ if not) and $Y$ represents the outcome which is observed only when $D$ is equal to $1$ (i.e., $Y$ is the observed wage when employed). The latent variables in the model are denoted as $(U, V)$, $X$ is a vector of observed exogenous covariates, and $Z \in \{0,1\}$ is a binary variable that respects the conditional independence assumption: $(U,V) \perp Z|X$ (i.e., job training, $Z$, is randomly assigned).  In the Heckman sample selection model, identification of the parameters in the outcome equation requires at least one variable that is independent of the latent variables but is excluded from the outcome equation. In model (\ref{eq: Heck1}, \ref{eq: Heck2}) this is equivalent to assuming $\alpha$ to be equal to $0$, implying that $Z$ is excluded from the outcome equation. In this context, $Z$ becomes a valid instrumental variable to consistently  estimate $\beta$, satisfying both the
independence and exclusion restriction conditions.  

However, as highlighted in \cite{lee_training_2009} and recognized more generally, the exclusion restriction can be violated in certain cases implying that $\alpha \neq 0$. Moreover, in such cases, $\alpha$ is potentially a parameter of primary interest. For instance, in the job training example, participating in training could boost an individual's human capital and directly affect their wage rate. Consequently, $\alpha$ can be interpreted as the causal effect of job training on the wage rate, which is the key parameter studied by \cite{lee_training_2009}. The primary methodological contribution of Lee is to provide a method that allows researchers to partially identify $\alpha$ in the presence of sample selection (i.e., when training can affect labor supply through $\eta$). 

A key assumption in the parametric model above is that the sample selection problem is binary: individuals are either employed or unemployed. Yet if job training affects not only whether an individual works but also which firm they work at, the sample selection problem becomes multilayered. 
We now generalize the seminal Heckman sample selection model (\ref{eq: Heck1}, \ref{eq: Heck2}) to allow for a richer model of labor supply where individuals choose layers, i.e., firms. We refer to this extended model as the ``\textit{parametric multilayered selection model}":

  \begin{eqnarray}
Y&=& 
\begin{cases}
\alpha_K Z+ \beta_K X^{'}+ U_K, \;\;\;\;\;\; \text{ if } D=K,\\
\;\;\;\;\;\; \;\;\;\;\;\; \vdots \;\;\;\;\;\; \;\;\;\;\;\; \qquad \qquad \qquad \vdots \\
\alpha_1 Z+ \beta_1 X^{'} + U_1, \;\;\; \;\;\;\;\;\;\; \text{ if } D=1,\\ \\
\text{unobserved}, \;\;\;\;\;\;\; \;\;\;\;\;\;\; \;\;\;\;\; \text{ if } D=0,\\
\end{cases} \label{eq: GHeck1}\\
D&=&\text{arg}\max_{d \in \{0,1,...,K\}}\{\eta_{d} Z + \theta_d X^{'}+V_{d}\} \label{eq: GHeck2}
\end{eqnarray}
where $\eta_{0} Z + \theta_0 X^{'}+V_{0}=0$. 
In this model, each layer ($D$) represents a distinct firm, with corresponding parameters $\alpha_d$, $\beta_d$, and latent variable $U_d$. Expected utility for a given firm $d$ is given by $\eta_{d} Z + \theta_d X' + V_{d}$. The utility of the outside option (i.e., unemployment) is $\eta_{0} Z + \theta_0 X' + V_{0}=0$. The worker selects the firm with the highest expected utility.

In the parametric multilayered sample selection model, 
$\alpha_d$ is the causal effect of job training on the wage rate at firm $d$.  We refer to this causal effect as the ``within-firm effect'' for layer $d$. The vector $(\eta_1,...,\eta_K)$ includes parameters that reflect the causal effect of job training on firm choice. 
In the next section, we demonstrate that Lee bounds combine both the within-firm effects $(\alpha_1,...,\alpha_K)$ with the sorting effects summarized by $(\eta_1,...,\eta_K)$. As discussed in the Introduction, it is interesting to separately identify these causal effects. For example, it is straightforward to show that the change in utility associated with a small increase in spending on job training is captured by the mechanical increase in wages. The sorting effect has only a second-order effect on worker utility due to the envelope theorem.



\subsection{Multilayered Sample Selection: Generalized version using the potential outcome model}

Let $(\Omega,\mathcal F, P)$ be a probability space, where we interpret $\Omega$ as the population of interest, and $\omega\in \Omega$ as a generic individual in the population.
Let $Y_{z,d}(\omega)$  be the potential outcome (i.e., potential wage)  if agent $\omega$ is externally assigned to the treatment group $z \in \{0,1\}$ (i.e., job training) and to a specific layer $d \in \{0,1,...,K\}$, where $d=0$ denotes the layer for which the outcome is not observed (i.e., $Y_{z,0}(\omega)$ is not observed).\footnote{In our empirical application, we will assume that the layer corresponds to a firm's type, where the type is constructed based on a firm's observable characteristics.}
$D_z$ denotes the potential layer that the individual selects if externally assigned to the treatment group $z \in \{0,1\}$.
We denote the realized outcome by $Y \in \mathcal Y \subseteq \mathbb R$ and the realized layer by $D \in \{0,1,...,K\}$. Finally, let $Z$ be the assigned treatment group and let $X$ be a vector of covariates. 
We assume that $(D,Z,X)$ is observed for every individual, but the realized outcome $Y$ is only observed if $D\neq 0$ (i.e., if the individual is employed). This implies the following model:\footnote{Strictly speaking, equation (\ref{GPOM1}) implies that the outcome $Y_{z,d}(\omega)=0$ when $D=0$, but it should really be interpreted as $Y_{z,d}(\omega)$ is unobserved.}
\begin{eqnarray}
Y&=& \sum_{d=1}^{K}\left[Y_{1,d}Z+Y_{0,d}(1-Z)\right]1\{D=d\}, \label{GPOM1}\\
D&=&D_1Z+D_0(1-Z),\label{GPOM2}
\end{eqnarray}
along with the following conditional independence assumption.
\begin{assumption}[Conditional Random Assignment] \label{Ass:RCT} Individuals are randomly assigned to a treatment group.
$\left\{(Y_{z,d}, D_z):  d \in \{0,1,...,K\}, z \in \{0,1\} \right\}  \perp Z |X.$ 
\end{assumption}

The outcome equation (\ref{GPOM1}) collapses to the outcome model of \cite{lee_training_2009} when there is no heterogeneity between the potential layered outcomes, that is, $Y_{z,d}=Y_{z}$ for $d \in \{1,...,K\}$.
In this case, we have:
\begin{multline*}
Y=\sum_{d=1}^{K}\left[Y_{1,d}Z+Y_{0,d}(1-Z)\right]1\{D=d\}=\left[Y_{1}Z+Y_{0}(1-Z)\right]\sum_{d=1}^{K}1\{D=d\}\\=\left[Y_{1}Z+Y_{0}(1-Z)\right]1\{D\neq 0\}.
\end{multline*}

What causal interpretation should be given to Lee bounds in the presence of multilayer sample selection, where $Y_{z,d}\neq Y_{z,d'}$, for $d,d' \in \{1,...,K\}$? Depending on the researcher's interest, various causal estimands of interests could be defined. Before defining our parameters of interest, we show that there is a link between the causal effects in our multilayered framework and those typically considered in the mediation analysis literature. We use this link to characterize our key estimands below.

\subsection{Direct and indirect effects in presence of sample selection}

In our model, particularly in equation (\ref{GPOM1}), there is a notable connection to the literature on mediation analysis, as discussed by \cite{pearl_direct_2001} and others. The graphical representation of the outcome equation in our model takes the form:\footnote{For the sake of clarity, this graph simplifies the discussion by omitting sample selection.}



\begin{figure}[!h]
    \centering
    \tikzset{every picture/.style={line width=0.75pt}} 

\begin{tikzpicture}[x=0.75pt,y=0.75pt,yscale=-1,xscale=1]
\draw    (35,60) -- (119.5,60) ;
\draw [shift={(122.5,60)}, rotate = 180] [fill={rgb, 255:red, 0; green, 0; blue, 0 }  ][line width=0.08]  [draw opacity=0] (8.93,-4.29) -- (0,0) -- (8.93,4.29) -- cycle    ;
\draw    (155,60) -- (239.5,60) ;
\draw [shift={(242.5,60)}, rotate = 180] [fill={rgb, 255:red, 0; green, 0; blue, 0 }  ][line width=0.08]  [draw opacity=0] (8.93,-4.29) -- (0,0) -- (8.93,4.29) -- cycle    ;
\draw    (192.5,135) -- (139.62,82.12) ;
\draw [shift={(137.5,80)}, rotate = 45] [fill={rgb, 255:red, 0; green, 0; blue, 0 }  ][line width=0.08]  [draw opacity=0] (8.93,-4.29) -- (0,0) -- (8.93,4.29) -- cycle    ; 
\draw    (192.5,135) -- (245.38,82.12) ;
\draw [shift={(247.5,80)}, rotate = 135] [fill={rgb, 255:red, 0; green, 0; blue, 0 }  ][line width=0.08]  [draw opacity=0] (8.93,-4.29) -- (0,0) -- (8.93,4.29) -- cycle    ;
\draw    (24,41) .. controls (79.65,2.58) and (197.88,2.98) .. (250.16,42.18) ;
\draw [shift={(252.5,44)}, rotate = 218.8] [fill={rgb, 255:red, 0; green, 0; blue, 0 }  ][line width=0.08]  [draw opacity=0] (8.93,-4.29) -- (0,0) -- (8.93,4.29) -- cycle    ;

\draw (182,143.4) node [anchor=north west][inner sep=0.75pt]    {$W$};
\draw (12,52.4) node [anchor=north west][inner sep=0.75pt]    {$Z$};
\draw (129,52.4) node [anchor=north west][inner sep=0.75pt]    {$D$};
\draw (248,52.4) node [anchor=north west][inner sep=0.75pt]    {$Y$};
\end{tikzpicture}
    \caption{DAG of causal relationships between variables in our model.}
    \label{fig:DAGmain}
\end{figure}


In the context of mediation analysis, where $Z$ represents the randomized treatment, $D$ is conceptualized as the ``mediator", and $Y$ denotes the outcome, our model allows the treatment (job training) to influence the outcome through two channels: a direct channel and an indirect channel that passes through the mediator. $W$ is a vector of latent unobserved variables often called ``confounding variables" which simultaneously affect $D$ and $Y$, making $D$ an endogenous variable. In the parametric model, $W\equiv (U_1,...,U_K,V_1,...,V_K)$.
In our framework, the mediator corresponds to the firm where the individual would be employed if they were externally assigned to job training.

In the mediation analysis literature, two categories of causal estimands have garnered attention: ``direct effects" and ``indirect effects". Focusing on the former, two types of ``direct effects" have been conceptualized. First, the \textit{control direct effect} (CDE) is defined as:
\begin{eqnarray}\label{CDE}
\text{CDE}(d)\equiv \mathbb E[Y_{1,d}-Y_{0,d}].
\end{eqnarray}
This captures the causal effect of job training on earnings within a specific firm $d$ when the firm is held fixed. It is equivalent to the within-firm wage effect for layer $d$. In \citeauthor{lee_training_2009}'s (\citeyear{lee_training_2009}) terminology, the CDE corresponds to the  causal impact of job training on the wage rate illustrated by the curved arrow in Figure \ref{fig:DAGmain}. This parameter is the primary focus of \cite{lee_training_2009}.

The CDE can vary significantly across firms ($d$), reflecting the potentially heterogeneous impact of job training on wages for different firms. CDEs are useful when the policymaker is primarily interested in the impact of job training on wages at a specific firm. More generally, policymakers may also be interested in understanding the overall impact of training on wages at firms that workers naturally choose when they receive training. The second type of direct effect -- the ``natural direct effect" (NDE) -- is introduced to capture this notion:
\begin{eqnarray}\label{DE}
\text{NDE}\equiv\mathbb E[Y_{1,D_1}-Y_{0,D_1}]=\sum_{d=0}^K\mathbb E[Y_{1,d}-Y_{0,d}|D_1=d]\times \mathbb P[D_1=d]
\end{eqnarray}

\noindent where $Y_{z,D_{z'}} \equiv\sum_{d=0}^{K} Y_{z,d}1\{D_{z'}=d\}$ for $z,z' \in \{0,1\}$, and $d \in \{0,...,K\}$. 
 The expression $Y_{1,D_1}(\omega)-Y_{0,D_1}(\omega)$ represents the causal impact of job training on wages for the specific firm that the worker $\omega$ would have selected if she had been externally assigned to receive job training. The NDE is essentially the average of these individual effects.

Turning to indirect effects, the ``natural indirect effect" (NIE) is defined as:
\begin{eqnarray}\label{IE}
\text{NIE}&\equiv& \mathbb E[Y_{0,D_1}-Y_{0,D_0}]\nonumber \\&=&\sum_{d=0}^{K} \sum_{d'=0:d'\neq d}^{K}
\mathbb E[Y_{0,d}-Y_{0,d'}| D_0=d',D_1=d]\times \mathbb P[D_0=d',D_1=d].
\end{eqnarray}

\noindent The term $Y_{0,d}-Y_{0,d'}$ represents the wage contrast between firms $d$ and $d'$ in the absence of job training. However, rather than specifying the pair of firms $(d, d')$, we can examine this wage difference at the ``natural representative" firms $D_1$ and $D_0$, resulting in $Y_{0,D_1}-Y_{0,D_0}$. The indirect effect aims to capture the causal impact of job training on the outcome purely due to the change in firms, and can therefore be seen as the influence of job training transitioning through a change in the firm. As highlighted by \cite{pearl_causal_2009}, the empirical relevance of the indirect effect estimand is controversial and questionable. Implementing an intervention that would suppress the direct effect of $Z$ on $Y$ while allowing the indirect channel through $D$ is not realistic. However, it remains a key parameter in the mediation analysis literature.

In the context of sample selection, the outcome is observable only when $D \neq 0$. We can categorize the population into four major groups: $\Omega=\{\omega: D_0(\omega)=0, D_1(\omega)=0\} \cup \{\omega: D_0(\omega)>0, D_1(\omega)=0\} \cup \{\omega: D_0(\omega)=0, D_1(\omega)>0\} \cup \{\omega: D_0(\omega)>0, D_1(\omega)>0\}$. For the first group we never observe outcomes, regardless of training status. For the second group only outcomes under job training are never observed, whereas for the third group only outcomes when not assigned to job training are never observed. If we are unwilling to assume that the outcome is missing at random (or selection on observables only) or impose parametric assumptions, the observed data cannot provide information on the causal effect of job training for individuals belonging to those groups. We refrain from imposing such stringent restrictions and focus solely on the causal effects for the final group, the subpopulation $\{\omega: D_0(\omega)>0, D_1(\omega)>0\}$.

It is useful to further divide our population of interest $\{\omega: D_0(\omega)>0, D_1(\omega)>0\}$ into finer groups, which we label \emph{response types}, i.e. $\{\omega: D_0(\omega)>0, D_1(\omega)>0\}=\cup_{\{d,d'\in \{1,..., K\}\}}\{\omega \in \Omega: D_1(\omega)=d, D_0(\omega)=d'\}$.\footnote{See \cite{heckman_unordered_2018} for a more detailed discussion on the advantages of such a partition.}
Response types are defined by the pair of firms that individual $\omega$ would choose to work for if externally assigned to the control group or the treatment group. Formally, the response type is defined as the random variable $T=(D_0, D_1)$ and $\mathcal T$ represents its support.

We now introduce two pivotal parameters, the Local Controlled Direct Effect (LCDE) and the Local Controlled Indirect Effect (LCIE):
\begin{eqnarray}\label{LCDE}
    \text{LCDE}(d|t)=\mathbb E[Y_{1,d}-Y_{0,d}|T=t], d \in \{1,...,K\}, \text{ and } t \in \mathcal T
\end{eqnarray}
and 
\begin{eqnarray}\label{LCIE}
    \text{LCIE}(z,d,d'|t)=\mathbb E[Y_{z,d}-Y_{z,d'}|T=t], d \in \{1,...,K\}, \text{ and } t \in \mathcal T
\end{eqnarray}

The individual CDEs may vary between individuals, i.e. $Y_{0,d}(\omega)-Y_{0, d'}(\omega) \neq Y_{0,d}(\omega')-Y_{0, d'}(\omega')$ for $\omega \neq \omega'$. By considering the LCDE, we allow for heterogeneity of the CDE  across response types. 
Note that in certain instances, a specific LCDE may be more policy-relevant than the CDE itself. Both the LCDE and CDE exhibit their own policy relevance, akin to the extensive debate in the instrumental variable (IV) literature regarding the empirical relevance between the average treatment effect (ATE) and local ATE (LATE). This analogy extends to the LCIE.

We now show that the ``sample selection" versions of the CDE, NDE, and NIE can be perceived as a weighted average of $\text{LCDE}(d|t)$ or $\text{LCIE}(z,d, d'|t)$. By ``sample selection'' version, we mean that the effects are defined conditionally on being always employed, i.e. $\{\omega: D_0(\omega)>0, D_1(\omega)>0\}$. 

\begin{multline*}
\hspace{-25pt} \mathbb E[Y_{1,d}-Y_{0,d}|D_0>0,D_1>0]=\sum_{l=1}^{K}\sum_{l'=1}^{K}\text{LCDE}(d|l,l')\times \mathbb P[T=(l,l')|D_0>0,D_1>0],
\\
\hspace{-25pt}\mathbb E[Y_{1,D_1}-Y_{0,D_1}|D_0>0,D_1>0]=\sum_{d=1}^{K}\sum_{d'=1}^{K}\text{LCDE}(d|d',d)\times \mathbb P[T=(d',d)|D_0>0,D_1>0], 
\\
\hspace{-25pt}\mathbb E[Y_{0,D_1}-Y_{0,D_0}|D_0>0,D_1>0]=\sum_{d=1}^{K}\sum_{d'=1:d\neq d'}^{K}\text{LCIE}(0,d,d'|d',d)\times \mathbb P[T=(d',d)|D_0>0,D_1>0].
\end{multline*}


As illustrated, $\text{LCDE}(d|t)$ or $\text{LCIE}(z,d, d'|t)$ represent more primitive parameters compared to CDE, NDE, and NIE. This paper will focus in particular on identifying the $\text{LCDE}(d|t)$. It should be noted that, in the absence of individual heterogeneity, whenever $Y_{z,d}(\omega)=Y_{z,d}(\omega')$ for $\omega \neq \omega'$, we have $\text{LCDE}(d|t)=\text{LCDE}(d)=\alpha_d$, as in the parametric version of the model. Moreover, since the outcome is never observed when $D=0$, hereafter we use the following notation $Y_{z,D_{z'}} \equiv\sum_{d=1}^{K} Y_{z,d}1\{D_{z'}=d\}$ for $z,z' \in \{0,1\}$.

\begin{remark}
    In \cite{lee_training_2009}, 
    training assignment $Z$ is a randomly assigned \textit{treatment} that fails to satisfy the exclusion restriction required to address sample selection using  standard methods. In our more general setting, while $Z$ remains a treatment of interest, it also plays the role of being an \textit{instrument} for $D$. In some settings, the impact of $D$ on the outcome may be of independent interest, and our results also apply to such settings. Moreover, since our results apply even when there is no sample selection (i.e. $Y$ is always observed, $P(D = 0) = 0$), our approach generalizes the IV model to settings where the instrument does not satisfy the exclusion restriction. 
\end{remark}

\section{The Causal Interpretation of Lee's bounds in the presence of Multilayered Sample Selection}\label{Sec: Causal}

First, notice that the generalized multilayered selection model, i.e., equations (\ref{GPOM1}, \ref{GPOM2}) implies the following:

\begin{eqnarray}
Y&=& \left[Y_{1,D_1}Z+ Y_{0,D_0}(1-Z)\right],\label{LPOM1}\\
1\{D>0\}&=&1\{D_1>0\}Z+1\{D_0>0\}(1-Z).\label{LPOM2}
\end{eqnarray}




Additionally, Lee imposes the following monotonicity assumption:
\begin{assumption}[Conditional Lee's Monotonicity Assumption]\label{Ass:LM} We impose the following restriction: 
$\mathbb P\left[1\{D_1>0\}\geq 1\{D_0>0\}\right|X]=1$ a.s.
\end{assumption}
This assumption means that being assigned to the treatment group can never lower employment, and this applies uniformly for all agents in the population. 
In our general framework with multilayered sample selection, this assumption requires that all agents are more likely to join an employment layer when assigned to treatment. 
Since Lee's monotonicity assumption is only required to hold conditional on $X$, 
it can be modified to allow the direction of monotonicity to vary across different values of $X$. Such a modification does not present a challenge for our identification analysis, which holds $X$ fixed throughout, but inference methods need to be adapted to accommodate such an assumption, especially when $X$ is continuous. 
For further details on such adaptations, see \cite{sloczynski_when_2020} and \cite{semenova_generalized_2020}, which provide inference methods that are valid under such assumptions.

All remaining analysis, results, and assumptions should be understood as implicitly conditioning on $X = x$ for some value $x$ of the vector of observed covariates, $X$, which will generally be suppressed in notation. 

\begin{lemma}[Lee Bounds]\label{lem:lee-bound}
Under Assumptions \ref{Ass:RCT} and \ref{Ass:LM} Lee bounds set identifies the following estimand $\mathbb E[Y_{1,D_1}-Y_{0,D_0}|D_0>0,D_1>0]$:
\begin{eqnarray}
\underline{\theta}^{\ell} \leq \mathbb E[Y_{1,D_1}-Y_{0,D_0}|D_0>0,D_1>0] \leq \overline{\theta}^{\ell}
\end{eqnarray}

where
\begin{enumerate}
\item [(i)] For continuous outcome:
\begin{eqnarray}
&&\underline{\theta}^{\ell}\equiv \mathbb E[Y|D>0,Z=1, Y\leq  F^{-1}_{Y|D>0,Z=1}(p)]-\mathbb E[Y|D>0,Z=0],\\
&&\overline{\theta}^{\ell} \equiv \mathbb E[Y|D>0,Z=1, Y\geq  F^{-1}_{Y|D>0,Z=1}(1-p)]-\mathbb E[Y|D>0,Z=0],
\end{eqnarray}

\item [(ii)] For binary outcome:

\begin{eqnarray}
&&\underline{\theta}^{\ell}\equiv \max\left\{0, 1-\frac{1}{p}P[Y=0|D>0,Z=1]\right\}-\mathbb E[Y|D>0,Z=0] ,\\
&&\overline{\theta}^{\ell} \equiv \min\left\{1, \frac{1}{p}\mathbb P[Y=1|D>0,Z=1]\right\}-\mathbb E[Y|D>0,Z=0],
\end{eqnarray}

\end{enumerate}

with $F_W^{-1}(u)\equiv \inf\{w \in \mathbb R: \mathbb P(W\leq w) \geq u\}$ for $u \in [0,1]$ 
and $p\equiv\frac{\mathbb P(D>0|Z=0)}{\mathbb P(D>0|Z=1)}$.
\end{lemma}

Lemma \ref{lem:lee-bound} shows that in the presence of heterogeneous firms, Lee's identification approach bounds the estimand $\mathbb E[Y_{1,D_1}-Y_{0,D_0}|D_0>0,D_1>0]$. 
What is the causal interpretation of this estimand? The following lemma sheds light on this. 

\begin{lemma}[Decomposition]\label{lem:decomp}
Assuming the generalized multilayered sample selection model, we have the following decomposition:
\begin{itemize}
\item [(i)] General decomposition:
\begin{multline}
\mathbb E[Y_{1,D_1}-Y_{0,D_0}|D_0>0,D_1>0]\\=\underbrace{ \sum_{d=1}^{K}\sum_{d'=1}^{K}\text{LCDE}(d|d',d)\times \mathbb P[T=(d',d)|D_0>0,D_1>0]}_{\mathbb E[Y_{1,D_1}-Y_{0,D_1}|D_0>0,D_1>0]} \\ + 
\underbrace{\sum_{d=1}^{K}\sum_{d'=1:d\neq d'}^{K}\text{LCIE}(0,d,d'|d',d)\times \mathbb P[T=(d',d)|D_0>0,D_1>0]}_{\mathbb E[Y_{0,D_1}-Y_{0,D_0}|D_0>0,D_1>0]}\label{eq:decomp}
\end{multline}
\item [(ii)]  No mediation effect (No firm-specific wage rate, i.e., $Y_{z,d}=Y_{z\bullet}$) or no sorting across firms, i.e., $\mathbb P[T=(d',d)|D_0>0,D_1>0]=0$ for $d \neq d'$. 


\begin{multline}
\mathbb E[Y_{1,D_1}-Y_{0,D_0}|D_0>0,D_1>0]\\=\sum_{d=1}^{K}\sum_{d'=1}^{K}\mathbb E[Y_{1\bullet}-Y_{0\bullet}|D_1=d,D_0=d']\mathbb P(D_1=d,D_0=d'|D_0>0,D_1>0)\\=\mathbb E[Y_{1\bullet}-Y_{0\bullet}|D_0>0,D_1>0]\label{eq:decomp1}
\end{multline}
\item [(iii)] No direct effect (i.e., $Y_{z,d}=Y_{\bullet d}$).

\begin{multline}
\mathbb E[Y_{1,D_1}-Y_{0,D_0}|D_0>0,D_1>0]\\=\sum_{d=1}^{K}\sum_{d'=1:d\neq d'}^{K}\mathbb E[Y_{\bullet d}-Y_{\bullet d'}| D_0=d',D_1=d]\times \mathbb P[D_0=d',D_1=d|D_0>0,D_1>0]\\=\mathbb E[Y_{\bullet D_1}-Y_{\bullet D_0}|D_0>0,D_1>0]
\end{multline}


\end{itemize}
\end{lemma}

Lemma \ref{lem:decomp} (i) shows that in the presence of firm heterogeneity, Lee's partial identification approach establishes bounds for a total effect. This total effect combines the sample selection version of the NDE and the NIE (i.e., conditional on $D_0>0$ and $D_1>0$), with each possessing distinct interpretations. Importantly, as discussed above, the NDE and NIE do not hold the mediator (firm) $D$ fixed. The NDE is an average of causal effects of job training at each firm weighted by fraction of workers choosing that firm under job training. The NIE is an average of causal effects of firm on wages (in the no-job training counterfactual scenario) weighted by the share of response types choosing those firms. Thus, without additional assumptions, this approach does not allow one to separately identify the CDEs (the within-firm wage effects) from the labor supply effects or sorting effects that transit through $D$. Lemma \ref{lem:decomp} (ii) shows that when there are no mediation effects (or no heterogeneity in wages between firms), i.e., $Y_{z,d}=Y_{z\bullet}$
as assumed in \cite{lee_training_2009} and illustrated in Figure \ref{fig:DAGnomed},
\begin{figure}[!h]
    \centering
    \tikzset{every picture/.style={line width=0.75pt}} 

\begin{tikzpicture}[x=0.75pt,y=0.75pt,yscale=-1,xscale=1]
\draw    (35,60) -- (119.5,60) ;
\draw [shift={(122.5,60)}, rotate = 180] [fill={rgb, 255:red, 0; green, 0; blue, 0 }  ][line width=0.08]  [draw opacity=0] (8.93,-4.29) -- (0,0) -- (8.93,4.29) -- cycle    ;
\draw    (192.5,135) -- (139.62,82.12) ;
\draw [shift={(137.5,80)}, rotate = 45] [fill={rgb, 255:red, 0; green, 0; blue, 0 }  ][line width=0.08]  [draw opacity=0] (8.93,-4.29) -- (0,0) -- (8.93,4.29) -- cycle    ; 
\draw    (192.5,135) -- (245.38,82.12) ;
\draw [shift={(247.5,80)}, rotate = 135] [fill={rgb, 255:red, 0; green, 0; blue, 0 }  ][line width=0.08]  [draw opacity=0] (8.93,-4.29) -- (0,0) -- (8.93,4.29) -- cycle    ;
\draw    (24,41) .. controls (79.65,2.58) and (197.88,2.98) .. (250.16,42.18) ;
\draw [shift={(252.5,44)}, rotate = 218.8] [fill={rgb, 255:red, 0; green, 0; blue, 0 }  ][line width=0.08]  [draw opacity=0] (8.93,-4.29) -- (0,0) -- (8.93,4.29) -- cycle    ;

\draw (182,143.4) node [anchor=north west][inner sep=0.75pt]    {$W$};
\draw (12,52.4) node [anchor=north west][inner sep=0.75pt]    {$Z$};
\draw (129,52.4) node [anchor=north west][inner sep=0.75pt]    {$D$};
\draw (248,52.4) node [anchor=north west][inner sep=0.75pt]    {$Y$};
\end{tikzpicture}
    \caption{DAG when there are no  mediation effects.}
    \label{fig:DAGnomed}
\end{figure}
the NIE vanishes while the NDE reduces to the CDE which is the target parameter in Lee's framework.\footnote{Notice that Figures \ref{fig:DAGnomed} and \ref{fig:DAGnomed2} are drawn for the subpopulation of always observed, i.e. $\{\omega: D_0(\omega)>0, D_1(\omega)>0\}$.} 
Finally, Lemma \ref{lem:decomp} (iii) reveals that in the absence of a direct effect of job training on wages (i.e., $Y_{z,d}=Y_{\bullet d}$) as depicted in Figure \ref{fig:DAGnomed2},  Lee bounds capture the effect of job training on wages coming exclusively from the sorting of individuals into different firms. 


 \begin{figure}[!h]
    \centering
    \tikzset{every picture/.style={line width=0.75pt}} 

\begin{tikzpicture}[x=0.75pt,y=0.75pt,yscale=-1,xscale=1]
\draw    (35,60) -- (119.5,60) ;
\draw [shift={(122.5,60)}, rotate = 180] [fill={rgb, 255:red, 0; green, 0; blue, 0 }  ][line width=0.08]  [draw opacity=0] (8.93,-4.29) -- (0,0) -- (8.93,4.29) -- cycle    ;
\draw    (155,60) -- (239.5,60) ;
\draw [shift={(242.5,60)}, rotate = 180] [fill={rgb, 255:red, 0; green, 0; blue, 0 }  ][line width=0.08]  [draw opacity=0] (8.93,-4.29) -- (0,0) -- (8.93,4.29) -- cycle    ;
\draw    (192.5,135) -- (139.62,82.12) ;
\draw [shift={(137.5,80)}, rotate = 45] [fill={rgb, 255:red, 0; green, 0; blue, 0 }  ][line width=0.08]  [draw opacity=0] (8.93,-4.29) -- (0,0) -- (8.93,4.29) -- cycle    ; 
\draw    (192.5,135) -- (245.38,82.12) ;
\draw [shift={(247.5,80)}, rotate = 135] [fill={rgb, 255:red, 0; green, 0; blue, 0 }  ][line width=0.08]  [draw opacity=0] (8.93,-4.29) -- (0,0) -- (8.93,4.29) -- cycle    ;

\draw (182,143.4) node [anchor=north west][inner sep=0.75pt]    {$W$};
\draw (12,52.4) node [anchor=north west][inner sep=0.75pt]    {$Z$};
\draw (129,52.4) node [anchor=north west][inner sep=0.75pt]    {$D$};
\draw (248,52.4) node [anchor=north west][inner sep=0.75pt]    {$Y$};
\end{tikzpicture}
    \caption{DAG when there is no direct effect.}
    \label{fig:DAGnomed2}
\end{figure}

This shows that, in general, interpreting Lee bounds as informative about the direct wage effect of job training is problematic, unless there is clear empirical evidence of the absence of mediation effects. Unfortunately, Lee's approach does not provide a means to assess this. Given these challenges, the next section introduces an alternative partial identification approach that is designed to overcome these limitations and aims to partially identify the true causal impact of job training on wage rates.

\section{Sharp bounds in the multilayered sample selection model}\label{Sec:Sharp}

In this section, we develop a partial identification strategy to recover the parameters $\text{LCDE}(d|t)$ and $\text{LCIE}(z,d, d'|t)$ that will allow us to isolate the within-firm effect of job training from the sorting effect.
Under Assumption \ref{Ass:RCT}, the response type $T$ is independent of $Z$. Assumption \ref{Ass:LM} restricts the response type support. For example, under Assumption \ref{Ass:LM}, $\mathbb P[T=(d,0)]=0$ for $d \in \{1,...,K\}$. We denote by $f_{Y_{z,d}|D,Z}(y|d',z')$ the conditional density of $Y_{z,d}$ given $\{D=d',Z=z'\}$ and assume that it is absolutely continuous with respect to a dominating measure $\mu$ on $Y_{z,d}$. We note that $f_{Y_{z,d},D|Z}(y,d|z)\equiv f_{Y_{z,d}|D,Z}(y|d,z)\mathbb P(D=d|Z=z)$.
 For  $d,d' \in \{1,...,K\}$ and  $z \in \{0,1\}$, and any $y \in \mathcal Y$ we have the following:
\begin{multline}
\hspace{-1.5em} f_{Y|D=d,Z=z}(y)=f_{Y_{z,d}|D_z}(y|d)=\sum_{d'=1}^K \frac{\mathbb P(D_z=d,D_{1-z}=d')}{\mathbb P(D=d|Z=z)}\times f_{Y_{z,d}|D_z,D_{1-z}}(y|d,d')
\end{multline}
where the first equality holds under Assumption \ref{Ass:RCT}. 

More precisely, under Assumption \ref{Ass:RCT}, the following system of equations characterizes the empirical content of the multilayered sample selection model.

\begin{eqnarray}
f_{Y,D=d|Z=1}(y)&=&\sum_{d'=0}^K \mathbb P[T=(d',d)]\times f_{Y_{1,d}|T}(y|d',d) \label{EC1}\\
f_{Y,D=d|Z=0}(y)&=&\sum_{d'=0}^K \mathbb P[T=(d,d')]\times f_{Y_{0,d}|T}(y|d,d')   \label{EC2}
\end{eqnarray}

\noindent and this holds for any $d,d' \in \{1,...,K\}$ and $y \in \mathcal Y$. 
The left-hand side of equations (\ref{EC1}) and (\ref{EC2}) are observed while the individual types, i.e., $\mathbb P[T=(d,d')]$ and the conditional potential outcome distributions, i.e.,  $f_{Y_{z,d}|T}(y|d',d)$  on the right-hand side of the equations are unknown. For a given $d$, the number of unknown quantities $2K+1 + 2(K+1)|\mathcal Y|$ is larger than the number of equations $2 |\mathcal Y|$. We therefore have an under-determined system of linear equations with unknown coefficients. As such, it is only possible to set identify these parameters. The identified set of unknown parameters could naturally shrink if the researcher is willing to impose additional assumptions, such as Assumption \ref{Ass:LM}. 
For example, under Assumption \ref{Ass:LM}, $\mathbb P[T=(d,0)]=0$ for $d \in \{1,...,K\}$ which implies that $\mathbb P[T=(d,0)] f_{Y_{z,d}|T}(y|d,0)=0$ for $d \in \{1,...,K\}$. Consequently, for a fixed $d$, this leads to a reduction of $|\mathcal Y|+1$ in the total number of unknown parameters while keeping the number of equations fixed. As a result, the system of equations becomes more tightly constrained.
When the support of $Y$, i.e., $\mathcal Y$, is finite, the system of equations (\ref{EC1})-(\ref{EC2}) could be solved using a linear programming method, with the drawback that the linear programming approach does not provide intuition about the source of identification power.\footnote{
If the researcher is interested in analyzing a discrete outcome and wishes to explore this avenue further, she could employ the inferential method developed by \cite{fang_inference_2023}.} 
More importantly, the linear programming approach can no longer be used when $Y$ is continuous, as is the case in our empirical application. To address this issue, we develop a two-step identification approach. The first step provides sharp bounds on the response types. This step involves only the distribution on $(D, Z)$ that has finite support in our framework and can therefore be solved using a linear programming approach, since it does not involve $Y$ which could have continuous support. The second step provides closed-form bounds on the treatment effects of interest, as functions of the sharp bounds on the response types computed in the first step. We show that these two steps provide sharp bounds on our parameters of interest. 

\subsection{Step 1: Sharp bounds on response type probabilities}
\label{sec: first step}
In this step, we focus on the partial identification of the distribution of response types. Integrating equations (\ref{EC1}) and (\ref{EC2}) over 
 $\mathcal Y$,  
we obtain the following system of equations, for each $d$:

\begin{eqnarray}
\mathbb P(D=d|Z=1)&=&\sum_{d'=0}^K \mathbb P[T=(d',d)] \label{T-EC1}\\
\mathbb P(D=d|Z=0)&=&\sum_{d'=0}^K \mathbb P[T=(d,d')] \label{T-EC2}
\end{eqnarray}

In general, in the standard IV model, the distribution of response types depends on the full joint distribution of the observed data $(Y, D, Z)$, not just on the distribution of $(D,Z)$.\footnote{This has been pointed out by \cite{huber_sharp_2017} and is also implicit in the results of \cite{kitagawa_identification_2021}. See Theorem 3 in \cite{vayalinkal_sharp_2024} for a result characterizing the relationship between outcome distributions and the identified set of response-type probabilities.}
This complexity occurs because in the IV framework, the exclusion restriction is imposed, i.e., $Y_{z,d}=Y_{\bullet d}$. In fact, when this restriction is imposed, the response-type conditional density of $Y_{\bullet d}$ appears in both equations (\ref{EC1}) and (\ref{EC2}), and integrating each equation separately can lead to a loss of information on the response-type probabilities (leading to non-sharp bounds). 
However, in the absence of the exclusion restriction, each response-type conditional density $f_{Y_{z,d}|T}$ in the system of equations (\ref{EC1}) and (\ref{EC2})  only appears in one equation, so the integration step can be performed without losing any information on response-type probabilities. Therefore, we show that in our model, the 
response-type probabilities are entirely characterized by the distribution of $(D,Z)$ which justifies proceeding in two steps. 
We will say that a vector $\mathbf{v}$ satisfies (\ref{T-EC1}, \ref{T-EC2}) if $\left(\mathbb P[T=(d,d')] : d,d' \in \{0,...,K\}\right) = \mathbf{v}$ is a solution to (\ref{T-EC1}, \ref{T-EC2}) for all $d$.  

\begin{lemma}\label{Lem: RT-C} Consider the model (\ref{GPOM1}, \ref{GPOM2}). Under Assumption \ref{Ass:RCT}, 
the (sharp) identified set of response-type probabilities 
is the set of non-negative vectors that satisfy 
(\ref{T-EC1}, \ref{T-EC2}).
\end{lemma}

The researcher may also seek to apply additional restrictions on the distribution of response types, including, but not limited to, Assumption \ref{Ass:LM}. For example, one could assume that there are more upward switchers than downward switchers or more stayers than downward switchers. We consider such restrictions as a possible auxiliary assumption. We designate the set of linear constraints that can be applied to the response types as $\mathcal R_T$.  

\begin{assumption}\label{Ass:Type-rest}[Restriction on response types]
Consider that the layers (i.e.,  firms) are ordered. 
\begin{enumerate}
    \item [(i)] [Strong Monotonicity] $1\{D_1(\omega)=d\}\geq 1\{D_0(\omega)=d'\}$ for $d\geq d'$, or equivalently $\mathbb P[T=(d',d)]=0$ for  $d\geq d'$. 
\item [(ii)] [More upward switchers than downward switchers]

$\mathbb P[T=(d,d')] \geq \mathbb P[T=(d',d)]$ for $d\geq d'$.

\item [(iii)] [More stayers than downward switchers]

$\mathbb P[T=(d,d)] \geq \mathbb P[T=(d',d)]$ for $d\geq d'$.
    
\end{enumerate}

\end{assumption}

A notable aspect of Assumptions \ref{Ass:LM} and \ref{Ass:Type-rest} is that these restrictions can seamlessly integrate into equations (\ref{T-EC1})-(\ref{T-EC2}) as supplementary linear constraints. Consequently, the process of recovering response types that conform to all these behavioral restrictions simplifies to a feasible linear programming problem. 
For example, researchers may choose $\mathcal R_T = \{\text{Assumption } \ref{Ass:LM}\}$, $\mathcal R_T = \{\text{Assumption } \ref{Ass:LM}$, $
\text{Assumption } \ref{Ass:Type-rest}(i)\}$, or $\mathcal R_T = \{\text{Assumption } \ref{Ass:LM}, \text{Assumption } \ref{Ass:Type-rest}\}$.
 
As mentioned in Lemma \ref{Lem: RT-C}, equations (\ref{T-EC1}, \ref{T-EC2}) sharply characterize the restrictions on the distribution of $T$ imposed by the model (\ref{GPOM1}, \ref{GPOM2}). Therefore, the identified set for response type probabilities under model (\ref{GPOM1}, \ref{GPOM2}), Assumptions \ref{Ass:RCT}, and responses type restrictions $\mathcal R_T$, denoted $\Theta_{I}(\mathcal R_T)$, is simply the set of non-negative vectors that jointly satisfy both $\mathcal R_T$ and (\ref{T-EC1}, \ref{T-EC2}), i.e., 
$$\Theta_{I}(\mathcal R_T)\equiv \left\{ \mathbf{v} \in \mathbb{R}_{\geq 0}^{(K+1)^2}:\begin{array}{l}
    \left(\mathbb P[T=(d,d')] : d,d' \in \{0,...,K\}\right) = \mathbf{v} \text{ satisfies }  \mathcal R_T \text{ and }
    (\ref{T-EC1}, \ref{T-EC2})
\end{array} 
\right\}.$$

We also have the following result:

\begin{lemma}\label{Lem: RT-restricted} Consider the model (\ref{GPOM1}, \ref{GPOM2}). Assumption \ref{Ass:RCT} and $\mathcal R_T$ are jointly rejected by the data if and only if $\Theta_{I}(\mathcal R_T)=\emptyset$.
\end{lemma}

Lemma \ref{Lem: RT-restricted} has significant practical implications. It shows that assessing the validity of model assumptions does not depend on knowledge of the outcome distribution, and therefore, can be reduced to testing the feasibility of a linear program. 
This property greatly simplifies the implementation of a falsification test for our model.\footnote{For example, implementing a (sharp) test of our model is much simpler than the sharp tests for instrument validity and monotonicity proposed by \cite{kitagawaTestInstrumentValidity2015} and \cite{mourifieTestingLocalAverage2017}.} In other words, once we can find a distribution of type that aligns with the assumptions of the model and the observed data on $(D,Z)$, it is always possible to find a corresponding distribution of potential outcomes $f_{Y_{z,d}|T}(y|d',d)$ that would rationalize the observed joint distribution of $(Y,D,Z)$. 

For simplicity, we introduce the shorthand notation, $p_{d,d'}\equiv \mathbb P[T=(d,d')]$, and $\gamma^z_{d,d'}\equiv\frac{p_{d,d'}}{\mathbb P(D=d|Z=z)}.$
When $\Theta_{I}(\mathcal R_T)\neq \emptyset$, 
let $\underline{p}^{ r}_{d,d'}$ denote the infimum over all 
values for $p_{d,d'}$ 
induced by distributions that 
belong to $\Theta_{I}(\mathcal R_T)$. 
Subsequently, we can define: $\underline{\gamma}^{z,r}_{d,d'}=\frac{\underline{p}^r_{d,d'}}{\mathbb P(D=d|Z=z)}.$ The “$r$" superscript is used to emphasize the 
dependence of $\underline{p}^{ r}_{d,d'}$ and $\underline{\gamma}^{z,r}_{d,d'}$ on $\mathcal R_T$, 
the set of assumptions imposed.



For all the potential choices of $\mathcal R_T$ considered above, $\Theta_{I}(\mathcal R_T)$ is the set of nonnegative solutions to a linear system. Therefore, $\underline{\gamma}^{z,r}_{d,d'}$ for $d,d' \in \{0,...,K\}$ and $z \in \{0,1\}$ can be obtained as a solution to a linear program. Since the linear system of interest here is generally small, it is also possible to obtain an analytic solution for $\underline{p}^r_{d,d'}$ (and therefore for $\underline{\gamma}^{z,r}_{d,d'}$) via Fourier-Motzkin elimination. The details of both the computational and analytical approaches are presented in Appendix \ref{app:add-results}.

\subsection{Step 2: Sharp bounds on the treatment effects} \label{sec:second step}

As evident from equations (\ref{EC1}) and (\ref{EC2}), the observed conditional distribution of earnings, $F_{Y|D,Z}(y|d,z)$, can be expressed as a finite mixture of the conditional potential outcome distributions given the response types, $F_{Y_{z,d}|T}(y|l,l')$. More precisely, we have:


\begin{eqnarray}
f_{Y|D=d,Z=1}(y)&=&\sum_{d'=0}^K \gamma^1_{d',d}\times f_{Y_{1,d}|T}(y|d',d) \label{M1}\\
f_{Y|D=d,Z=0}(y)&=&\sum_{d'=0}^K \gamma^0_{d,d'}\times f_{Y_{0,d}|T}(y|d,d') \label{M2}
\end{eqnarray}

The unknowns in this mixture are the weights, $\gamma^z_{d,d'}$ for $z \in \{0,1\}$, and $d,d' \in \{0,...,T\}$.
In \cite{lee_training_2009}, Assumption \ref{Ass:LM} implies that the weights are point identified, and establishing identification reduces to recovering the mean average of the mixture components. However, in our scenario, the mixture weights are not point identified under Assumption \ref{Ass:LM}. Even when we 
strengthen Assumption \ref{Ass:LM} with Assumption \ref{Ass:Type-rest}, point identification is still not achieved. This underidentification issue arises primarily due to the presence of numerous unobserved types stemming from the multiple layers, i.e., firms. Nevertheless, we can still derive informative bounds on these weights, as elaborated in the previous subsection. 

\cite{horowitz_identification_1995} proposed sharp bounds on the distribution of mixture components in a single-equation mixture model with two components, where the weights are unknown but researchers possess non-trivial bounds for these weights, and \cite{cross_regressions_2002} extended these results to single-equation models with many components. However, the empirical content of our model is characterized by a set of systems of mixture equations, one system for each $d \in \left\{1, \dots, K\right\}$, each with up to $(K+1)^2$ components. Importantly, in our setting, the weights are unknown and shared between these systems, introducing a cross-equation dependence that is not present in \cite{horowitz_identification_1995} or \cite{cross_regressions_2002}. We derive the identified set for the weights and then extend the approaches of \cite{horowitz_identification_1995} and \cite{cross_regressions_2002} to this more general case, deriving closed-form bounds on our key parameters of interest. 
Moreover, Lee demonstrated that for continuous outcomes, the bounds proposed by \cite{horowitz_identification_1995} can be equivalently expressed as the mean of a truncated distribution. We extend Lee’s results by introducing a generalized truncated mean representation that applies regardless of the outcome distribution, whether it is continuous, discrete or mixed.

Hereafter, to simplify our notation and enhance readability, we introduce the following notation.
For any $d$ and $z$, define: $\underline{\mathbb E}_{F^{-1}_{Y|D,Z}}(\gamma;d,z)\equiv \mathbb E[F^{-1}_{Y|D=d,Z=z}(U) | U \leq \gamma]$, and $\overline{\mathbb E}_{F^{-1}_{Y|D,Z}}(\gamma;d,z)\equiv \mathbb E[F^{-1}_{Y|D=d,Z=z}(U) | U \geq 1-\gamma]$, for some $U\sim \text{Uniform}(0,1)$.
We define $y_L$ as the lower bound of the support of $Y$ and $y_U$ as the upper bound.\footnote{Note that these bounds need not be finite.} 

Before stating the main result, we note the following. When the outcome is continuously distributed 
$\underline{\mathbb E}_{F^{-1}_{Y|D,Z}}(\gamma;d,z)$ is exactly equal to the truncated mean used in \cite{lee_training_2009} i.e., $$\mathbb E[F^{-1}_{Y|D=d,Z=z}(U) | U \leq \gamma]=\mathbb E[Y|D=d,Z=z, Y\leq  F^{-1}_{Y|D=d,Z=z}(\gamma)].$$ When the outcome is binary we have$$\mathbb E[F^{-1}_{Y|D=d,Z=z}(U) | U \leq \gamma]=\max\left\{0, 1-\frac{1}{\gamma}P[Y=0|D=d,Z=z]\right\}.$$
This formulation provides a general truncation formula that applies to any type of outcomes, continuous, discrete, or mixed.

\begin{theorem}\label{Theorem:Main}
Suppose that Assumptions \ref{Ass:RCT}, and restrictions $\mathcal R_T$ hold.
Whenever $\Theta_{I}(\mathcal R_T)\neq \emptyset$, then the  following bounds are pointwise sharp:
\begin{enumerate}
    \item [(i)] Local Controlled Direct Effect (LCDE):
\begin{eqnarray}
&& \hspace{-2em}\underline{\mathbb E}_{F^{-1}_{Y|D,Z}}(\underline{\gamma}^{1,r}_{d,d};d,1) 
-\overline{\mathbb E}_{F^{-1}_{Y|D,Z}}(\underline{\gamma}^{0,r}_{d,d};d,0) 
\leq  \text{LCDE}(d|d,d) \leq 
\overline{\mathbb E}_{F^{-1}_{Y|D,Z}}(\underline{\gamma}^{1,r}_{d,d};d,1) -
\underline{\mathbb E}_{F^{-1}_{Y|D,Z}}(\underline{\gamma}^{0,r}_{d,d};d,0), \nonumber\\ \nonumber \\
&& \hspace{-2em}\underline{\mathbb E}_{F^{-1}_{Y|D,Z}}(\underline{\gamma}^{1,r}_{d',d};d,1) 
-y_U 
\leq  \text{LCDE}(d|d',d) \leq 
\overline{\mathbb E}_{F^{-1}_{Y|D,Z}}(\underline{\gamma}^{1,r}_{d',d};d,1) -
y_L, \text{ for } d\neq d', \nonumber\\ \nonumber\\
&& \hspace{-2em}y_L 
-\overline{\mathbb E}_{F^{-1}_{Y|D,Z}}(\underline{\gamma}^{0,r}_{d,d'};d,0) 
\leq  \text{LCDE}(d|d,d') \leq 
y_U -
\underline{\mathbb E}_{F^{-1}_{Y|D,Z}}(\underline{\gamma}^{0,r}_{d,d'};d,0) \text{ for } d\neq d'. \nonumber
\end{eqnarray}
\item [(ii)] Local Controlled Indirect Effect (LCIE). 
\begin{eqnarray}
&& \hspace{-2em}\underline{\mathbb E}_{F^{-1}_{Y|D,Z}}(\underline{\gamma}^{1,r}_{l,d};d,1) 
-y_U 
\leq  \text{LCIE}(1,d, d'|l,d) \leq 
\overline{\mathbb E}_{F^{-1}_{Y|D,Z}}(\underline{\gamma}^{1,r}_{l,d};d,1) -
y_L, \text{ for } d\neq d'\text{ and any } l, \nonumber\\ \nonumber\\
&& \hspace{-2em}\underline{\mathbb E}_{F^{-1}_{Y|D,Z}}(\underline{\gamma}^{0,r}_{d,l};d,0) 
-y_U 
\leq  \text{LCIE}(0,d, d'|d,l) \leq 
\overline{\mathbb E}_{F^{-1}_{Y|D,Z}}(\underline{\gamma}^{0,r}_{d,l};d,0) -
y_L, \text{ for } d\neq d', \text{ and any } l.\nonumber
\end{eqnarray}

\item [(iii)] Aggregate LCDE: 
\begin{multline}
     \hspace{-2em}
\inf_{\substack{\left\{(p_{d,d}: d \in \{1,...,K\})\right.\\\left. \in \Theta_I(\mathcal R_T)\vphantom{(p_{d,d}: d \in \{1,...,K\})}\right\}}}\sum_{d=l}^{l'} \frac{p_{d,d}}{\sum _{d'=l}^{l'}p_{d',d'}}\left[ \underline{\mathbb E}_{F^{-1}_{Y|D,Z}}\left(\frac{p_{d,d}}{\mathbb P(D=d|Z=1)};d,1 \right) 
-\overline{\mathbb E}_{F^{-1}_{Y|D,Z}}\left(\frac{p_{d,d}}{\mathbb P(D=d|Z=0)};d,0 \right)\right]  \\ \leq \sum _{d=l}^{l'} \frac{p_{d,d}}{\sum _{d'=l}^{l'}p_{d',d'}}\text{LCDE}(d|d,d) \leq \\ 
\hspace{-2em}
\sup_{\substack{\left\{(p_{d,d}: d \in \{1,...,K\})\right.\\\left. \in \Theta_I(\mathcal R_T)\vphantom{(p_{d,d}: d \in \{1,...,K\})}\right\}}}
\sum _{d=l}^{l'} \frac{p_{d,d}}{\sum _{d'=l}^{l'}p_{d',d'}}\left[ \overline{\mathbb E}_{F^{-1}_{Y|D,Z}}\left(\frac{p_{d,d}}{\mathbb P(D=d|Z=1)};d,1 \right) 
-\underline{\mathbb E}_{F^{-1}_{Y|D,Z}}\left(\frac{p_{d,d}}{\mathbb P(D=d|Z=0)};d,0 \right)\right].\nonumber
\end{multline}
\end{enumerate}
\end{theorem}
The derivation of the bounds in Theorem \ref{Theorem:Main} comes from extending the \cite{horowitz_identification_1995} bounding approach summarized in Lemma \ref{mix-lem} in Appendix \ref{app:add-results}.
However, demonstrating their sharpness presents a considerably more complex challenge. This involves showing that solving equations (\ref{EC1}) to (\ref{EC2}) for all $y \in \mathcal{Y}$ and $d \in \{1,...,K\}$ while imposing the restrictions defined in $\mathcal{R}_T$ consistently produces the same information as the closed-form bounds presented in Theorem \ref{Theorem:Main}. As explained above, the absence of the IV exclusion restrictions facilitates this result.

Theorem \ref{Theorem:Main} (i) indicates that, without additional assumptions on the potential outcome distributions, the derived bounds can best determine the direction (sign) of the within-firm effect at layer $d$ solely for individuals who remain with firm $d$ under any treatment assignment $Z$, i.e., $\mathbb E[Y_{1,d}-Y_{0,d}|T=(d,d)]$.
This finding is somewhat intuitive given that these ``stayers" are equivalent to the so-called ``always-employed" in Lee's model, where firm heterogeneity is not taken into account. On the other hand, the bounds for those who switch firms due to treatment (``switchers''), such as $\mathbb E[Y_{1,d}-Y_{0,d}|T=(d, d')]$ and $\mathbb E[Y_{1,d}-Y_{0,d}|T=(d',d)]$ for $d\neq d'$, always include $0$.
This is because the observed data $(Y,D,Z)$ do not reveal any information on the following unobserved counterfactuals $\mathbb E[Y_{0,d}|T=(d', d)]$ and $\mathbb E[Y_{1,d}|T=(d, d')]$.

Similarly, Theorem \ref{Theorem:Main} (ii) reveals that in the absence of additional restrictions on the potential outcome distributions, it is impossible to identify the sign of $\text{LCIE}(z,d,d'|t)=\mathbb E[Y_{z,d}-Y_{z,d'}|T=t]$. This underscores the inherent challenges of identifying some specific treatment effects without imposing further assumptions on the outcome distributions.

Finally, Theorem \ref{Theorem:Main} (iii) presents the closed form bounds that correspond to the weighted average of $\text{LCDE}(d|d,d)$, $\sum _{d=l}^{l'} \frac{p_{d,d}}{\sum _{d=l}^{l'}p_{d,d}}\text{LCDE}(d|d,d)$. These bounds are sharp and take into account the interdependence between equations (\ref{M1}) and (\ref{M2}). 

\begin{remark}\label{rmk:aggbnds}
   To derive bounds on the aggregate LCDE, defined as 
$\sum_{d=l}^{l'} \frac{p_{d,d}}{\sum_{d'=l}^{l'} p_{d',d'}} \text{LCDE}(d \mid d,d),$
one might be tempted to adopt a na\"ive approach by taking a weighted average of the pointwise sharp bounds derived in Theorem \ref{Theorem:Main} (i). However, this approach not only fails to provide sharp bounds on the aggregate quantity but may also yield invalid bounds. We discuss this in more detail in Appendix \ref{App:Agg}. The difference between the sharp bounds and na\"ive ``bounds'' is illustrated using simulations (\Cref{fig:res1agg,fig:res2agg}) in \Cref{sec:num}, and using our empirical applications (\Cref{fig:healthbnds,fig:ms_bounds,fig:te_bounds})
in  
\Cref{sec:emp_jobcorps}.
\end{remark}

To better illustrate our theoretical results, in Appendix \ref{sec:num},  we examine the special case with two firm types, which follows our empirical applications below.
 We derive closed-form expressions for the bounds on response-type probabilities. Plugging the resulting expression(s) into the bounds given in Theorem 1 (i)-(ii) yields analytic expressions for the bounds on the LCDEs and LCIEs.
Through a numerical illustration, we show that our bounds can be sufficiently informative to distinguish cases with a positive within-firm effect from those with no within-firm effect, even when both yield strictly positive Lee bounds.





\subsection{Inference} Inference on the causal parameters considered in \Cref{Theorem:Main} can often be performed using existing methods for inference on parameters bounded by truncated conditional expectations. The three sets of bounds in \Cref{Theorem:Main} are known functions of the $(D,Z)$ conditional expectations of $Y$ truncated above or below at particular quantiles. Inference in such settings is complicated by the need to estimate two nuisance parameters: the conditional quantile functions themselves and the truncation quantile level. 

In parts (i) and (ii) of  \Cref{Theorem:Main}, the truncation quantile levels are of the form $\underline{\gamma}^{z,r}_{d,d'}=\frac{\underline{p}^r_{d,d'}}{\mathbb P(D=d|Z=z)}$ and these can often be estimated at a faster rate than the second step (truncated conditional expectation). 
Moreover, each bound involves either only one truncated conditional expectation, or truncated conditional expectations that can be independently estimated.  Therefore, in such cases, inference on the parameters considered in \Cref{Theorem:Main}(i)-(ii) can be performed by plugging in the estimators of
\cite{lee_training_2009}, \cite{semenova_generalized_2020}, or \cite{olma_nonparametric_2021} for the truncated conditional expectation(s) and adapting the inference approaches discussed there (subject to the regularity conditions outlined there).\footnote{However, $\underline{\gamma}^{z,r}_{d,d'}$ can often be only directionally differentiable with respect to the propensity score vector, as is evident in the closed form expressions for the 2 firms type case provided in  \Cref{sec:num}. 
In such cases, inference procedures that depend on the bootstrap will need to be adapted to remain valid (see \cite{fangInferenceDirectionallyDifferentiable2019} for details).} These approaches allow for conditioning on, and aggregating over the covariates $X$, which can lead to much tighter bounds than the unconditional case, as noted by \cite{lee_training_2009}. The approach proposed in \cite{lee_training_2009} applies when $X$ is finitely supported, whereas the approaches developed by \cite{semenova_generalized_2020} and \cite{olma_nonparametric_2021} allow $X$ to be continuous. 

Inference on the ``aggregate'' parameters considered in part (iii) of \Cref{Theorem:Main} is more complicated since the truncation quantile level to be estimated depends on the solution to a higher-dimensional optimization problem involving the outcome distributions. We defer the development of estimation and inference methods for this case to future work. 






\section{Empirical Applications: Job Corps Study and WorkAdvance RCT}\label{sec:emp_jobcorps}

\subsection{Application \#1: Job Corps Study}\label{sec:jc_prog_desc}
Job Corps is the largest residential career training program in the U.S. It is free for participants and targets disadvantaged people aged 16 to 24  with the aim of helping these people become more responsible, employable, and productive citizens (\citeauthor{johnson_national_1999} \citeyear{johnson_national_1999}).
Most participants live at a local Job Corps center and complete 440 hours of academic instruction and 700 hours of vocational training. Job Corps also provides job search assistance upon participant completion of the program.
The typical participant completes the program over a span of 30 weeks. Job Corps has trained more than two million individuals since its inception under the Economic Opportunity Act of 1964. The program trains over 60,000 enrollees per year, at roughly $130$ Job Corps centers nationwide, with an estimated cost of 34,301 USD per enrollee and 57,312 USD per graduate (\citeauthor{liu_estimating_2020} \citeyear{liu_estimating_2020}).


During the mid- to late 1990s, the U.S. Department of Labor funded a randomized evaluation of Job Corps, which was implemented by Mathematica Policy Research, Inc. Existing evaluations of the Job Corps Study include \cite{schochet_national_2001}, \cite{schochet_does_2008}, \cite{lee_training_2009}, and \cite{blanco_effects_2013}. The Job Corps Study randomized 80,883 eligible individuals who applied to Job Corps for the first time between November 1994 and December 1995 into two groups: (i) 5,977 individuals into the control group who were embargoed from participating in Job Corps for three years and (ii) 74,906 individuals into the treatment group.
Of the 74,906 individuals assigned to treatment, 9,409 individuals were randomly selected for data collection and all control individuals were selected for data collection. The final sample therefore consists of 15,386 participants who were interviewed at the time of random assignment and then subsequently 12, 30, and 48 months after random assignment.

We use the publicly available data from the National Job Corps Study (\citeauthor{schochet_national_2003} \citeyear{schochet_national_2003}). 
We impose two sample restrictions to address missing values due to interview non-response and sample attrition over time. The first sample restriction, which follows \cite{lee_training_2009}, is to keep individuals who have nonmissing values for weekly earnings and hours for every week following random assignment. Restricting the sample in this manner decreases the sample size to 9,145 individuals (= 3,599 control units + 5,546 treated units). 

The second sample restriction is to keep individuals who have non-missing values of health insurance for the weeks of interest (90, 135, 180 and 208). This comes from our classification of firm type based on the provision of health insurance.\footnote{Another potentially useful classification of firm type is based on industry since Job Corps targets certain sectors. However, industry is not observed in the public use of the Job Corps data. We consider this type of classification in the next empirical application.} 
This is motivated by \cite{dey_flinn_2005} who consider a setting where workers and firms bargain over wages and health insurance and find that, on average, better productivity matches lead to higher wages and the provision of health insurance.\footnote{At the time of the National Job Corps study, there were no legal requirements for firms to provide any of these amenities and, conditional on firm provision, federal law generally prohibited discriminatory provision across workers (\citeauthor{united_states_equal_employment_opportunity_commission_federal_2009} \citeyear{united_states_equal_employment_opportunity_commission_federal_2009}). The relevant federal laws at the time of the National Job Corps Study included the following. Title VII of the Civil Rights Act of 1964; the Age Discrimination in Employment Act of 1967; Title I and Title V of the Americans with Disabilities Act of 1990 (\citeauthor{united_states_equal_employment_opportunity_commission_federal_2009} \citeyear{united_states_equal_employment_opportunity_commission_federal_2009}).} 
This restriction results in a final sample size of 6,403 individuals (= 2,540 control units + 3,863 treated units).

The three key variables of interest are employment, hourly wage, and provision of health insurance for employed individuals. We follow \cite{lee_training_2009} by defining employment in a week based on whether an individual has positive earnings in that week and defining the hourly wage in a week by dividing weekly earnings by weekly hours worked. In Appendix C, we present summary statistics for our final sample and demonstrate that, on average, firms that offer these amenities pay higher wages than firms that do not. We also show
that the wage distribution for firms that offer amenities stochastically dominates the wage distribution for firms that do not, in both the treatment and control groups. Finally, we show that being randomly assigned to Job Corps led individuals to work at firms with better job amenities compared to the control group. This combined evidence suggests that sample selection is multilayered and motivates the implementation of our sharp bounds to these data. 



\subsubsection{Multilayered Bounds for Job Corps Study}
As a first step, we replicate the bounds reported in \cite{lee_training_2009} which, as discussed above, target the parameter $\mathbb E[Y_{1,D_1}-Y_{0,D_0}|D_0>0,D_1>0]$.\footnote{In this section, certain empirical results only presented are for week 90 which follows the preferred specification in \citet{lee_training_2009}. In these cases, tables and figures for other weeks of interest (135, 180 and 208) are presented in \Cref{app:jc_app}} In \Cref{app:leebounds_rep_app} we report Lee's bounds for weeks 90, 135, 180 and 208 along with the trimming proportion $p\equiv\mathbb P(AE)$, e.g., the share of the always-employed among individuals receiving job training. Lee focuses on week 90 which produces the tightest bounds $[0.0468, 0.0484]$.\footnote{As we detail in \Cref{app:leebounds_rep_app}, these are Lee's bounds when treating $ln(\text{hourly wage})$ as a continuous variable, as we do throughout this paper. \cite{lee_training_2009} uses vingtiles of $ln(\text{hourly wage})$ that produce bounds $[0.0423,0.0428]$.}


We now consider the scenario with two firm types, denoted as $L$ and $H$. Firms are classified based on the provision of health insurance with $H$ denoting firms that offer health insurance and $L$ denoting firms that do not.\footnote{Results for classifying firms based on the provision of pension/retirement benefits and paid vacation are presented in \Cref{app:pscores,app:mlbounds}.
} As in Section 4, we always impose Assumptions 1-2 and then sequentially impose the restrictions in \Cref{Ass:Type-rest}: (i) $p_{H,H} \geq p_{H,L}$ (more stayers than downward switchers) (ii) $\displaystyle\min_t \mathbb P[T=t] = p_{H,L}$ (more upward switchers than downward switchers) and (iii) $p_{H,L} = 0$ (strong monotonicity).

\Cref{tab:healthpscores} presents the estimated propensity scores for each week of interest from the National Job Corps Study.\footnote{Our sample restriction to keep observations with non-missing amenity values for the weeks of interest drops only employed individuals. This restriction mechanically reduces the propensity scores. To ensure comparability with \cite{lee_training_2009}, we rescale our estimated propensity scores so that the probabilities of employment by treatment status are the same as those reported in \cite{lee_training_2009}.} As expected, in all weeks $\mathbb P[D>0|Z=1] > \mathbb P[D>0|Z=0]$ the treated individuals are more likely to be employed. The table also shows that $\mathbb P[D=H|D>0,Z=1] > \mathbb P[D=H|D>0,Z=0]$ so that individuals who receive Job Corps training have a higher propensity to be employed in firms that offer health insurance than individuals who do not receive Job Corps training, conditional on employment, consistent with the evidence presented in Appendix \Cref{schoct390}.

\begin{table}[htbp]
    \centering
    \caption{Job Corps: Propensity scores, by week. Health insurance amenity. \label{tab:healthpscores}}
    \begin{tabular}{l*{4}{c}}
        \toprule
        &  $\mathbb P[D=H|Z=0]$ & $\mathbb P[D=H|Z=1]$&  $\mathbb P[D=L|Z=0]$&  $\mathbb P[D=L|Z=1]$\\
        \midrule
        \input empiric_tbls_figs/pscores/health_pscores_weighted.tex
        \bottomrule
    \end{tabular}
\end{table}

\begin{figure}[htbp]
    \centering
    \includegraphics[width=\textwidth]{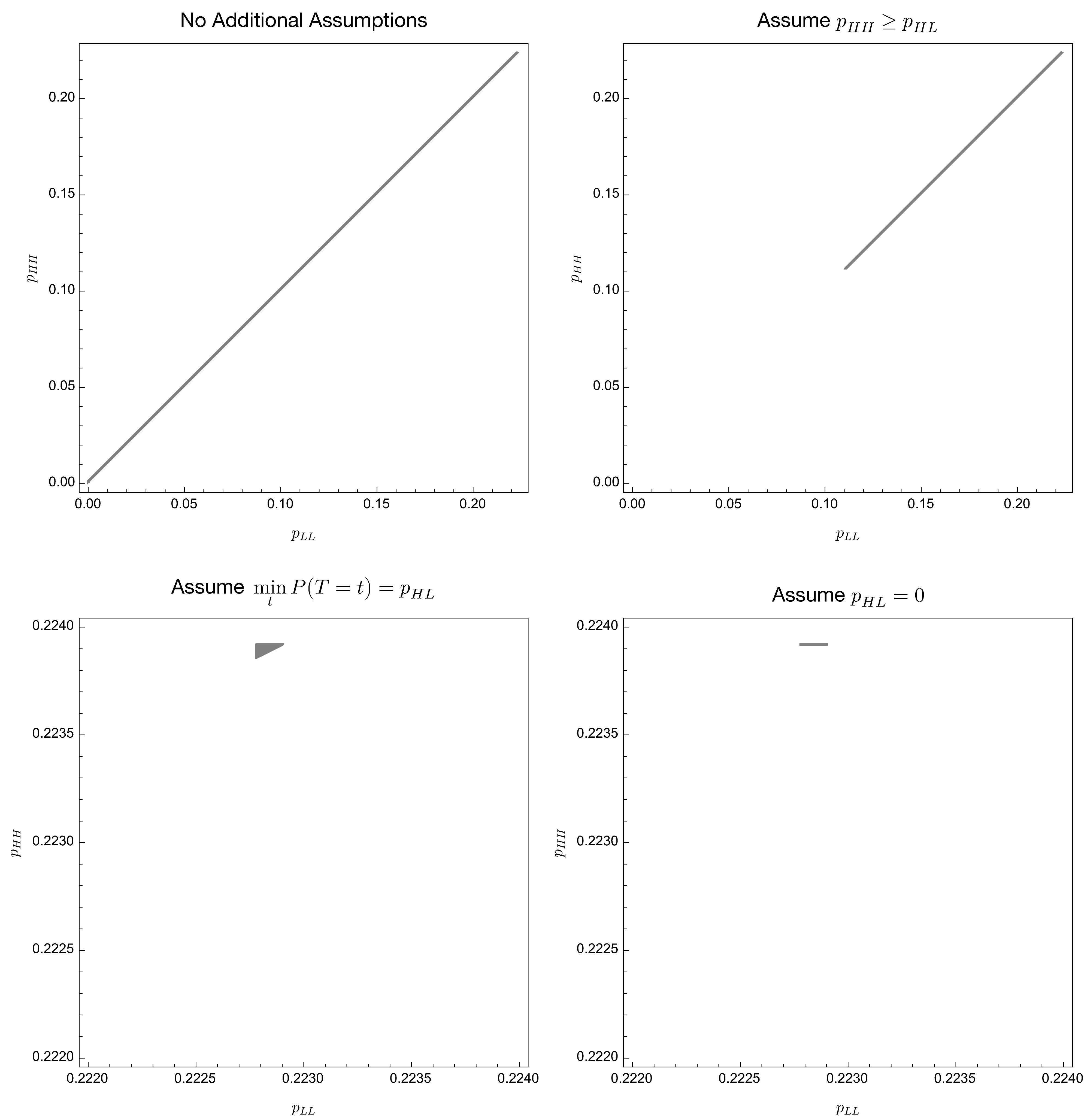}
    \vspace{0.1em}
    
               \begin{minipage}{\textwidth}
    \footnotesize
    {Notes: The first panel illustrates the identified set when $\mathcal R_T = \{\text{Assumption } \ref{Ass:LM}\}$. The remaining panels illustrate the identified set of additional assumptions imposed in addition to \Cref{Ass:LM}. In the two bottom panels of the figure, we shrink the scale of the axes to make the identified set visible.}
    \end{minipage}    
    \caption{Job Corps: Identified set for $(p_{L,L},p_{H,H})$, week 90. Health insurance amenity. 
    }
    \label{fig:health90idset}
\end{figure}

Using the week 90 propensity scores from \Cref{tab:healthpscores}, \Cref{fig:health90idset} presents the identified set for $\left(p_{L,L},p_{H,H}\right)$. Naturally, incorporating additional restrictions on the response types leads to a tightening of the identified sets. Having characterized the identified set of response-type probabilities, we now present our multilayered bounds. 

Recall that under \Cref{Ass:LM}, the support of possible response types is as follows:
\begin{align*}
\mathcal T := \left\{ \left(0,0\right), \left(0,L\right),\left(0,H\right),\left(L,L\right), \left(H,H\right),\left(L,H\right),\left(H,L\right)\right\}.
\end{align*}

\noindent The always-employed (AE) definition used in \cite{lee_training_2009} therefore combines four different response types: $\{D_0>0, D_1>0\}=\left\{\left(L, L\right), \left(H, H\right),\left(L, H\right),\left(H, L\right)\right\}\equiv AE$. We focus on the bounds for stayers, defined as the response types $(H,H)$ and $(L,L)$.

\begin{figure}[hbtp]
     \centering
\includegraphics[width=\textwidth]{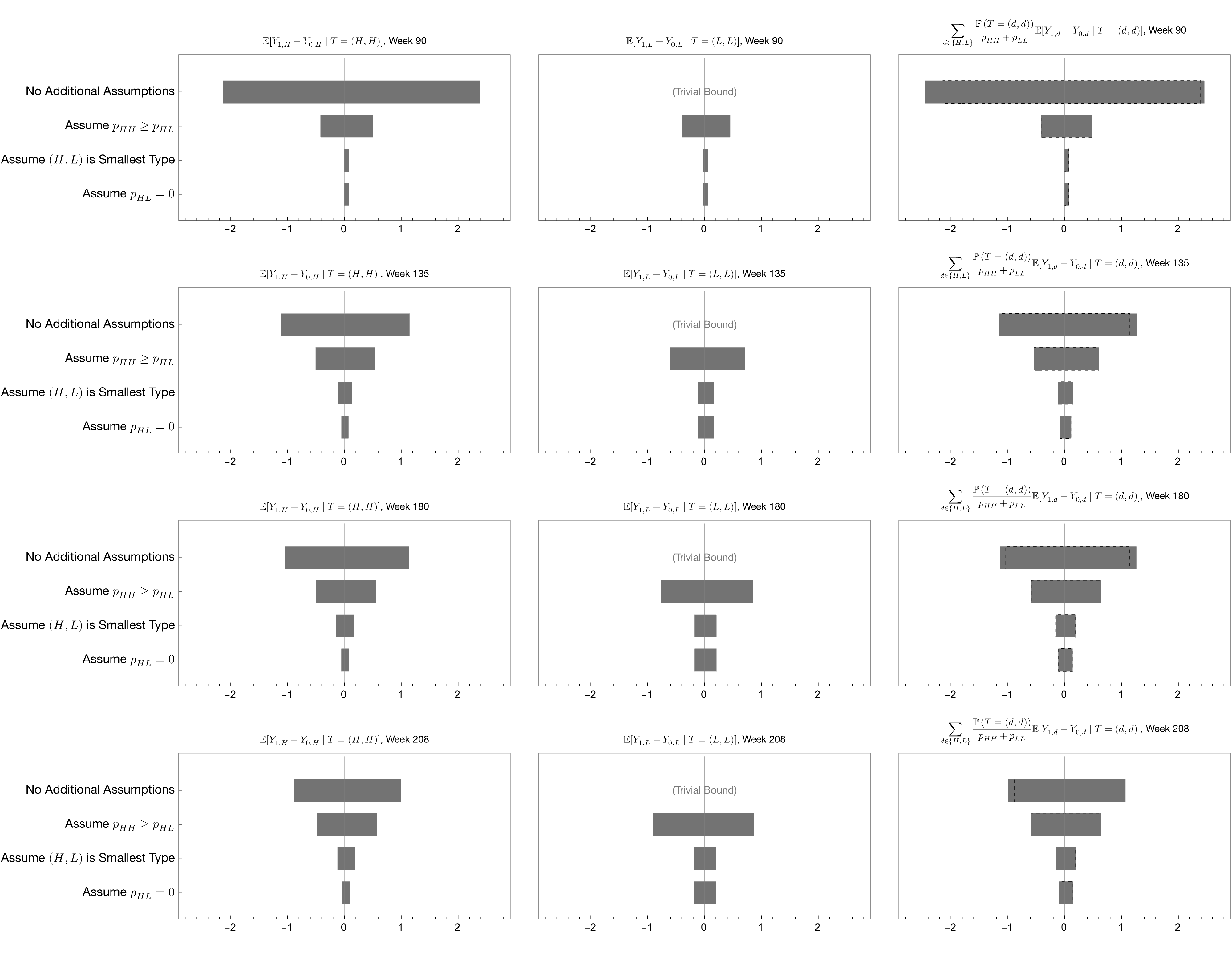}
\begin{minipage}{\textwidth}
    \footnotesize
    {Notes: Outcome is ln(hourly wage); hourly wage calculated as weekly earnings divided by  weekly hours for the employed. The first two columns provide the bounds for each firm-level effect and the last column provides bounds on the weighted average of firm-level effects, by week.  In all panels, shaded areas indicate sharp bounds; in the last column, thin dashed lines indicate the result of the ``na\"ive approach'' of taking the weighted average of firm-level  bounds (see \Cref{rmk:aggbnds}).}
    \end{minipage}
    \caption{Job Corps: Multilayered bounds  
    based on health insurance amenity. 
    }
    \label{fig:healthbnds}
\end{figure}

\Cref{fig:healthbnds} presents our multilayered bounds for $\mathbb E[Y_{1,H}-Y_{0,H}|T=\left(H, H\right) ]$ and $\mathbb E[Y_{1,L}-Y_{0,L}|T=\left(L, L\right) ]$, along with our aggregate bounds, for weeks 90, 135, 180 and 208. We illustrate the bounds in the baseline case (only \Cref{Ass:RCT,Ass:LM} is imposed) and also when we sequentially impose the following restrictions: (i) more stayers than downward switchers, (ii) more upward switchers than downward switchers, and (iii) strong monotonicity. As discussed above, the Lee bounds for week 90 are $[0.0468, 0.0484]$. Focusing on the type $H$ firms (firms that offer health insurance), our estimates indicate $\mathbb E[Y_{1,H}-Y_{0,H}|T=\left(H, H\right) ] \in [-2.1415,2.3907]$. Assuming more stayers than downward switchers tightens these bounds to $[-0.4214, 0.5020]$. Moreover, assuming that $(H,L)$ is the smallest response type, further tightens them to $[-0.0023,0.0754]$. Finally, assuming strong monotonicity narrows these bounds to $[-0.0018,0.0750]$. 

We find a similar pattern of results for the $L$-type bounds. These patterns persist across all weeks of interest and also when classifying firms based on the provision of alternative amenities (paid vacation and retirement/pension benefits) as shown in Appendix \Cref{fig:pensionbnds,fig:vacatbnds}. \Cref{tab:healthbnds} reports our estimated bounds across weeks.\footnote{Appendix \Cref{tab:pensionbnds,tab:vacatbnds} provide our estimated within-firm-type bounds when classifying firms based on the provision of paid vacation and retirement/pension benefits, respectively.} The aggregate multilayered bounds are reported in Appendix \Cref{tab:agg_healthbnds}.\footnote{Appendix \Cref{tab:agg_vacatbnds,tab:agg_pensionbnds} provide our estimated aggregate bounds when classifying firms based on the provision of paid vacation and retirement/pension benefits, respectively.}

\begin{table}[!htbp]
    \centering
    \caption{Job Corps: Multilayered bounds based on health insurance amenity.\label{tab:healthbnds}}
    \begin{tabular}{l*{7}{c}}
        \toprule
        & & & \multicolumn{2}{c}{\footnotesize $\mathbb E(Y_{1,H}-Y_{0,H}|T = (H,H))$} & \multicolumn{2}{c}{\footnotesize $\mathbb E(Y_{1,L}-Y_{0,L}|T = (L,L))$} \\
        \cmidrule(lr){4-5}
        \cmidrule(lr){6-7}
        \textbf{Week 90} &  $ p^{*}_{H,H} $&  $ p^{*}_{L,L} $&       lower &       upper &       lower&       upper
       \input empiric_tbls_figs/bounds/health_wk90_bounds_weighted.tex
        \midrule
        \textbf{Week 135} & & & & & & 
        \input empiric_tbls_figs/bounds/health_wk135_bounds_weighted.tex
        \midrule
        \textbf{Week 180} & & & & & & 
        \input empiric_tbls_figs/bounds/health_wk180_bounds_weighted.tex
        \midrule
        \textbf{Week 208} & & & & & & 
        \input empiric_tbls_figs/bounds/health_wk208_bounds_weighted.tex
        \bottomrule
        &&&&&&\\
    \end{tabular}
    \begin{minipage}{\textwidth}
    \footnotesize
    {Notes: Treatment bounds are for ln(hourly wage); hourly wage calculated as weekly earnings divided by  weekly hours for the employed. $p^{*}_{t}$ is the minimum value of $p_t$ over $\Theta_I\left(\mathcal R_T\right)$, for the corresponding $\mathcal R_T$.
    }
    \end{minipage}
\end{table}

Thus, while the conventional Lee bounds are strictly positive, our bounds for the within-firm effects include 0 even under strong assumptions on the response types. This suggests that Lee bounds may capture a pure sorting response to job training rather than a direct wage effect. 

\subsection{Application \#2: WorkAdvance RCT}

As a second empirical application, we evaluate MDRC's WorkAdvance sectoral employment program. WorkAdvance trains disadvantaged adults with the goal of matching them to high-quality jobs in specific industries with strong labor demand. The MDRC WorkAdvance demonstration featured four different community-based providers in three different locations (New York City, Tulsa, and Northeast Ohio). In this section, we apply our bounds to the Madison Strategies RCT in Tulsa, Oklahoma, and the Towards Employment RCT in Northeast Ohio.\footnote{Our primary dataset 
is state-level administrative Unemployment Insurance (UI) data, obtained via a confidential data use agreement with MDRC. The administrative UI data for Oklahoma (Madison Strategies RCT) and Ohio (Towards Employment RCT) provide the two-digit North American Industry Classification System (NAICS) code for the industry in which the participant worked, while the New York data (Per Scholas and St. Nicks Alliance RCTs) do not. Therefore, we focus on the Oklahoma and Ohio evaluations.}\textsuperscript{,}\footnote{For more details on the WorkAdvance RCT and an evaluation of the impacts, see \cite{katz_why_2022}. For Madison Strategies, the reported impact is $12.4$ percent on earnings two years after the program. For Towards Employment, the impact is $14$ percent.}

The Madison Strategies RCT targeted high-quality jobs in Transportation and Manufacturing. The duration of the program ranged from 4 to 32 weeks. The curriculum focused on pre-employment career readiness and sector-specific skills training. It also involved sector-specific placement services, post-employment retention, and advancement services. Upon completion of the program, participants received the appropriate certification. 

The Madison Strategies RCT enrollment period was from June 2011 to June 2013. It targeted low-income adults who met the following skill requirements: (i) tested at the eighth grade level; (ii) passed a behavioral assessment; (iii) passed mechanical aptitude and manual dexterity exams;  (iv) had a driver's license. The control group was not eligible to receive WorkAdvance services. Eligible potential participants (697 units) were randomly assigned to treatment (353 units) and control (344 units). 

The Towards Employment RCT targeted jobs in Health Care and Manufacturing. The target participants were low-income adults who: (i) tested at the sixth to tenth grade level; (ii) passed background and drug tests. The curriculum, program duration, and enrollment period were the same as Madison Strategies. Eligible potential participants (698 units) were randomly assigned to treatment (349 units) and control (349 units).

The key variables for our analysis of both RCTs are quarterly wages along with employment status and sector. We define quarterly wages as quarterly earnings subject to UI and follow both \cite{lee_training_2009}'s and our evaluation of the Job Corps RCT in defining quarterly employment status based on whether an individual has positive earnings in a quarter.\footnote{To interpret our estimates as reflecting wage effects, this requires that job training does not affect hours of work.} We assume that $d=H$ if the firm of employment is in the target sector and $d=L$ if the firm is not. All of our results focus on 8 quarters post-random assignment.

Appendix \Cref{tab:ms_summ_stats,tab:te_summ_stats} show summary statistics at baseline, as well earnings and employment outcomes 8 quarters post-randomization, for Madison Strategies and Towards Employment, respectively. For Madison Strategies, the employment rates 8 quarters post randomization are quite similar between the treatment group ($0.67$) and the control group ($0.66$). However, the composition of employment by target sector differs meaningfully by treatment status: among the treatment group, $44$ percent are employed in the target sector whereas, among the control group, only $31$ percent are. Towards Employment increased overall employment from $0.62$ to $0.69$ and the share of employment in the target sector from $0.47$ to $0.51$. \Cref{tab:wa_pscores} shows the propensity score estimates for both Madison Strategies and Towards Employment which further illustrate the sorting effects of both RCTs. As was the case for the Job Corps study, this combined evidence suggests multilayered sample selection and again motivates the use of our bounds.

Appendix \Cref{fig:hist_wa} shows that average wages are higher in the target sector and Appendix \Cref{fig:ms_cdf} (Madison Strategies) and Appendix \Cref{fig:te_cdf} (Towards Employment) show that the wage distributions in the target sector stochastically dominate the wage distributions in non-targeted sectors, conditional on treatment status. 

\begin{table}[htbp]
    \centering
    \caption{Propensity scores for WorkAdvance RCTs\label{tab:wa_pscores}}
    \begin{tabular}{l*{4}{c}}
        \toprule
        &  $\mathbb P[D=H|Z=0]$ & $\mathbb P[D=H|Z=1]$&  $\mathbb P[D=L|Z=0]$&  $\mathbb P[D=L|Z=1]$\\
        \midrule
        \input empiric_tbls_figs/workadv_bounds/pscores.tex
        \bottomrule
    \end{tabular}
\end{table}

\begin{figure}
    \centering
    \includegraphics[width=\textwidth]{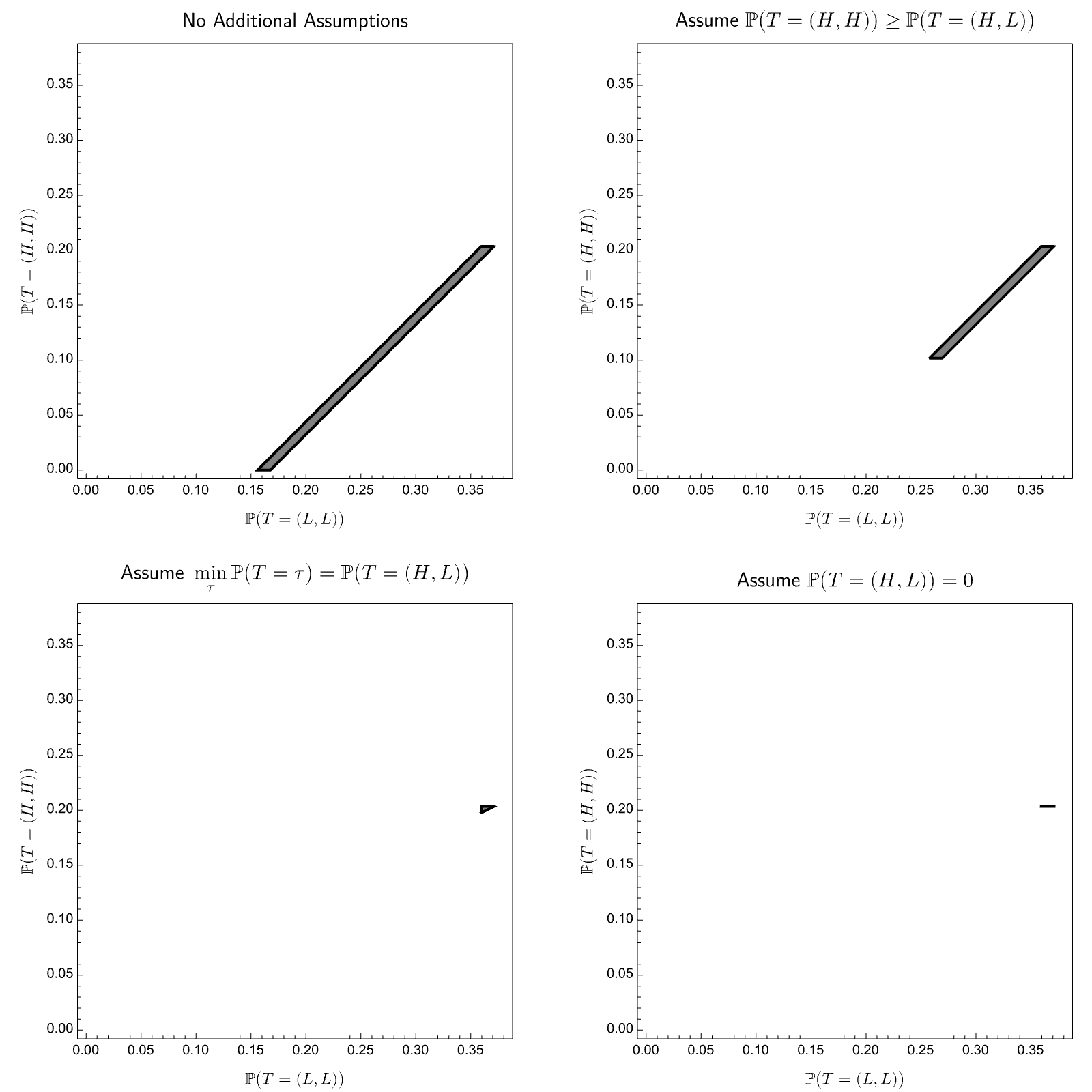}
    \vspace{0.1em}
    
               \begin{minipage}{\textwidth}
    \footnotesize
    {Notes: The first panel illustrates the identified set when $\mathcal R_T = \{\text{Assumption } \ref{Ass:LM}\}$. The remaining panels illustrate the identified set of additional assumptions imposed in addition to \Cref{Ass:LM}.}
    \end{minipage}    
    \caption{Madison Strategies RCT: Identified set for $(p_{L,L},p_{H,H})$.}
    \label{fig:ms_idset}
\end{figure}

\begin{figure}
    \centering
    \includegraphics[width=\textwidth]{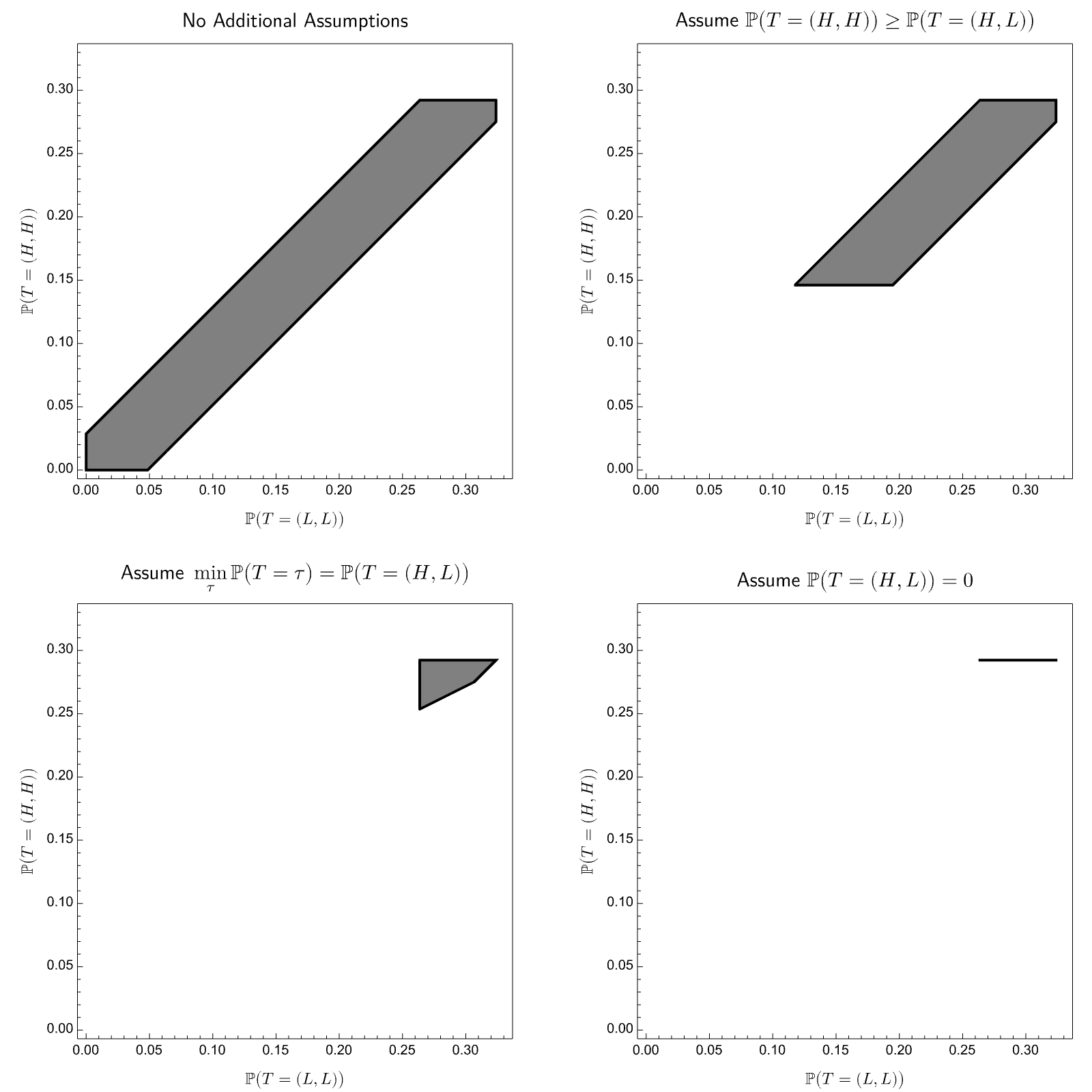}
    \vspace{0.1em}
    
               \begin{minipage}{\textwidth}
    \footnotesize
    {Notes: The first panel illustrates the identified set when $\mathcal R_T = \{\text{Assumption } \ref{Ass:LM}\}$. The remaining panels illustrate the identified set of additional assumptions imposed in addition to \Cref{Ass:LM}.}
    \end{minipage}    
    \caption{Towards Employment RCT: Identified set for $(p_{L,L},p_{H,H})$.}
    \label{fig:te_idset}
\end{figure}

\Cref{fig:ms_idset} and \Cref{fig:te_idset} show the identified sets for $\left(p_{L,L},p_{H,H}\right)$ for Madison Strategies and Towards Employment, respectively. Although sorting appears to be stronger in the WorkAdvance RCT than in the Job Corps study, $p^*_{H,H}$ and $p^*_{L,L}$ are fairly similar. This is driven by differences in $p_{0,0}$, which is a point identified by $P(D=0|Z=1)$. In Job Corps, $p_{0,0}=0.54$. By comparison, $p_{0,0}=0.33$ in Madison Strategies and $p_{0,0}=0.31$ in Towards Employment. Thus, the reason for the similarities in $p^*_{H,H}$ and $p^*_{L,L}$ is driven by much stronger sorting (larger $p_{LH}$) in WorkAdvance. All else being equal, this attenuates $p^*_{d,d}$ and increases the bounds.

\begin{table}
    \centering
    \caption{Multilayered bounds for WorkAdvance RCT.\label{tab:wa_bnds}}
    \begin{tabular}{l*{7}{c}}
        \toprule
        & & & \multicolumn{2}{c}{\footnotesize $\mathbb E(Y_{1,H}-Y_{0,H}|T = (H,H))$} & \multicolumn{2}{c}{\footnotesize $\mathbb E(Y_{1,L}-Y_{0,L}|T = (L,L))$} \\
        \cmidrule(lr){4-5}
        \cmidrule(lr){6-7}
        \textbf{Madison Strategies} &  $ p^{*}_{H,H} $&  $ p^{*}_{L,L} $&       lower &       upper &       lower&       upper 
        \input empiric_tbls_figs/workadv_bounds/bounds_okla_weighted.tex
        \midrule
        \textbf{Towards Employment} & & & & & & 
        \input empiric_tbls_figs/workadv_bounds/bounds_ohio_weighted.tex
        \bottomrule
    \end{tabular}
        \vspace{0.1em}
        
        \begin{minipage}{\textwidth}
        \footnotesize
            Notes: 
            Outcome is quarterly wages. $p^{*}_{t}$ is the minimum value of $p_t$ over the identified set for response-types under the given assumption.
        \end{minipage}
\end{table}
\clearpage
\begin{figure}[H]
    \centering
        \includegraphics[width=\textwidth]{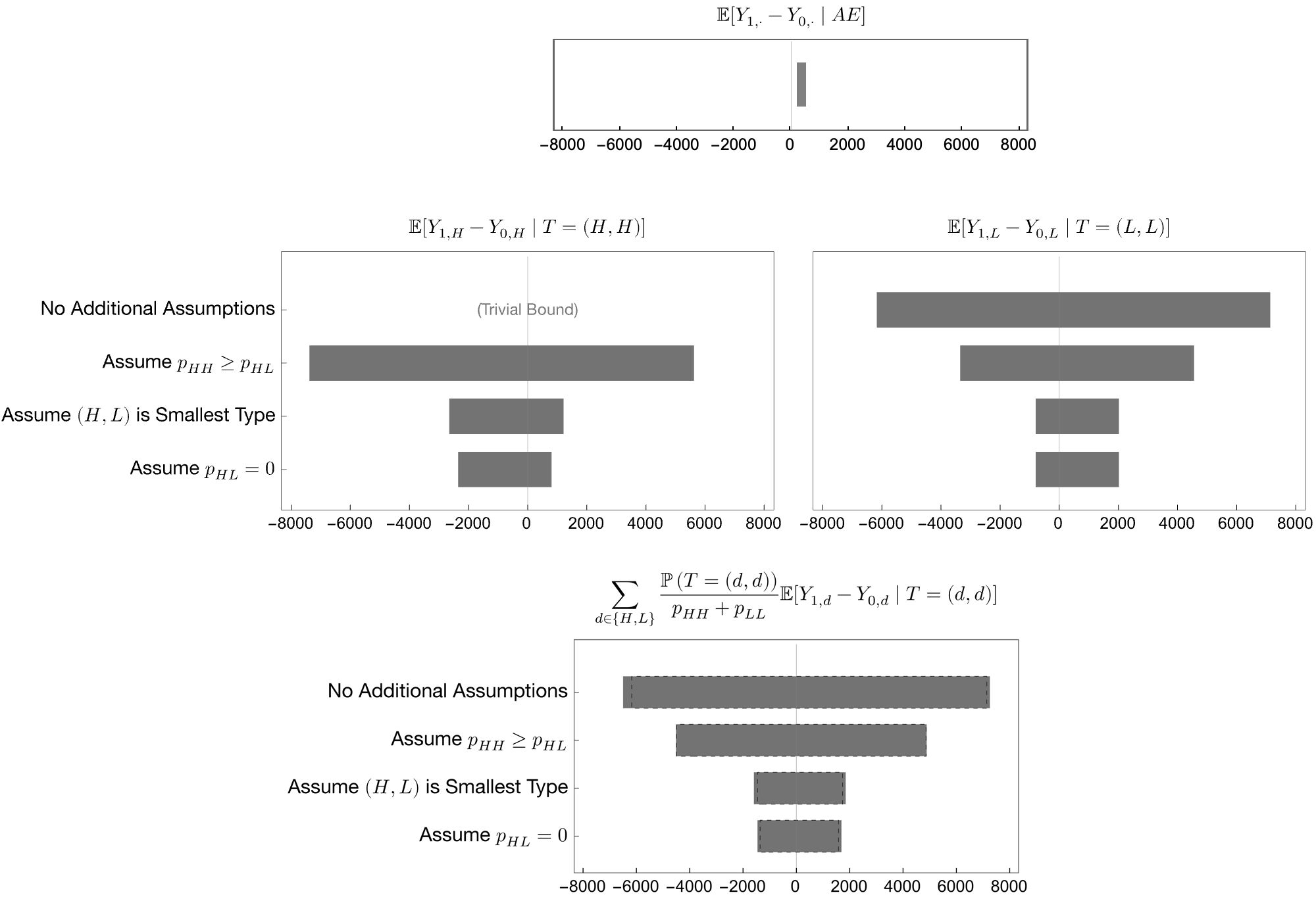}
        \vspace{0.1em}
        
        \begin{minipage}{\textwidth}
        \footnotesize
            Notes: Top panel illustrates \cite{lee_training_2009} bounds, under \Cref{Ass:RCT,Ass:LM}. Remaining panels illustrate the multilayered bounds under various sets of additional assumptions: the middle panels provide the bounds for each firm-level effect, and the bottom panel provides bounds on the weighted average of firm-level effects.
    In all panels, shaded areas indicate sharp bounds; in the bottom panel, thin dashed lines indicate the result of the ``na\"ive approach'' of taking the weighted average of firm-level  bounds (see \Cref{rmk:aggbnds}).
        \end{minipage}
    \caption{
    Bounds for the  Madison Strategies RCT.}
    \label{fig:ms_bounds}
\end{figure}
\begin{figure}[H]
    \centering
    \includegraphics[width=\textwidth]{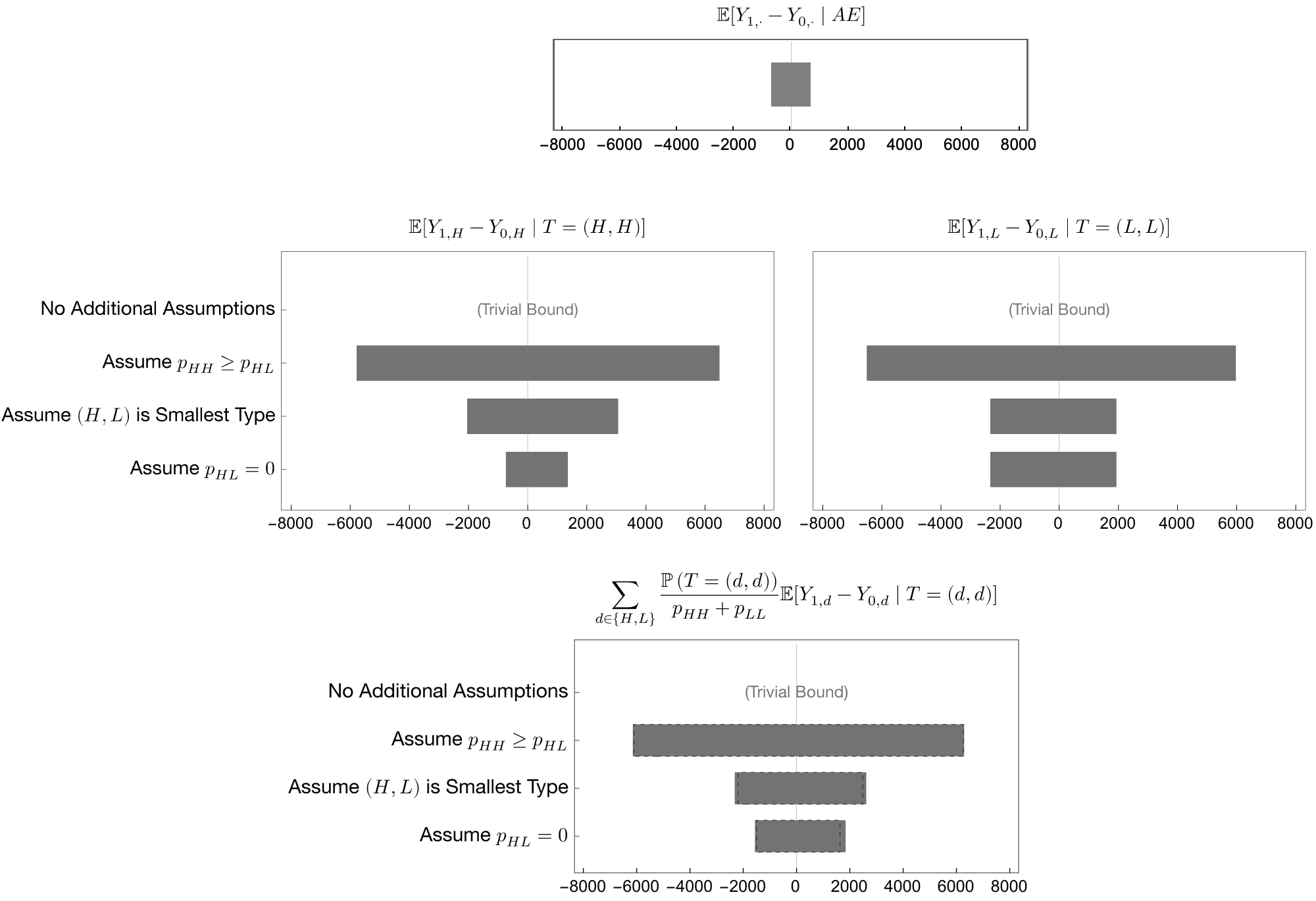}
     \vspace{0.1em}
        
        \begin{minipage}{\textwidth}
        \footnotesize
            Notes: Top panel illustrates \cite{lee_training_2009} bounds, under \Cref{Ass:RCT,Ass:LM}. Remaining panels illustrate the multilayered bounds under various sets of additional assumptions: the middle panels provide the bounds for each firm-level effect, and the bottom panel provides bounds on the weighted average of firm-level effects.
    In all panels, shaded areas indicate sharp bounds; in the bottom panel, thin dashed lines indicate the result of the ``na\"ive approach'' of taking the weighted average of firm-level  bounds (see \Cref{rmk:aggbnds}).
        \end{minipage}
    \caption{
    Bounds for the  Towards Employment RCT.}
    \label{fig:te_bounds}
\end{figure}

\Cref{fig:ms_bounds,fig:te_bounds} present Lee bounds, our multilayered bounds for  $\mathbb E[Y_{1,H}-Y_{0,H}|T=\left(H, H\right) ]$ and $\mathbb E[Y_{1,L}-Y_{0,L}|T=\left(L, L\right) ]$, along with our aggregate bounds, for Madison Strategies and Towards Employment, respectively. As before, we present our multilayered bounds in the baseline case (only \Cref{Ass:RCT,Ass:LM} are imposed) and then sequentially impose the following restrictions: (i) more stayers than downward switchers, (ii) more upward switchers than downward switchers, and (iii) strong monotonicity. Even under the most restrictive assumption of strong monotonicity, the within-firm bounds include 0 for both sites, suggesting that the impact of WorkAdvance on wages is primarily driven by sorting to the target sector. Notably, this contrasts with Lee bounds in the case of Madison Strategies, which are positive and tight. \Cref{tab:wa_bnds} presents the within-firm-type multilayered bounds for Madison Strategies (top panel) and Towards Employment (bottom panel). Appendix \Cref{tab:agg_workadvance,tab:wa_leebounds} present the aggregate multilayered bounds and Lee bounds, respectively, for both sites.


\section{Conclusion}

This paper develops a new methodology to partially identify the causal effect of job training on wages in the presence of multilayered sample selection. We define new treatment effects that operate within and between firms and provide a new identification approach that extends the \citet{horowitz_identification_1995} bounds. As a proof of concept, we show how to empirically implement these bounds by considering applications to the Job Corps Study and the WorkAdvance RCT.

Although we consider our approach in the context of job training where a layer corresponds to a firm, we view it as naturally extending to other settings. In particular, it applies to any setting where sample selection is multilayered. As an example, consider a setting where a researcher is interested in estimating the causal effect of a tuition subsidy on labor market outcomes.\footnote{\cite{bettinger_long-run_2019} evaluate the impact of California's state-based financial aid on long-run earnings.} The subsidy may have an effect on the type of institution that an individual enrolls in and graduates from. If earnings depend on institutional quality, part of the earnings effects of the subsidy could reflect the value-added of institutions that are affected by the subsidy. 

Although our framework has focused mainly on nonparametric (partial) identification, we are currently reexamining the classic parametric sample selection approach of \citet{heckman_sample_1979} in the context of multilayered sample selection, as well as its semiparametric version discussed in \cite{honore_selection_2020}. By imposing additional structure on the unobservables, this approach has the potential to significantly tighten the bounds and may achieve point identification of causal effects.

\clearpage
\bibliographystyle{ecta-fullname}
\bibliography{gen_leebnds_sisub}
\clearpage

\appendix

\section{2 Firm Types Case: A numerical illustration}\label{sec:num}

To provide intuition for our bounding approach, in this section we consider a scenario involving two types of firms. We assume that the firms are identical within each type. Firms are categorized as high type ($H$) or low type ($L$). Our objective is to investigate whether the instrument has a causal effect on wages within each firm type. Under Assumption \ref{Ass:LM}, the support of possible response types is:
\begin{align*}
\mathcal T := \left\{ \left(0,0\right), \left(0,L\right),\left(0,H\right),\left(L,L\right), \left(H,H\right),\left(L,H\right),\left(H,L\right)\right\}.
\end{align*}

Thus, the always-employed (AE) encompasses four distinct response types:$$\{D_0>0, D_1>0\}=\left\{\left(L, L\right), \left(H, H\right),\left(L, H\right),\left(H, L\right)\right\}\equiv AE.$$
Lee bounds applied to this setting correspond to:

\begin{multline*}
\underline{\theta}^{\ell} \leq 
\frac{p_{L,L}+p_{H,L}}{\mathbb P(AE)} \mathbb E[Y_{1,L}-Y_{0,L}|T\in \left\{\left(L, L\right), \left(H, L\right) \right\} ] \\+ \frac{p_{H,H}+p_{L,H}}{\mathbb P(AE)} \mathbb E[Y_{1,H}-Y_{0,H}|T\in \left\{\left(H, H\right), \left(L, H\right) \right\} ] + \frac{p_{L,H}}{\mathbb P(AE)} \mathbb E[Y_{0,H}-Y_{0,L}|T= \left(L, H\right)\} ] + \\ \frac{p_{H,L}}{\mathbb P(AE)} \mathbb E[Y_{0,L}-Y_{0,H}|T=\left(H, L\right)\} ]
\leq \overline{\theta}^{\ell}.
\end{multline*}
To establish bounds on our causal effects of interest, we first characterize identification of the response-type probabilities:
$\{p_{t}: t \in \mathcal T\}$. 
Using information on $(D,Z)$ only, $\{p_{t}: t \in \mathcal T\}$ has to satisfy equations $(\ref{T-EC1},\ref{T-EC2})$ from step 1 above. In addition, if we impose Assumption \ref{Ass:LM}, i.e.,
$\mathcal R_T = \{\text{Assumption } \ref{Ass:LM}\}$,
we can show that the identified set for the response types in this simple case is characterized 
by non-negative solutions to 
the following set of (in)equalities:
\begin{eqnarray}
&&\hspace{-2em}p_{0,0}= 1 - \mathbb P\left(D = H \given Z = 1\right) - \mathbb P\left(D = L \given Z = 1\right),\label{EX:1}\\
&&\hspace{-2em} p_{0,L} =  \mathbb P\left(D = L \given Z = 1\right) - \mathbb P\left(D = H \given Z = 0\right) + p_{H,H} -  p_{L,L},\\
&& \hspace{-2em} p_{0,H} = \mathbb P\left(D = H \given Z = 1\right) - \mathbb P\left(D = L \given Z = 0\right)+ p_{L,L} -  p_{H,H},\\
&& \hspace{-2em} p_{L,H} = \mathbb P\left(D = L \given Z = 0\right) -p_{L,L},\\
&&\hspace{-2em} p_{H,L} = P\left(D = H \given Z = 0\right)-p_{H,H},\label{EX:5}\\
&& \hspace{-2em} \max\{0,\mathbb P(D=H|Z=0)-\mathbb P(D=L|Z=1)\} \nonumber \leq 
\\ &&  \qquad \qquad \qquad \qquad
p_{H,H}  \leq 
\min\{\mathbb P(D=H|Z=0),\mathbb P(D=H,Z=1)\} \label{EX:6}\\
&& \hspace{-2em} \max\{0,\mathbb P(D=L|Z=0)-\mathbb P(D=H|Z=1)\} \nonumber \leq  \\ &&  \qquad \qquad \qquad \qquad p_{L,L}  \leq \min\{\mathbb P(D=L|Z=0),\mathbb P(D=L,Z=1)\}.\label{EX:7}
\end{eqnarray}
More precisely, we can show that 
$$\Theta_{I}(\mathcal R_T)=\left\{
\left(p_{t}: t \in \mathcal T\right) \geq 0: \text{ such that eqs } (\ref{EX:1}) \text{ to } (\ref{EX:7}) \text{ are satisfied}  \right\}.$$

In this case, 
it is also straightforward to show that 
$\Theta_{I}(\mathcal R_T)\neq\emptyset$ if and only if the propensity scores satisfy 
\begin{eqnarray}
    &&\mathbb P(D=0|Z=1) \leq \mathbb P(D=0|Z=0). \label{EX:8} 
\end{eqnarray}
Having defined the identified set for response types, we proceed to construct bounds on our treatment effects of interest. In this particular case, we have:

\begin{eqnarray*}
&&\underline{\gamma}^{z}_{H,H}=\frac{\max\{0,\mathbb P(D=H|Z=0)-\mathbb P(D=L|Z=1)\}}{\mathbb P(D=H|Z=z)}, \text{ for } z \in \{0,1\},\\
&&\underline{\gamma}^{z}_{L,L}=\frac{\max\{0,\mathbb P(D=L|Z=0)-\mathbb P(D=H|Z=1)\}}{\mathbb P(D=L|Z=z)}, \text{ for } z \in \{0,1\},
\end{eqnarray*}
One can then apply the closed-form formula from Theorem \ref{Theorem:Main} to establish bounds on the treatment effects of interest. A unique aspect of our methodology is that imposing additional restrictions on response types does not alter the form of the bounds in Theorem \ref{Theorem:Main}. The bounds remain valid (and sharp) for the new set of restrictions, as long as $\underline{\gamma}^{z}_{d,d}$ is adjusted 
according to the new restrictions.\footnote{That is, if $\underline{\gamma}^{z}_{d,d}$ is updated to be the result of optimizing over the new $\Theta_{I}(\mathcal R_T)$.}

In the numerical illustration below, we describe $\Theta_{I}(\mathcal R_T)$ and demonstrate how imposing further assumptions on response types can significantly refine $\Theta_{I}(\mathcal R_T)$. More precisely, we consider the following additional assumptions.

\begin{enumerate}
\item [(i)] $p_{H,H} \geq p_{H,L}$: remaining within a high-type firm is more probable than transitioning from a high-type to a low-type firm as a consequence of the treatment.
\item [(ii)] $\min_{t} p_t= p_{H,L}$: the smallest proportion of response type consists of individuals moving from a high-type to a low-type firm due to the treatment. 
\item [(iii)] $p_{H,L}=0$: the absence of transitions from high-type to low-type firms as a result of treatment.
\end{enumerate}

In this case, as shown by \eqref{EX:1}-\eqref{EX:5}, the response-type probability $p_t$, for each $t \in \{ (0,0), (0,L), (0,H), (L,H), (H,L) \}$, 
can be represented as a linear function of 
$p_{H,H}$ and $p_{L,L}$, which are themselves only set identified; in other words, $\Theta_{I}(\mathcal R_T)$ is parameterized by $(p_{H,H},p_{L,L})$. 
Hence, our forthcoming discussion will focus primarily on illustrating the projection of $\Theta_{I}(\mathcal R_T)$ with respect to the coordinates of $(p_{H,H},p_{L,L})$.

\subsubsection{Data Generating Process}
We first generate propensity scores to be consistent with the data in \cite{lee_training_2009}. This is reported in \Cref{table:conditional_probabilities}.

\begin{table}[ht]
    \centering
    \caption{Propensity scores for numerical illustrations
    }
    \label{table:conditional_probabilities}
    \begin{tabular}{c c c}
        \hline 
        &&\\
        Job Training & $P(D = L | Z = z)$ & $P(D = H | Z = z)$ \\ 
        \hline
        &&\\
        $Z = 1$ & 0.302886 & 0.408114 \\  
        $Z = 0$ & 0.313959 & 0.373041 \\
        \hline
        &&\\
    \end{tabular}
\end{table}
The data are generated such that the true values for $p_{H,H}$, and  $p_{L,L}$ are:
 \begin{eqnarray*}
p_{H,H} &=&\mathbb P\left(D = H \given Z = 0\right) = 0.373041,\\
p_{L,L} &=& 0.278886.
\end{eqnarray*}


Next, we randomly generate the outcomes.\footnote{By construction, the outcomes (wages) simulated here are independent of the dataset used in \cite{lee_training_2009}.} For each type $t \in \mathcal T$, denote by $D_z$ their employment status when externally assigned $Z =z$. The conditional distributions of $\exp(Y_{z,D_z})\given T = t$ for each $t \in \mathcal T$ are assumed to follow $\operatorname{Lognormal}(\mu_{z|t}, \sigma_{z|t})$ distributions. Here, $\sigma_{z|t} = 1$ for all combinations of $z$ and $t$, indicating that variability only arises through $\mu_{z|t}$ between types. We present two distinct potential earnings distribution models as described in \Cref{tab:sim1} and \Cref{tab:sim2}.

\begin{table}[h]
\renewcommand*{\arraystretch}{2}
\caption{Design 1}\label{tab:sim1}
\begin{tabular}
{@{}lll@{}}
\toprule
$t$ & Dist. of $\exp( Y_{1,D_1})| T =t$ &Dist. of $\exp(Y_{0,D_0})| T =t$
\\
\midrule
$(0,L)$ & $\operatorname{Lognormal}(9.5,1)$ & $\operatorname{Lognormal}(9.5,1)$ 
\\
$(0,H)$ & $\operatorname{Lognormal}(11.5,1)$ & $\operatorname{Lognormal}(9.5,1)$ 
\\
$(L,H)$ & $\operatorname{Lognormal}(16.5,1)$ & $\operatorname{Lognormal}(9.5,1)$ 
\\
$(H,L)$ & $\operatorname{Lognormal}(9.75,1)$ & $\operatorname{Lognormal}(9.6,1)$
\\
$(L,L)$ & $\operatorname{Lognormal}(9.5,1)$ & $\operatorname{Lognormal}(9.5,1)$ 
\\
$(H,H)$ & $\operatorname{Lognormal}(14.5,1)$ & $\operatorname{Lognormal}(14.5,1)$
\\
\bottomrule
\end{tabular}
\end{table}

\begin{table}[h]
\caption{Design 2}\label{tab:sim2}
\renewcommand*{\arraystretch}{2}
\begin{tabular}
{lll}
\toprule
$t$ & Dist. of $\exp( Y_{1,D_1})| T =t$ &Dist. of $\exp(Y_{0,D_0})| T =t$
\\
\midrule
$(0,L)$ & $\operatorname{Lognormal}(10.5,1)$ & $\operatorname{Lognormal}(9.5,1)$ 
\\
$(0,H)$ & $\operatorname{Lognormal}(12.5,1)$ & $\operatorname{Lognormal}(9.5,1)$
\\
$(L,H)$ & $\operatorname{Lognormal}(14.5,1)$ & $\operatorname{Lognormal}(9.5,1)$
\\
$(H,L)$ & $\operatorname{Lognormal}(10.5,1)$ & $\operatorname{Lognormal}(10.5,1)$
\\
$(L,L)$ & $\operatorname{Lognormal}(10.5,1)$ & $\operatorname{Lognormal}(9.5,1)$
\\
$(H,H)$ & $\operatorname{Lognormal}(14,1)$ & $\operatorname{Lognormal}(12,1)$
\\
\bottomrule
\end{tabular}
\end{table}

\subsubsection{Simulations Results}

We begin by exploring the geometry of $\Theta_{I}(\mathcal R_T)$, and demonstrate how incorporating further assumptions regarding response types can significantly shrink its shape. 

\begin{figure}
    \centering
    \includegraphics[width=\textwidth]{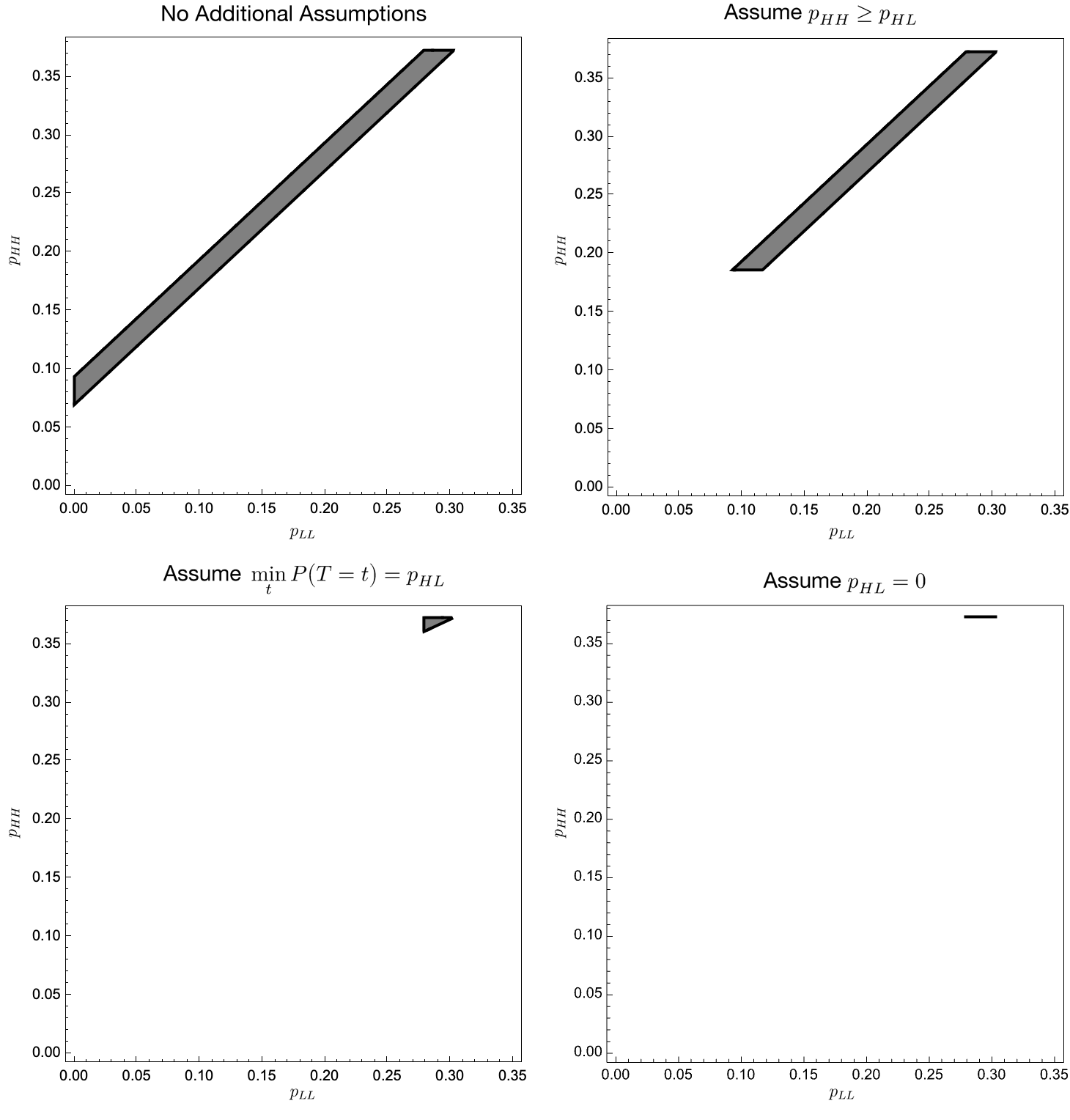}
    \vspace{0.1em}
    
               \begin{minipage}{\textwidth}
    \footnotesize
    {Notes: The first panel illustrates the identified set when $\mathcal R_T = \{\text{Assumption } \ref{Ass:LM}\}$. The remaining panels illustrate the identified set of additional assumptions imposed in addition to \Cref{Ass:LM}. In the two bottom panels of the figure, we shrink the scale of the axes to make the identified set visible.}
    \end{minipage}    
    \caption{Identified set for $(p_{L,L},p_{H,H})$ in both simulated DGPs. }
    \label{fig:enter-label}
\end{figure}

\begin{table}[!h]
\renewcommand*{\arraystretch}{3}
\centering
\caption{Results for Design 1.}\label{tab:res1}
\begin{tabular}
{@{}L{5cm}L{1.25cm}L{7cm}d{1.4}d{1.4}@{}}
\toprule
\multirow{2}{5cm}{Parameter}&\multicolumn{1}{c}{\multirow{2}{1.25cm}{True value}}&\multirow{2}{7cm}{Assumptions}&\multicolumn{2}{c}{\multirow{1}{*}{Bounds}}\\
\cmidrule{4-5}
&&  &\multicolumn{1}{c}{\multirow{1}{*}{Lower}} & \multicolumn{1}{c}{\multirow{1}{*}{Upper}}\\
\midrule
$E(Y_{1,D_1}-Y_{0,D_0}|AE)$&$0.3574$
&Assumptions 1-2&0.0785&0.4131\\
\midrule
\multirow{8}{5cm}{$E(Y_{1,H}-Y_{0,H}|T = (H,H))$}&\multirow{8}{*}{$0.000$}
&Assumptions 1-2&-2.8752&3.4005\\
&&Assumptions 1-2 and $P(T=(H,H)) \geq P(T=(H,L))$ & -1.5958&1.9590\\
&&Assumptions 1-2 and $\displaystyle\min_\tau P(T=\tau) = P(T=(H,L))$ &-0.1699&0.4874\\
&&Assumptions 1-2 and $P(T=(H,L)) = 0$ &-0.0385&0.3588\\
\midrule
\multirow{8}{5cm}{$E(Y_{1,L}-Y_{0,L}|T = (L,L))$}&\multirow{8}{*}{$0.000$}&Assumptions 1-2&\multicolumn{2}{c}{\textit{Trivial Bounds}}\\
&&Assumptions 1-2 and $P(T=(H,H)) \geq P(T=(H,L))$ &-2.3443&2.2676\\
&&Assumptions 1-2 and $\displaystyle\min_\tau P(T=\tau) = P(T=(H,L))$ &-0.3960&0.3460\\
&&Assumptions 1-2 and $P(T=(H,L)) = 0$ &-0.3960&0.3460\\
\bottomrule
\end{tabular}
\vspace{0.25em}
\end{table}

\begin{figure}
    \centering
    \includegraphics[width=\textwidth]{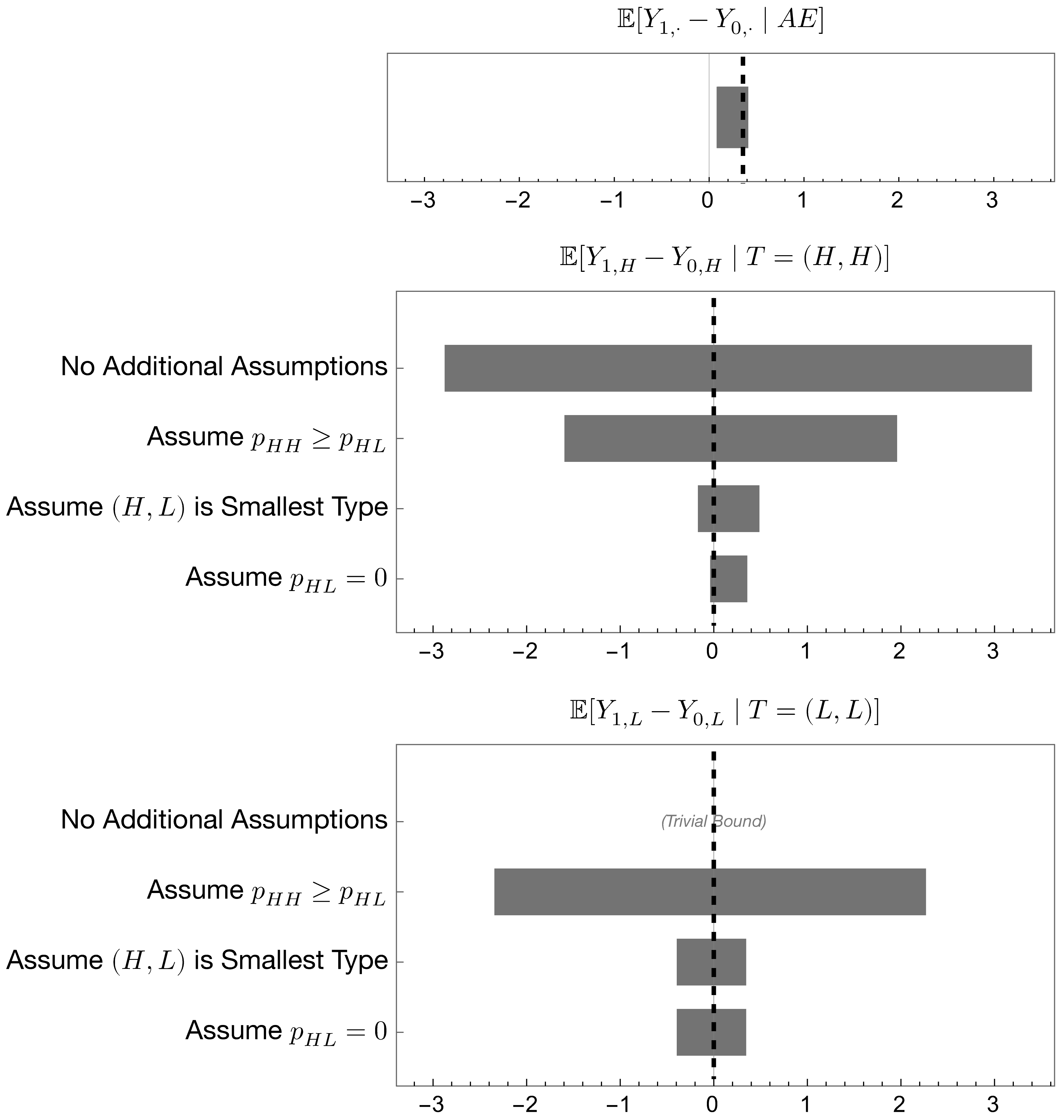}
        \begin{minipage}{\textwidth}
    \footnotesize
    Notes: Dashed black line indicates the true value of the parameter.
    \end{minipage}
    \caption{\cite{lee_training_2009} Bounds and Multilayered Bounds in Design 1.}
    \label{fig:res1}
\end{figure}

\begin{figure}
    \centering
    \includegraphics[width=\textwidth]{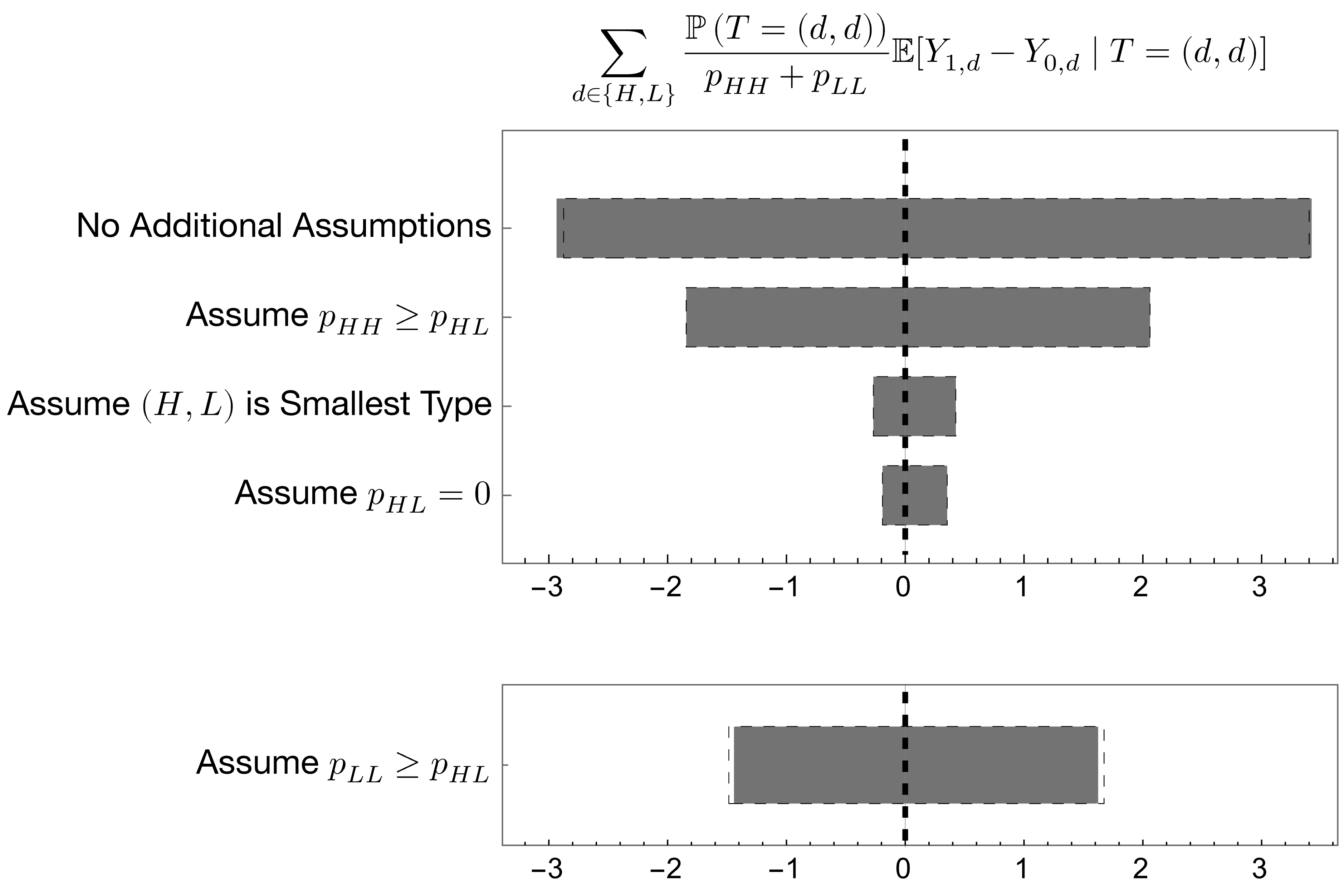}
   \begin{minipage}{\textwidth}
    \footnotesize
        Notes: Thick dashed black line indicates true value of the aggregate treatment effect.  Shaded areas indicate sharp bounds and thin dashed lines indicate the result of the ``na\"ive approach'' of taking the weighted average of firm-level  bounds (see \Cref{rmk:aggbnds}).
    \end{minipage}
    \caption{Aggregate Multilayered Bounds in Design 1.}
    \label{fig:res1agg}
\end{figure}

\begin{table}[h]
\renewcommand*{\arraystretch}{3}
\centering
\caption{Results of Design 2.}\label{tab:res2}
\begin{tabular}
{@{}L{5cm}L{1.25cm}L{7cm}d{1.4}d{1.4}@{}}
\toprule
\multirow{2}{5cm}{Parameter}&\multicolumn{1}{c}{\multirow{2}{1.25cm}{True value}}&\multirow{2}{7cm}{Assumptions}&\multicolumn{2}{c}{\multirow{1}{*}{Bounds}}\\
\cmidrule{4-5}
&&  &\multicolumn{1}{c}{\multirow{1}{*}{Lower}} & \multicolumn{1}{c}{\multirow{1}{*}{Upper}}\\
\midrule
$E(Y_{1,D_1}-Y_{0,D_0}|AE)$&$1.7472$
&Assumptions 1-2&1.5435&1.8029\\
\midrule
\multirow{8}{5cm}{$E(Y_{1,H}-Y_{0,H}|T = (H,H))$}&\multirow{8}{*}{$2.000$}
&Assumptions 1-2&-0.8958&4.9802\\
&&Assumptions 1-2 and $P(T=(H,H)) \geq P(T=(H,L))$ & 0.3698&3.7244\\
&&Assumptions 1-2 and $\displaystyle\min_\tau P(T=\tau) = P(T=(H,L))$ &1.7503&2.3440\\
&&Assumptions 1-2 and $P(T=(H,L)) = 0$ &1.8741&2.2194\\
\midrule
\multirow{8}{5cm}{$E(Y_{1,L}-Y_{0,L}|T = (L,L))$}&\multirow{8}{*}{$1.000$}&Assumptions 1-2&\multicolumn{2}{c}{\textit{Trivial Bounds}}\\
&&Assumptions 1-2 and $P(T=(H,H)) \geq P(T=(H,L))$ &-1.3443&3.2676\\
&&Assumptions 1-2 and $\displaystyle\min_\tau P(T=\tau) = P(T=(H,L))$ &0.6040&1.3460\\
&&Assumptions 1-2 and $P(T=(H,L)) = 0$ &0.6040&1.3460\\
\bottomrule
\end{tabular}
\end{table}

\begin{figure}
    \centering
    \includegraphics[width=\textwidth]{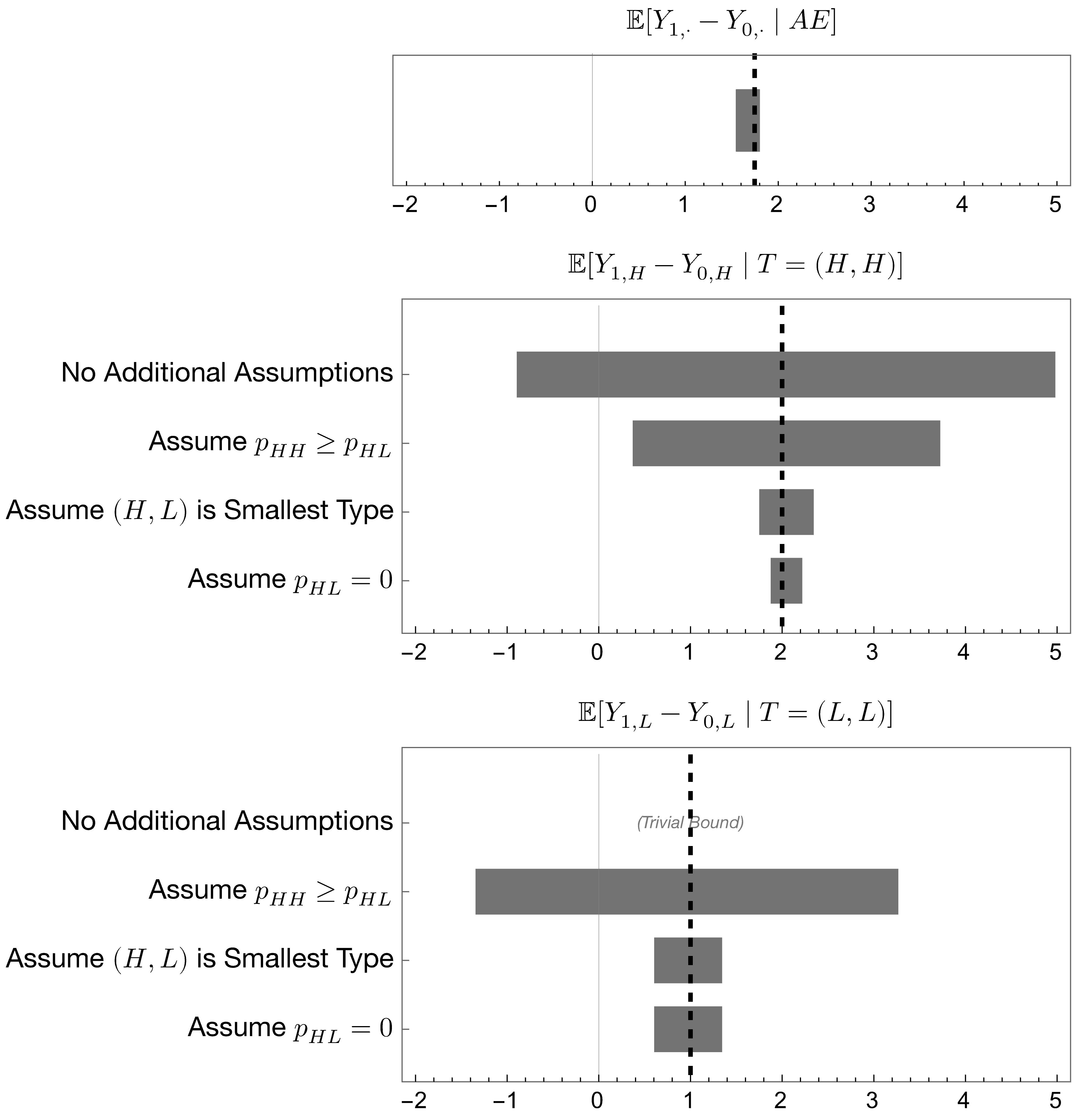}
        \begin{minipage}{\textwidth}
    \footnotesize
    Notes: Dashed black line indicates the true value of the parameter.
    \end{minipage}
    \caption{\cite{lee_training_2009} Bounds and Multilayered Bounds in Design 2. }
    \label{fig:res2}
\end{figure}
\begin{figure}
    \centering
    \includegraphics[width=\textwidth]{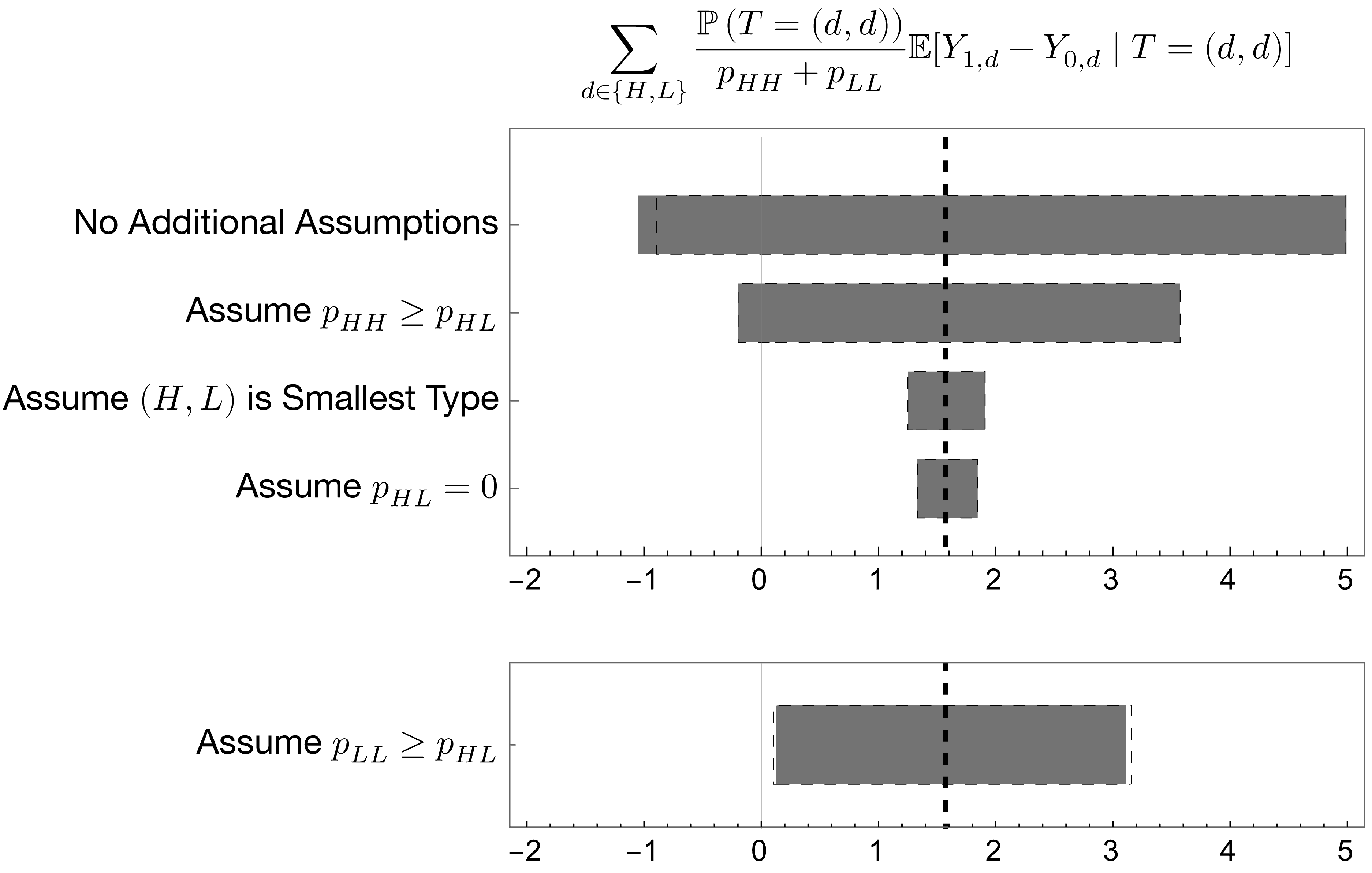}
    \begin{minipage}{\textwidth}
    \footnotesize
        Notes: Thick dashed black line indicates true value of the aggregate treatment effect.  Shaded areas indicate sharp bounds and thin dashed lines indicate the result of the ``na\"ive approach'' of taking the weighted average of firm-level  bounds (see \Cref{rmk:aggbnds}).
    \end{minipage}
    \caption{Aggregate Multilayered Bounds in Design 2.}
    \label{fig:res2agg}
\end{figure}

Figures \ref{fig:res1} and \ref{fig:res2}, along with Tables \ref{tab:res1} and \ref{tab:res2}, present the outcomes for designs 1 and 2, respectively. Initially, we compute Lee bounds, which set identifies the total effect $\mathbb E(Y_{1,D_1}-Y_{0,D_0}|D_0>0,D_1>0)$, as in Lemma \ref{lem:decomp}.
 
Within the framework of design 1, these bounds lie entirely within the positive quadrant, excluding $0$. When ignoring firm heterogeneity in wages, these bounds suggest that job training increases wage rates for the always-employed, i.e.,
 $ AE=\left\{\left(L,L\right), \left(H,H\right),\left(L,H\right),\left(H,L\right)\right\}$. However, this DGP is consistent with the fact that the true within-firm causal effect for $t \in \left\{\left(L,L\right), \left(H,H\right),\left(L,H\right),\left(H,L\right)\right\}$ is $0$, i.e. 
 $\mathbb E(Y_{1,H}-Y_{0,H}|T = t)= \mathbb E(Y_{1,L}-Y_{0,L}|T = t)=0 \; \forall \;  t \in \left\{\left(L,L\right), \left(H,H\right),\left(L,H\right),\left(H,L\right)\right\}.$  
 That is, the positive effect recovered by Lee bounds purely captures the effect of job training on sorting rather than a change in wages.  This distinction is further clarified through the equation below, which illustrates that whenever $\mathbb E(Y_{1,H}-Y_{0,H}|T = (H,H))= \mathbb E(Y_{1,L}-Y_{0,L}|T = (L,L))=0$, the observed changes in wages captured by Lee's bounds are primarily a consequence of sorting induced by job training: 
 


\begin{multline*}
\mathbb E(Y_{1,D_1}-Y_{0,D_0}|D_0>0,D_1>0)=
\frac{p_{L,H}}{\mathbb P(AE)} \mathbb E[Y_{1,H}-Y_{0,L}|T= \left(L, H\right)\} ] \\+ \frac{p_{H,L}}{\mathbb P(AE)} \mathbb E[Y_{1,L}-Y_{0,H}|T=\left(H, L\right)\}].
\end{multline*}




Applying our bounds on the within-firm effects, we can provide a more accurate assessment of the impact of job training on wages. Our bounds for $\mathbb E(Y_{1,H}-Y_{0,H}|T = (H,H))$ and $\mathbb E(Y_{1,L}-Y_{0,L}|T = (L,L))$ always include $0$ across the different scenarios which correspond to different assumptions on response types, suggesting that there is not enough evidence in the data to support the hypothesis that job training has a causal effect on wages.

In design 2, the Lee bounds also lie entirely within the positive quadrant, excluding $0$. However, in this case, our DGP is consistent with the fact that the true within-firm causal effects for $t \in \left\{\left(L,L\right), \left(H,H\right),\left(L,H\right),\left(H,L\right)\right\}$  are strictly positive. Interestingly, our bounds for $\mathbb E(Y_{1,H}-Y_{0,H}|T = (H,H))$ and $\mathbb E(Y_{1,L}-Y_{0,L}|T = (L,L))$ lie entirely within the positive quadrant, excluding $0$, when restricting the response types. This shows that our bounds are informative enough to reveal that job training directly influences wages.

\section{Additional Results}\label{app:add-results}

\subsection{Generalization of  \cite{horowitz_identification_1995} Bounds}

\begin{lemma}\label{mix-lem}
Consider a vector of random variables $(W, W_1,...,W_K)$ satisfying the following condition for all $w$:
\begin{eqnarray}
    F_{W}(w)=\sum_{k=0}^K\gamma_kF_{W_k}(w)
\end{eqnarray}
where $\gamma_k\geq 0$, $F_{W_k}(\cdot)$ represents the cumulative distribution function (CDF) of $W_k$, and $F_{W}(\cdot)$ represents the CDF of $W$. 

The results are as follows:
\begin{enumerate}
   \item [(i)]The bounds for a fixed component of the mixture are point-wise sharp and given by:
\begin{eqnarray}
\mathbb E[F^{-1}_{W}(U) | U \leq \gamma_k] \leq \mathbb E[W_k] \leq \mathbb E[F^{-1}_{W}(U) | U \geq 1-\gamma_k], 
\end{eqnarray}
and this  for  $k \in \{1,...,K\}$ where $U \sim \text{Uniform}[0,1]$.

\item [(ii)]The bounds for a weighted average of fixed components of the mixture are point-wise sharp and given by:
\begin{eqnarray}
\mathbb E\left[F^{-1}_{W}(U) | U \leq \sum_{k=l}^{l'}\gamma_k\right] \leq \sum_{k=l}^{l'}\frac{\gamma_k}{\sum_{k=l}^{l'}\gamma_k}\mathbb E[W_k] \leq \mathbb E\left[F^{-1}_{W}(U) | U \geq 1- \sum_{k=l}^{l'}\gamma_k\right], 
\end{eqnarray} 
\end{enumerate}
\end{lemma}  

Lemma \ref{mix-lem}(i) extends the \cite{horowitz_identification_1995} bounds to scenarios involving mixtures with more than two components. Lemma \ref{mix-lem}(ii) introduces novel bounds on the weighted average of certain mixture components spanning $l$ to $l'$. It is noteworthy that
 $$\mathbb E\left[F^{-1}_{W}(U) | U \leq \sum_{k=l}^{l'}\gamma_k\right] \geq \sum_{k=l}^{l'}\frac{\gamma_k}{\sum_{k=l}^{l'}\gamma_k} \mathbb E[F^{-1}_{W}(U) | U \leq \gamma_k],$$ 
indicating that utilizing the point-wise bounds from Lemma \ref{mix-lem}(i) (Horowitz and Manski bounds) to establish bounds on the weighted average incurs some information loss.

\subsection{Bounds on Aggregate Local  Control Direct Effect (LCDE)}\label{App:Agg}
To derive bounds on the aggregate LCDE, defined as 
$\sum_{d=l}^{l'} \frac{p_{d,d}}{\sum_{d'=l}^{l'} p_{d',d'}} \text{LCDE}(d \mid d,d),$
one might be tempted to adopt a naive approach by taking a weighted average of the pointwise sharp bounds derived in \ref{Theorem:Main} (i). However, this approach not only fails to provide sharp bounds on the aggregate quantity but may also yield invalid bounds. 

The primary issue lies in the fact that both the weights and the LCDE quantities are only set-identified. Consequently, the naive aggregation does not preserve the sharpness or validity of the bounds. To better illustrate this problem, let us provide more details:
\begin{enumerate}
    \item  

The sharp bounds on the aggregate LCDE can be tighter than the weighted average of the marginal bounds on $\text{LCDE}(d \mid d,d)$ for $d = l, \dots, l'$ when these marginal bounds are achieved at values of $p_{d,d}$ for $d = l, \dots, l'$ that are not \textit{jointly} attainable. Specifically, this occurs whenever $\left(\underline{p}_{d,d} : d = l, \dots, l'\right) 
$does not lie within the identified set for 
$\left(p_{d,d} : d = l, \dots, l'\right).$
As a result, the sharp bounds, which are obtained by optimizing over the identified set for the \textit{vector} 
$\left(p_{d,d} : d = l, \dots, l'\right),$
may be tighter.

\item   The sharp bounds on the LCDE can be wider than the weighted average of the marginal LCDE bounds when there exist $d, d' \in \{l, \dots, l'\}$ such that the lower bound for $\text{LCDE}(d \mid d,d)$ is significantly smaller than the lower bound for $\text{LCDE}(d' \mid d',d')$, and $\underline{p}_{d,d}$ is also much smaller than $\underline{p}_{d',d'}$. 

In such cases, the choice of $p_{d,d}$ directly affects the relative weight of $\text{LCDE}(d \mid d,d)$ in the aggregate LCDE. Consequently, the value of the objective function in the optimization problem defining the lower bound in Theorem \ref{Theorem:Main} (iii) may actually decrease as $p_{d,d}$ increases.

To illustrate this more clearly, consider a simple example with only two groups, $a$ and $b$. For each $d \in \{a, b\}$, let $L_d(p)$ denote the sharp lower bound for $\text{LCDE}(d \mid d,d)$ under the assumption that $p_{d,d} = p$. Suppose $\underline{p}_{b,b} > \underline{p}_{a,a} = 0$, so that the lower bound for $\text{LCDE}(a \mid a,a)$, given by $L_a(\underline{p}_{a,a}) = L_a(0) = y_L$, is the trivial bound. Additionally, assume the lower bound for $\text{LCDE}(b \mid b,b)$ is non-trivial, i.e., $L_b(\underline{p}_{b,b}) > y_L$.

In this scenario, the weighted average of these lower bounds simplifies to:
\[
\frac{\underline{p}_{a,a} L_a(\underline{p}_{a,a}) + \underline{p}_{b,b} L_b(\underline{p}_{b,b})}{\underline{p}_{a,a} + \underline{p}_{b,b}} = L_b(\underline{p}_{b,b}),
\]
since $\underline{p}_{a,a} = 0$.

Now, consider the case where there exists a point $p > 0$ such that $(p, \underline{p}_{b,b})$ lies within the identified set for $(p_{a,a}, p_{b,b})$, and $L_a(p) < L_b(\underline{p}_{b,b})$. In this case, we have:
\[
\frac{p L_a(p) + \underline{p}_{b,b} L_b(\underline{p}_{b,b})}{p + \underline{p}_{b,b}} < L_b(\underline{p}_{b,b}).
\]

By definition, the sharp lower bound given in Theorem \ref{Theorem:Main} (iii) will be at least as small as the left-hand side above, resulting in a smaller lower bound for the aggregate LCDE. Similarly, it is also possible for the sharp upper bound to be larger than the weighted average of the marginal upper bounds.

\end{enumerate}

\section{Proofs of the Main results}\label{app:proof}

\subsection{Proof of \Cref{lem:decomp}}
\begin{proof}
Note that (ii) and (iii) follow immediately from (i), so it suffices to show (i). First, notice that 
\begin{align}\label{eq:lem1pf}
   \mathbb E\left[Y_{1,D_1} - Y_{0, D_0} | D_0 > 0, D_1 > 0 \right] =&\  \mathbb E\left[Y_{1,D_1} - Y_{0, D_1} | D_0 > 0, D_1 > 0 \right]\\ &\ +  \mathbb E\left[Y_{0,D_1} - Y_{0, D_0} | D_0 > 0, D_1 > 0 \right]~.
\end{align}

Next, we have that 
\begin{align*}
   &\ \mathbb E\left[Y_{1,D_1} - Y_{0, D_1} | D_0 > 0, D_1 > 0 \right]\\ 
   = &\  \sum_{d = 1, d' = 1}^{K,K}    \mathbb E\left[Y_{1,D_1} - Y_{0, D_1} | D_0 = d', D_1 = d \right]  \mathbb P\left(D_0 =d',  D_1 = d | D_0 > 0, D_1 > 0\right)\\ 
   = &\  \sum_{d = 1, d' = 1}^{K,K}    \mathbb E\left[Y_{1,d} - Y_{0, d} | D_0 = d', D_1 = d \right]  \mathbb P\left(D_0 =d',  D_1 = d | D_0 > 0, D_1 > 0\right)\\
   =&\ \sum_{d = 1, d' = 1}^{K,K}    LCDE\left(d|d',d\right) \mathbb P\left(T = \left(d', d\right) | D_0 > 0, D_1 > 0\right)~,
\end{align*}
and similarly, 
\begin{align*}
   &\ \mathbb E\left[Y_{0,D_1} - Y_{0, D_0} | D_0 > 0, D_1 > 0 \right]\\ 
   = &\  \sum_{d = 1, d' = 1}^{K,K}    \mathbb E\left[Y_{0,D_1} - Y_{0, D_0} | D_0 = d', D_1 = d \right]  \mathbb P\left(D_0 =d',  D_1 = d | D_0 > 0, D_1 > 0\right)\\ 
   = &\  \sum_{d = 1, d' = 1}^{K,K}    \mathbb E\left[Y_{0,d} - Y_{0, d'} | D_0 = d', D_1 = d \right]  \mathbb P\left(D_0 =d',  D_1 = d | D_0 > 0, D_1 > 0\right)\\ 
   = &\  \sum_{d = 1, d' = 1, d \neq d'}^{K,K}    \mathbb E\left[Y_{0,d} - Y_{0, d'} | D_0 = d', D_1 = d \right]  \mathbb P\left(D_0 =d',  D_1 = d | D_0 > 0, D_1 > 0\right)\\
   =&\  \sum_{d = 1, d' = 1, d \neq d'}^{K,K}   LCIE\left(0,d,d'|d',d\right) \mathbb P\left(T = \left(d',d\right) | D_0 > 0, D_1 > 0\right)~,
\end{align*}
and plugging these values back into \eqref{eq:lem1pf} immediately implies the result. 
\end{proof}
\subsection{Proof of \Cref{Lem: RT-restricted}}
\begin{proof}
	By the definition of $T$, it is clear that the response-type probabilities satisfy the eqs (\ref{T-EC1}, \ref{T-EC2}). It suffices to show that given any solution $\left(\boldsymbol{p}_{(d,d')} : (d', d) \in \left\{0, \dots, K\right\}^2\right) \geq 0$ such that $\sum_{d = 0, d'=0}^{K,K} \boldsymbol{p}_{(d,d')} = 1$ and satisfying, for every $d \in \left\{0, \dots, K\right\}$,
\begin{eqnarray}
\mathbb P(D=d|Z=1)&=&\sum_{d'=0}^K \boldsymbol{p}_{(d,d')}\label{eq:lem3pf1}\\
\text{ and }\mathbb P(D=d|Z=0)&=&\sum_{d'=0}^K \boldsymbol{p}_{(d,d')}\label{eq:lem3pf2}~,
\end{eqnarray}
there exists a joint distribution $Q$ of $\left(\left(Y_{0, d} : d \in \left\{0, \dots, K\right\}\right), \left(Y_{1, d} : d \in \left\{0, \dots, K\right\}\right), D_0, D_1, Z \right)$ such that 
\begin{align*}
\left(\mathbb P_Q \left[T = (d,d')\right] : (d', d) \in \left\{0, \dots, K\right\}^2\right) =&\  \left(\boldsymbol{p}_{(d,d')} : (d', d) \in \left\{0, \dots, K\right\}^2\right)~,
\end{align*}
and $Q$ induces a distribution of $\left(Y,D,Z\right)$ under (\ref{GPOM1}, \ref{GPOM2}) and \Cref{Ass:RCT}, that is consistent with the observed data. Since $Y$ is not observed when $D = 0$, set $Y | D = 0, Z = 1$ and $Y | D = 0, Z = 0$ to arbitrary distributions, so that we can treat $\left(Y, D, Z\right)$ as observed. We will now construct a $Q$ that induces this distribution.

Define $\mathcal Y_{z,d} := \operatorname{supp}\left(Y | D = d, Z = z\right)$ and, for each $d \in \left\{0, \dots, K\right\}$, $z \in \left\{0,1\right\}$, and $\left(d', d''\right) \in \left\{0, \dots, K\right\}^2$ define the CDF $F^{(z,d)}_{(d',d'')}$ as 
\begin{align*}
 F^{(z,d)}_{(d',d'')}\left(y\right) = P\left(Y \leq y |D = d, Z = z\right)~,
\end{align*}
and note that it does not depend on $(d', d'')$. Next, for every $\left(d', d''\right) \in \left\{0, \dots, K\right\}^2$, let $C_{(d', d'')}$ be an arbitrary copula of dimension $|\{0,1\} \times \{0, \dots, K\}|$. Finally, define $Q$ as
\begin{align*}
    Q\left(y,t,z\right) := 
    C_{t}\left(\left(F^{(z,d)}_t\left(y_{(z,d)}\right) : (z,d) \in \{0,1\} \times \{0, \dots, K\}\right)\right)\times \boldsymbol{p}_{(t)} \times P\left(Z = z\right)~,
    \end{align*}
    for any $y \in \prod\limits_{(z,d) \in \{0,1\} \times \{0, \dots, K\}} \mathcal Y_{z,d}$, $t \in \left\{0, \dots, K\right\}^2$, and $z \in \left\{0,1\right\}$, where   $ Q\left(y,t,z\right)$ is shorthand for 
    \begin{align*}
  Q\left(\left(Y_{z,d} : (z,d) \in \{0,1\} \times \{0, \dots, K\}\right) \leq y, \left(D_0, D_1\right) = t, Z = z\right)
\end{align*}

Next,  for all $\left(d', d''\right) \in \left\{0, \dots, K\right\}^2$, let the conditional joint distribution
\begin{align*}
 Q\left(Y_{z,d} \leq y_{z,d} | T = \left(d', d''\right) \right)~.
\end{align*}

By construction, $Q$ satisfies \Cref{Ass:RCT} and $\left(\mathbb P_Q \left[T = (d,d')\right] : (d', d) \in \left\{0, \dots, K\right\}^2\right) =  \left(\boldsymbol{p}_{(d,d')} : (d', d) \in \left\{0, \dots, K\right\}^2\right)$. The result now follows immediately by noting that $Q$ induces the observed data distribution under  (\ref{GPOM1}, \ref{GPOM2}) since,  for any $y \in \mathcal Y_{z,d}, d \in \left\{0, \dots, K\right\}$ and $z \in \left\{0,1\right\}$, we have that
\begin{align*}
  Q\left(Y_{z,d} \leq y, D_z = d, Z = z\right) =&\  Q\left(Y_{z,d} \leq y | D_z = d, Z = z\right) Q\left(D_z = d | Z =z\right) Q\left(Z = z\right)\\
  =&\ Q\left(Y_{z,d} \leq y | D_z = d\right) Q\left(D_z = d \right) P\left(Z = z\right)\\
  =&\  P\left(Z = z\right) \sum_{d' = 1}^{K,K} Q\left(Y_{z,d} \leq y | D_z = d, D_{1-z}= d'\right) \left((1-z)\boldsymbol{p}_{(d,d')} + z\boldsymbol{p}_{(d',d)}\right)\\
  =&\  P\left(Z = z\right) \sum_{d' = 1}^{K,K} \left((1-z) F^{(z,d)}_{(d,d')}\left(y\right)\boldsymbol{p}_{(d,d')} + z F^{(z,d)}_{(d',d)}\left(y\right)\boldsymbol{p}_{(d',d)}\right)\\
  =&\  P\left(Z = z\right)  P\left(Y \leq y |D = d, Z = z\right)  \sum_{d' = 1}^{K,K} \left((1-z) \boldsymbol{p}_{(d,d')} + z \boldsymbol{p}_{(d',d)}\right)\\
  =&\   P\left(Z = z\right) P\left(Y \leq y |D = d, Z = z\right)P\left(D = d | Z =z\right) \\
  =&\   P\left(Y \leq y, D = d, Z = z\right)
\end{align*}
where the second equality follows from $Q$ satisfying \Cref{Ass:RCT}, and the penultimate equality follows from $\boldsymbol{p}$ satisfying \eqref{eq:lem3pf1} and \eqref{eq:lem3pf2}.
\end{proof}

\subsection{Proof of \Cref{Lem: RT-restricted}}
\begin{proof}
	Given \Cref{Lem: RT-C} above, this result follows immediately from Theorem 3 in \cite{vayalinkal_sharp_2024}. Since our proof of \Cref{Lem: RT-C} is constructive, however, we can also argue directly, as follows. Note that both parts below proceed by showing the contrapositive. 
	
	($\impliedby$) If Assumption \ref{Ass:RCT} and $\mathcal R_T$ are consistent with the data, then there exists a joint distribution $Q$ of $\left(\left(Y_{0, d} : d \in \left\{0, \dots, K\right\}\right), \left(Y_{1, d} : d \in \left\{0, \dots, K\right\}\right), D_0, D_1, Z \right)$ that is consistent with the observed data distribution such that the response-type probabilities induced by $Q$ is in $\Theta_{I}(\mathcal R_T)$ and so $\Theta_{I}(\mathcal R_T) \neq\emptyset$. 
	
($\implies$) Suppose that $\Theta_{I}(\mathcal R_T) \neq \emptyset$, then there exists $\boldsymbol{p} = \left(\boldsymbol{p}_{(d,d')} : (d', d) \in \left\{0, \dots, K\right\}^2\right) \in \Theta_{I}(\mathcal R_T)$. Now, our proof of \Cref{Lem: RT-C} shows that we can construct a joint distribution $Q$ of $\left(\left(Y_{0, d} : d \in \left\{0, \dots, K\right\}\right), \left(Y_{1, d} : d \in \left\{0, \dots, K\right\}\right), D_0, D_1, Z \right)$ such that $Q$ satisfies \Cref{Ass:RCT} and induces the observed data distribution under (\ref{GPOM1}, \ref{GPOM2}), and 
  \begin{align*}
\left(\mathbb P_Q \left[T = (d,d')\right] : (d', d) \in \left\{0, \dots, K\right\}^2\right) =&\  \left(\boldsymbol{p}_{(d,d')} : (d', d) \in \left\{0, \dots, K\right\}^2\right)~,
\end{align*}
which implies that $Q$ also satisfies $\mathcal R_T$, as required, since $\boldsymbol{p} \in \Theta_{I}(\mathcal R_T)$.  
\end{proof}

\subsection{Proof of \Cref{mix-lem}}
\begin{proof}
Since (i) is just a special case of (ii), it suffices to show (ii). Let $W$ and $\left\{W_k\right\}_{k=0}^K$ be defined on the common probability space $\left(\Omega, \Sigma, \mathrm P\right)$, and take values in the set $\mathcal W \subseteq \mathbb R$, equipped with the Borel sigma algebra and a probability measure $\mu$. Moreover, let $\mu$ be such that $W$ and $\left\{W_k\right\}_{k=0}^K$ are $\mu$-integrable and have densities with respect to $\mu$. Denote the $\mu$-density of $W$ by $f_W$, and denote the $\mu$-density of $W_k$ by $f_{W_k}$, for $k \in \left\{0, \dots, K\right\}$.  

First, we show that the bounds are valid. Define $\bar \gamma := \sum_{k=l}^{l'}\gamma_k$ and let $U \sim \text{Uniform}[0,1]$.  
Now,  suppose that 
\begin{align*}
  \mathbb E\left[F^{-1}_{W}(U) | U \leq \bar \gamma \right] > \sum_{k=l}^{l'}\frac{\gamma_k}{\bar \gamma }\mathbb E[W_k]~,
\end{align*}
then there must exist $w$ such that 
\begin{align*}
  \mathbb P\left(F^{-1}_{W}(U)  \leq w | U \leq \bar \gamma \right) <  \sum_{k=l}^{l'}\frac{\gamma_k}{\bar \gamma } \int_{(-\infty, w]} f_{W_k}\left(x\right)\ d\mu\left(x\right) = \frac{1}{\bar \gamma } \int_{(-\infty, w]}\sum_{k=l}^{l'} \gamma_k f_{W_k}\left(x\right)\ d\mu\left(x\right)~.
\end{align*}

Now, note that 
\begin{align*}
    \mathbb P\left(F^{-1}_{W}(U)  \leq w | U \leq \bar \gamma \right) =&\  \mathbb P\left(F^{-1}_{W}(\bar \gamma U)  \leq w \right) =  \mathbb P\left(\bar \gamma U  \leq F_W\left(w\right) \right)\\ 
    =&\ \frac{F_W\left(w\right)}{\bar \gamma} = \frac{\int_{(-\infty, w]} f_W\left(x\right)\ d\mu\left(x\right)}{\bar \gamma} ~,
\end{align*}
but this implies that 
\begin{align*}
 \int_{(-\infty, w]}\sum_{k=0}^{K} \gamma_k f_{W_k}\left(x\right)\ d\mu\left(x\right) =  \int_{(-\infty, w]} f_W\left(x\right)\ d\mu\left(x\right) < \int_{(-\infty, w]}\sum_{k=l}^{l'} \gamma_k f_{W_k}\left(x\right)\ d\mu\left(x\right)~,
\end{align*}
which, in turn, implies that 
\begin{align*}
 \int_{(-\infty, w]}\sum_{k=0}^{l-1} \gamma_k f_{W_k}\left(x\right) + \sum_{k=l'+1}^{K} \gamma_k f_{W_k}\left(x\right)\ d\mu\left(x\right) < 0~,
\end{align*}
a contradiction. Therefore, the lower bound is valid. The validity of the upper bound follows from the analogous argument.

Define $\underline \gamma : = \bar \gamma +  \sum_{k'=0}^{l-1} \gamma_{k'}$. Now, sharpness of the lower bound follows immediately by defining each $F_{W_k}$ as follows 
\begin{align*}
  F_{W_k}\left(w\right) :=   \left\{\begin{array}{ll} \mathbb P\left(F^{-1}_{W}(U)  \leq w | U \in \left(\sum_{k'=l}^{k-1} \gamma_{k'}, \sum_{k'=l}^{k} \gamma_{k'}\right)\right)&\text{ if } k \in \left\{l, \dots, l'\right\}\\
\mathbb P\left(F^{-1}_{W}(U)  \leq w | U \in \left(\bar \gamma + \sum_{k'=0}^{k-1} \gamma_{k'}, \bar \gamma + \sum_{k=0}^{k} \gamma_{k'}\right)\right)&\text{ if } k \in \left\{0, \dots, l-1\right\}\\
\mathbb P\left(F^{-1}_{W}(U)  \leq w | U \in \left(\underline \gamma +  \sum_{k'=l'+1}^{k-1} \gamma_{k'}, \underline \gamma  +  \sum_{k'=l'+1}^{k} \gamma_{k'}\right)\right)&\text{ all other } k
   \end{array}\right.	~.
\end{align*}
Sharpness of the upper bound follows from the analogous construction. 
\end{proof}
\subsection{Proof of \Cref{Theorem:Main}}
\begin{proof}
Define
\begin{align*}
   {\bf f}_{(z,d)|d,d'}(y) \equiv  \mathbb P[T=(d',d)] f_{Y_{z,d}|T}(y|d',d)~,
\end{align*}
and also define the shorthand notation
\begin{align*}
   {\bf f}_{z|d,d'}(y) \equiv  \mathbb P[T=(d',d)] f_{Y_{z,T(z)}|T}(y|d',d)~.
\end{align*}
Consider the vector of weighted response-type conditional densities
\begin{align*}
{\bf f}(y) \equiv&\ \left( {\bf f}_{(z,d'')|d,d'}(y) : d,d',d'' \in \left\{0, \dots, K\right\}, z \in \left\{0,1\right\}\right)
~.\end{align*}
First, note that ${\bf f}$ must satisfy \eqref{EC1} and \eqref{EC2}, i.e. we must have that 
\begin{eqnarray*}
f_{Y,D=d|Z=1}(y)&=& \sum_{d'=0}^K \mathbb P[T=(d',d)] f_{Y_{1,d}|T}(y|d',d) = \sum_{d'=0}^K   {\bf f}_{1|d',d}(y) \\
f_{Y,D=d|Z=0}(y)&=& \sum_{d'=0}^K \mathbb P[T=(d,d')] f_{Y_{0,d}|T}(y|d,d') = \sum_{d'=0}^K {\bf f}_{0|d,d'}(y) 
\end{eqnarray*}
for all $d \in \left\{1, \dots, K\right\}$. Second, we must also have that there exists a $p:=\left(p_{t} : t \in \operatorname{supp}\left(T\right)\right) \in \Theta_I(\mathcal R_T)$ such that $\int_{y_L}^{y_U}  {\bf f}_{(z,d'')|d',d}(y)\ d\mu\left(y\right)  = p_{d',d}$ for all $z \in \left\{0,1\right\}$ and $d,d',d'' \in \left\{0, \dots, K\right\}$. Finally, we must have that each component of $\bf f$ is a non-negative function supported on (a subset of) $[y_L, y_U]$

The remainder of the proof proceeds in two parts. We first show that these conditions are sharp (i.e. they define the identified set of $\bf f$). We then show that this allows us to complete the proof using the sharp bounds on mixture components given in \Cref{mix-lem}.

 \noindent\textit{A Preliminary: Identified set of $\bf f$.} Note that for any type $t$ and $z \in \left\{0,1\right\}$,  $f_{Y_{z,d''}|T}(y|t)$ is independent of the data whenever $d'' \neq t\left(z\right)$, and so, is only constrained to be a density that is supported over (a subset of) $\left[y_L, y_U\right]$, the support of $Y_{z, d''}$; this immediately implies that the sharp identification region for the expectation of any such component is simply $\left[y_L, y_U\right]$. Given \Cref{Lem: RT-C} above, it now follows from Theorem 3.2 in \cite{vayalinkal_sharp_2024} that these conditions are sharp, i.e. any ${\bf f}$ satisfying these conditions is consistent with the data. We summarize the argument here, as follows.

First, note that the observed data depends only on the (i) the distribution of $Z$ ($F_Z$), (ii) the marginal  distribution of $D_z$ for each $z \in \{0,1\}$, and (iii) the conditional marginal distribution of $Y_{z,d}$ given $D_z = d, Z = z$ for all $d \in \{1, \dots, K\}$ and $z \in \{0,1\}$. For any joint distribution $\left(\left(Y_{z,d}: d \in \{0, \dots, K\}, z \in \{0,1\}\right),T,Z\right) \sim Q$, let ${\bf f}_Q$ be the vector of weighted response-type conditional densities implied by $Q$. Given ${\bf f}$ satisfying the conditions above, we construct a $Q$ with ${\bf f}_Q = {\bf f}$ as follows: define $Q_Z = F_Z$, $Q\left(T = t\right) = \int_{\mathcal Y} {\bf f}_{1|t}(y)d\mu(y)$, and define 
\begin{align*}
&Q\left(Y_{0,0} \leq y_{0,0}, \dots, Y_{0,K} \leq y_{0,K},Y_{1,0} \leq y_{1,0}, \dots, Y_{1,K} \leq y_{1,K}, T = t, Z = z \right)\\ 
=& \left(\prod_{k=0}^K \int_{y_L}^{y_{0,k}} {\bf f}_{(0,k)|t}(y) d\mu(y)\right)\left(\prod_{k=0}^K \int_{y_L}^{y_{1,k}} {\bf f}_{(0,k)|t}(y) d\mu(y)\right)Q\left(T = t\right)Q\left(Z = z\right)~.
\end{align*}
The above construction assumes that the potential outcome distributions are independent given $T$, but any dependence structure (copula) can be used, after conditioning on a value of $T$. 
Suppose we are given a $Q$ such that ${\bf f}_Q$ satisfies the conditions above. By construction, $Q_Z = F_Z$, $Q\left(D_z = d\right) = \sum_{t: t(z) = d} Q\left(T = t\right) = \sum_{t: t(z) = d} \int_{y_L}^{y_U} {\bf f}_{z|t}(y)d\mu(y) = P(D=d | Z = z)$ for all $d \in \{1, \dots, K\}$ and $z \in \{0,1\}$. This also implies $Q\left(D_z = 0\right) = P(D=0 | Z = z)$ by the definition of $\Theta_I$ and, finally, we have that for any $z \in \{0,1\}, d \in \{1, \dots, K\}$
\begin{align*}
Q\left(Y_{z,d} \leq y | D_z =d, Z = z\right) =& \sum_{t: t(z) = d} \int_{y_L}^{y} {\bf f}_{z|t}(y)d\mu(y) =  \int_{y_L}^{y} \sum_{t: t(z) = d} {\bf f}_{z|t}(y)d\mu(y)\\ 
=& \int_{y_L}^{y} f_{Y,D = d|Z =z}(y) d\mu(y) = P(Y \leq y | D = d, Z = z)~,
\end{align*}
as required, showing that the above conditions define the identified set for ${\bf f}$. 


 \noindent\textit{Remainder of proof.}
Since the two mixtures given by \eqref{EC1} and \eqref{EC2} do not share any components, the above result reduces the problem of finding bounds on the conditional expectation of $Y_{d,z}$ given $T$ to the problem of finding sharp bounds on expectations of mixture components. Therefore, we now complete the proof using the results given in \Cref{mix-lem}, as follows. 

\textit{Proof of (i) and (ii):} For any type $\left(d', d\right)$ and any $z \in \left\{0,1\right\}$, the weighted conditional density $\mathbb P[T=(d',d)]\times f_{Y_{z,T(z)}|T}(y|d',d)$ only appears in at most one of \eqref{EC1} and \eqref{EC2}. Therefore, sharp bounds on the expectation of any such component can be obtained as the bounds of \cite{horowitz_identification_1995} (HM), which are the bounds provided in \Cref{mix-lem}(i) evaluated at the smallest feasible value of $\gamma_k$. This immediately implies the validity and sharpness of (ii) and the last two parts of (i) above. 
This also implies that the bounds in the first part of (i) are valid, as follows: (I) LCDE($d|d,d$) consists of the difference in expectation of two such components, and (II) the lower (upper) bound is given by the HM lower (upper) bound of the first component minus the HM upper (lower) bound of the second component. For sharpness, first note that LCDE($d|d,d$) consists of the difference in expectation of two components, each of which belongs to a \textit{different} mixture of the two defined by \eqref{EC1}  and \eqref{EC2}. Since these two mixtures do not share any components (only weights), the two HM bounds can be attained jointly whenever the weights $\underline{\gamma}^{1,r}_{d,d}$ and $\underline{\gamma}^{0,r}_{d,d}$ are jointly feasible. The result now follows by noting that  $\underline{\gamma}^{1,r}_{d,d}$ and $\underline{\gamma}^{0,r}_{d,d}$ are jointly feasible if and only if $\mathbb P(D=d|Z=1)\underline{\gamma}^{1,r}_{d,d} = \mathbb P(D=d|Z=0)\underline{\gamma}^{0,r}_{d,d} = \underline{p}^r_{d,d'}$ belongs to the identified set for $p_{d,d'}$ which is true by definition of  $\underline{p}^r_{d,d'}$. 

\textit{Proof of (iii):} Finally, for (iii), note that 
\begin{align*}
 &\  \sum _{d=l}^{l'} \frac{p_{d,d}}{\sum _{d=l}^{l'}p_{d,d}}\text{LCDE}(d|d,d)
  =
  \sum _{d=l}^{l'} \frac{p_{d,d}}{\sum _{d=l}^{l'}p_{d,d}}\mathbb E[Y_{1,d}-Y_{0,d}|T=(d,d)]\\
  =&\ \sum _{d=l}^{l'} \frac{p_{d,d}}{\sum _{d=l}^{l'}p_{d,d}}\mathbb E[Y_{1,d}|T=(d,d)]- \sum _{d=l}^{l'} \frac{p_{d,d}}{\sum _{d=l}^{l'}p_{d,d}}\mathbb E[Y_{0,d}|T=(d,d)]\\
  =&\ \sum _{d=l}^{l'} \frac{p_{d,d}}{\sum _{d=l}^{l'}p_{d,d}}\int_\mathcal Y y {\bf f}_{1|d,d}(y)\ d\mu\left(y\right)- \sum _{d=l}^{l'} \frac{p_{d,d}}{\sum _{d=l}^{l'}p_{d,d}}\int_\mathcal Y y {\bf f}_{0|d,d}(y)\ d\mu\left(y\right)~,
\end{align*}
where the first term is the expectation of the aggregation of components of the mixture  \eqref{EC1} and the second term is the expectation of the aggregation of components of the mixture \eqref{EC2}; we can now use the same argument as above, but now based on \Cref{mix-lem}(ii), as follows. First, suppose that $\left\{\left(p_{d,d}: l \leq d \leq l'\right)\ \middle|\ p \in \Theta_{I}(\mathcal R_T) \right\}$ is a singleton, so that the weights are known and there is no optimization required: sharpness now follows immediately from \Cref{mix-lem}(ii) since the two mixtures do not share any components. Finally, when this is not the case, the sharp bounds are obtained by maximizing (resp. minimizing) the pointwise (in $\left(p_{d,d}: l \leq d \leq l'\right)$) upper (resp. lower) bound over the identified set for $\left(p_{d,d}: l \leq d \leq l'\right)$, which is exactly the bounds given in (iii). 
\end{proof}

\clearpage

\section{Additional Empirical Tables and Figures for Job Corps RCT}\label{app:jc_app}

\subsection{Summary Statistics}
Table \ref{sumstats} presents summary statistics for the sample of individuals who have non-missing values for weekly earnings and hours for every week following random assignment. Means and standard deviations for a number of baseline and post-randomization variables are reported separately by treatment status. Consistent with successful randomization in the National Job Corps Study, the table shows that there are no statistical differences in the means of demographic, education, background and baseline employment/income variables across treatment and control groups. This finding aligns with the previous evaluations of the National Job Corps Study.

Table \ref{sumstats} also shows economically and statistically significant differences in employment and earnings outcomes by treatment status, post randomization. We see that 52 weeks after randomization, treatment group hours and earnings are lower than the control group, but 104 weeks after randomization, treatment group hours and earnings exceed the control group. After 208 weeks, the hours and earnings of the treatment group are approximately 8\% and 14\% higher, respectively, than those of the control group. These differences are statistically significant and are consistent with previous evaluations of the National Job Corps Study.

\begin{table}[htbp]
    \centering
    \resizebox{\textwidth}{!}{
    \begin{tabular}{l*{3}{cc}}
        \toprule &
        \multicolumn{2}{c}{Control} & \multicolumn{2}{c}{Treated} & \multicolumn{2}{c}{Difference} \\
        \cmidrule(lr){2-3}
        \cmidrule(lr){4-5}
        \cmidrule(lr){6-7}
        \input empiric_tbls_figs/sum_stats/sumstats_weighted_1.tex 
        \textbf{Household income} & & & & & \\
        \input empiric_tbls_figs/sum_stats/sumstats_weighted_2.tex 
        \textbf{Personal income} & & & & & \\
        \input empiric_tbls_figs/sum_stats/sumstats_weighted_3.tex 
        \textbf{At baseline} & & & & & \\
        \input empiric_tbls_figs/sum_stats/sumstats_weighted_4.tex 
        \textbf{Post randomization} & & & & & \\
        \input empiric_tbls_figs/sum_stats/sumstats_weighted_5.tex

\input empiric_tbls_figs/sum_stats/sumstats_weighted_6.tex 
        \midrule
        \input empiric_tbls_figs/sum_stats/sumstats_weighted_7.tex 
        \bottomrule
        \multicolumn{7}{l}{\footnotesize Notes: Weekly earnings calculated as the sum of total earnings in a given week and are not conditional on}\\
        \multicolumn{7}{l}{\footnotesize employment (i.e., includes 0s for the unemployed).} \\
    \end{tabular}}
    \caption{Summary statistics by treatment status \label{sumstats}}
\end{table}

\subsection{Joint Distribution of Wages and Amenities}

\begin{figure}[h!]
    \centering
    \includegraphics[width=.8\linewidth]{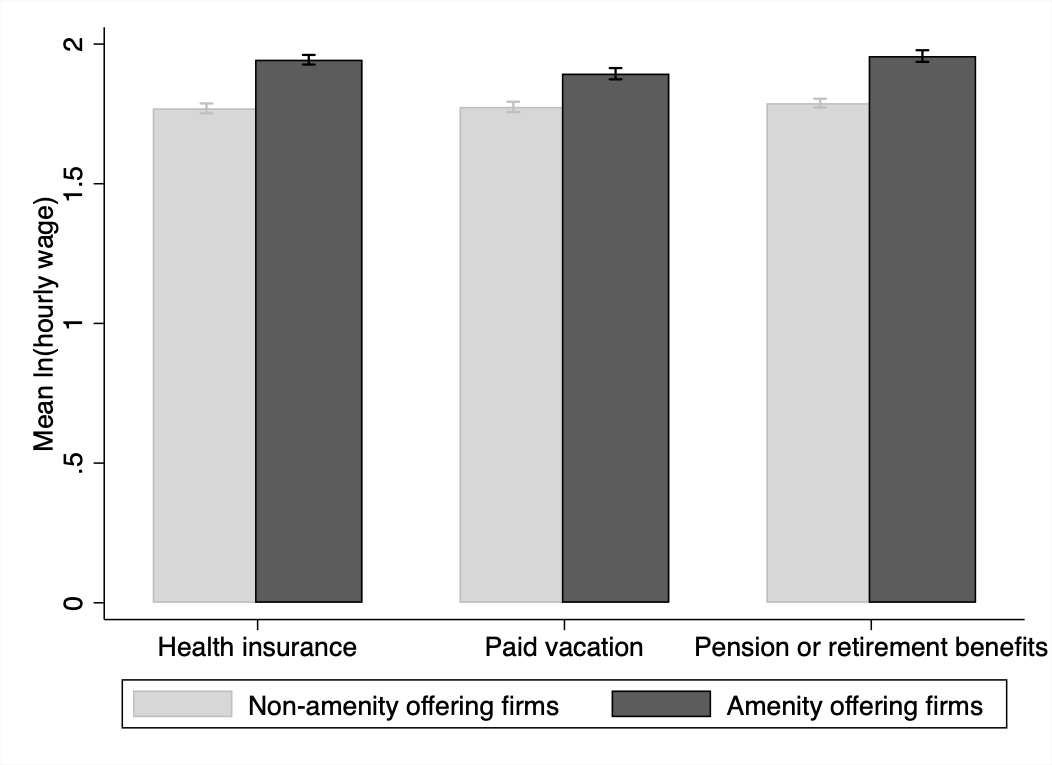}
    \begin{tabular}{p{.8\linewidth}}
    \footnotesize{Notes: Hourly wage calculated as weekly earnings divided by weekly hours for the employed.} 
    \end{tabular}
    \caption{Mean ln(hourly Wage) by firm amenity provision at week 90}
    \label{fig:hist_wk90}
\end{figure}

\begin{figure}[h!]
    \centering
    \subfigure[control units]{\includegraphics[width=.49\linewidth]{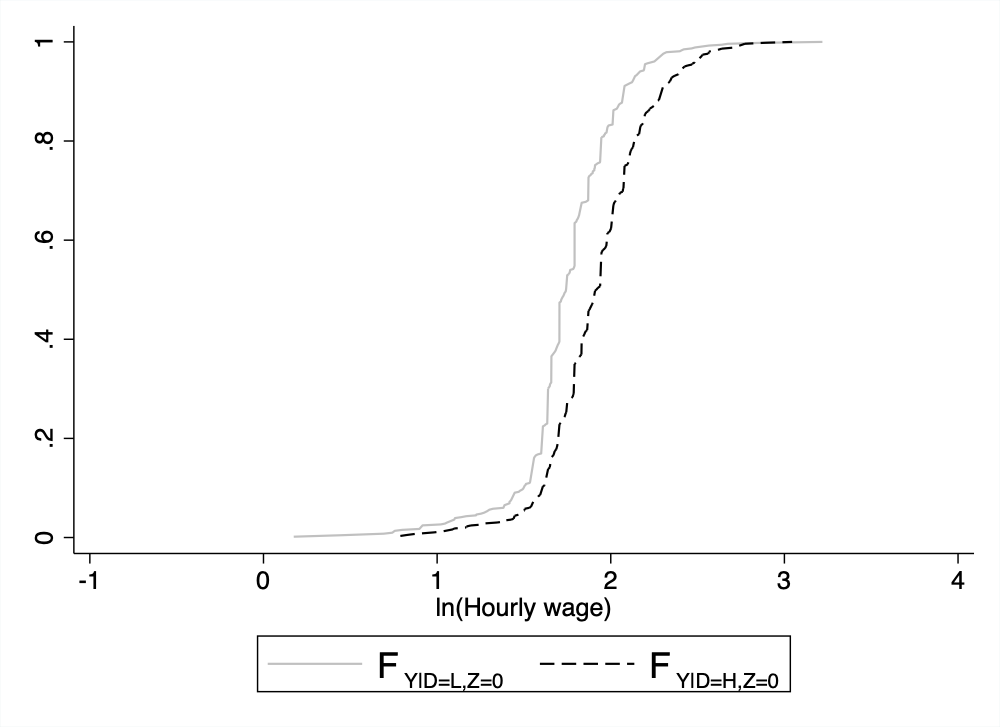}}
    \subfigure[treated units]{\includegraphics[width=.49\linewidth]{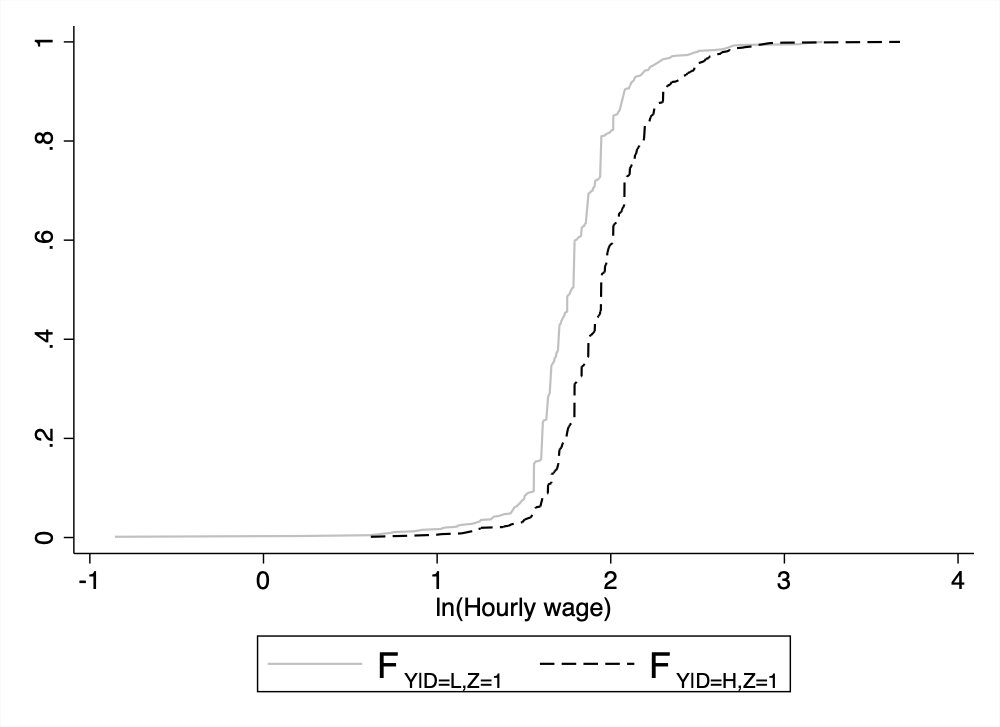}}
    \begin{tabular}{p{.8\linewidth}}
    \footnotesize{Notes: Hourly wage calculated as weekly earnings divided by weekly hours for the employed.} 
    \end{tabular}
    \caption{Cumulative distribution function by firm type at week 90, amenity=health}
    \label{fig:ecdf_health90}
\end{figure}

Figure \ref{fig:hist_wk90} presents the mean log wage at week 90 according to whether firms provide amenities. At week 90 (208), mean wages at firms that offer amenities are approximately 15\% (18\%) higher compared to firms that do not. \Cref{fig:ecdf_health90} presents the empirical cumulative distribution functions of log wage by firm type, classifying firms according to the provision of health insurance, for treatment and control groups at week 90. For both treated and control units, the distribution of log wages for firms that provide health insurance stochastically dominates the distribution of log wages for firms that do not. This evidence suggests that firms providing amenities pay higher wages than firms that do not.\footnote{Of course, it is possible that firms pay compensating differentials which causally reduce wages. The evidence presented here shows that the variation between firms dominates the variation within firms. This is consistent with evidence in \cite{lamadon_imperfect_2022} who show that the high-amenities firms are also the more productive firms.} A natural follow-up question is whether Job Corps affects the sorting of workers into amenity-providing firms.

\subsection{Differential Sorting of Treatment and Control Workers}\label{sec:diff_sort_treat_ctl}
\cite{lee_training_2009} focuses on the potential for job training to affect labor supply along the extensive margin but ignores an additional margin of labor supply: firm choice. If there is scope for job training to affect worker sorting to firms, sample selection is multilayered. Table \ref{schoct390} presents the probability of working in a firm (conditional on employment) that provides observable amenities, at week 90, according to the status of treatment. The evidence shows that treated individuals are more likely to work at firms with job amenities in all but one case (and we also find that this trend persists across all weeks). This is consistent with the evidence presented in \cite{schochet_does_2008}.\footnote{\label{fn:type_dist_discussion}In \Cref{app:type_dist_appendix} we focus on the amenities we use to classify firm types -- health benefits $H$, retirement/pension benefits $R$ and paid vacation $V$ -- and categorize jobs into eight mutually exclusive categories based on the amenities available. Appendix Table \ref{typedist90} presents the distribution of workers by the amenity category in which their job falls at week 90. Treated workers are approximately 15\% more likely to work in jobs that offer all amenities and 10\% less likely to work in jobs that offer no amenities. Reinforcing the finding from Figure \ref{fig:hist_wk90}, hourly wages are approximately 22\% higher in jobs with all amenities compared to jobs without amenities. These trends persist qualitatively throughout all weeks.}

\begin{table}[htbp]
    \centering
    \begin{tabular}{l*{3}{cc}}
        \toprule &
        \multicolumn{2}{c}{Control} & \multicolumn{2}{c}{Treated} & \multicolumn{2}{c}{Difference} \\
        \cmidrule(lr){2-3}
        \cmidrule(lr){4-5}
        \cmidrule(lr){6-7}
        \input empiric_tbls_figs/amen_dist/scho_t3_wk90_weighted_A_slides.tex 
        \midrule
        \input empiric_tbls_figs/amen_dist/scho_t3_wk90_weighted_B_slides.tex
        \bottomrule
        \multicolumn{7}{l}{\footnotesize Notes: Control and treatment probabilities have interpretation as $P[D=H|D>0,Z=z]$ for $z \in \{0,1\}$,}\\
        \multicolumn{7}{l}{\footnotesize respectively, when classifying firms as type $H$ if they provide a given amenity.}\\
    \end{tabular}
    \caption{Probability of working at amenity-providing firm at week 90 (conditional on employment) \label{schoct390}}
\end{table}



Taken together, \Cref{fig:hist_wk90,fig:ecdf_health90} and Table \ref{schoct390} show: (i) firms that provide amenities pay higher wages than firms that do not and (ii) Job Corps affects the sorting of workers into amenity-providing firms. We therefore conclude that Job Corps training not only affects labor supply along the extensive margin but also along the additional margin of firm choice. As a result, sample selection is multilayered motivating the use of our bounds.

\subsection{Additional descriptive figures and tables}\label{app:appdescript}
\Cref{app:appdescript} provides descriptive figures and tables from \Cref{sec:emp_jobcorps} for weeks 135, 180 and 208. 

\begin{figure}[h!]
    \centering
    \includegraphics[width=.8\linewidth]{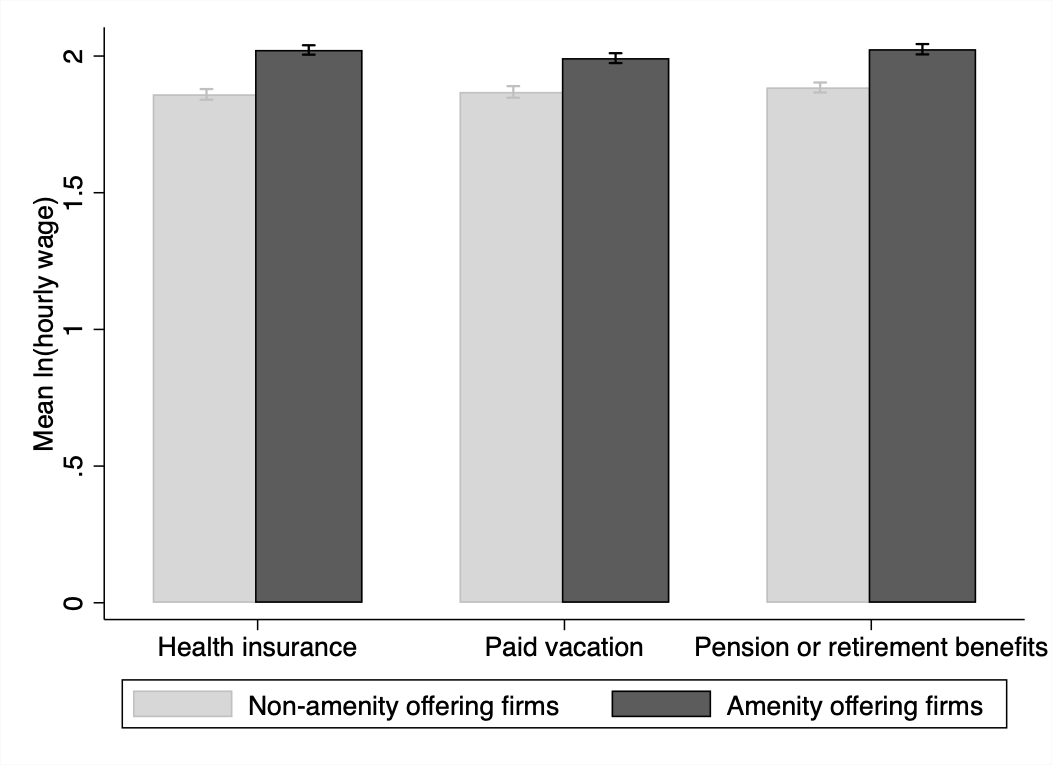}
    \begin{tabular}{p{.8\linewidth}}
    \footnotesize{Notes: Hourly wage calculated as weekly earnings divided by weekly hours for the employed.} 
    \end{tabular}
    \caption{Mean ln(hourly Wage) by firm amenity provision at week 135}
    \label{fig:hist_wk135}
\end{figure}

\begin{figure}[h!]
    \centering
    \includegraphics[width=.8\linewidth]{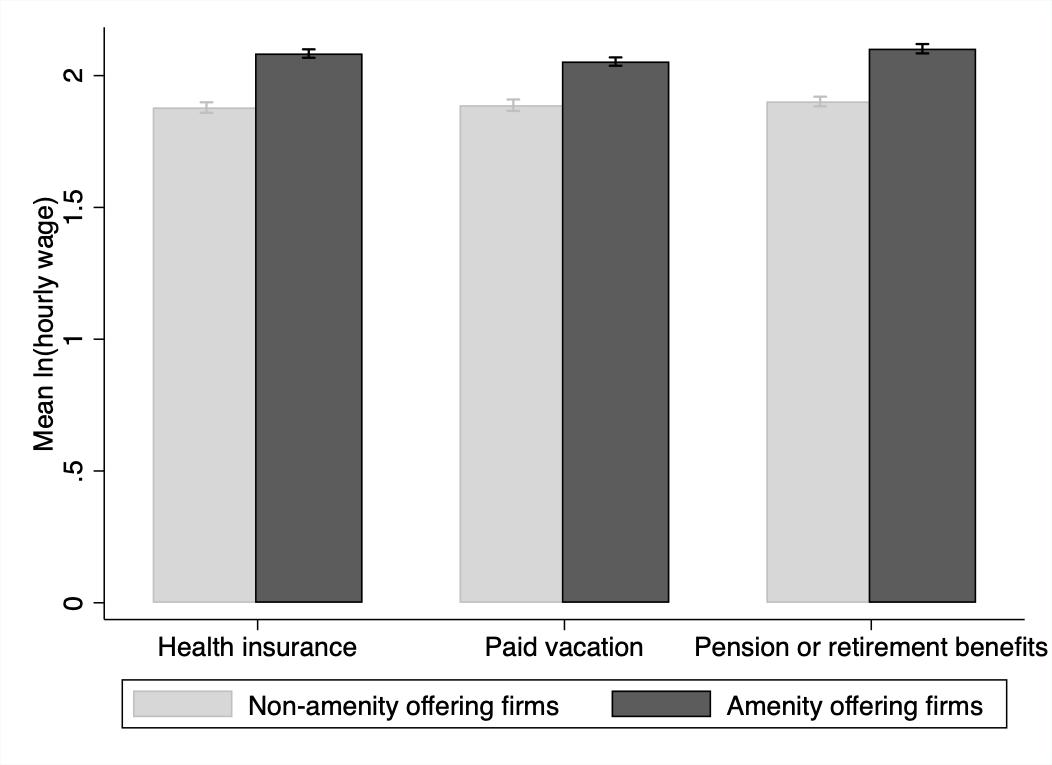}
    \begin{tabular}{p{.8\linewidth}}
    \footnotesize{Notes: Hourly wage calculated as weekly earnings divided by weekly hours for the employed.} 
    \end{tabular}
    \caption{Mean ln(hourly Wage) by firm amenity provision at week 180}
    \label{fig:hist_wk180}
\end{figure}

\begin{figure}[h!]
    \centering
    \includegraphics[width=.8\linewidth]{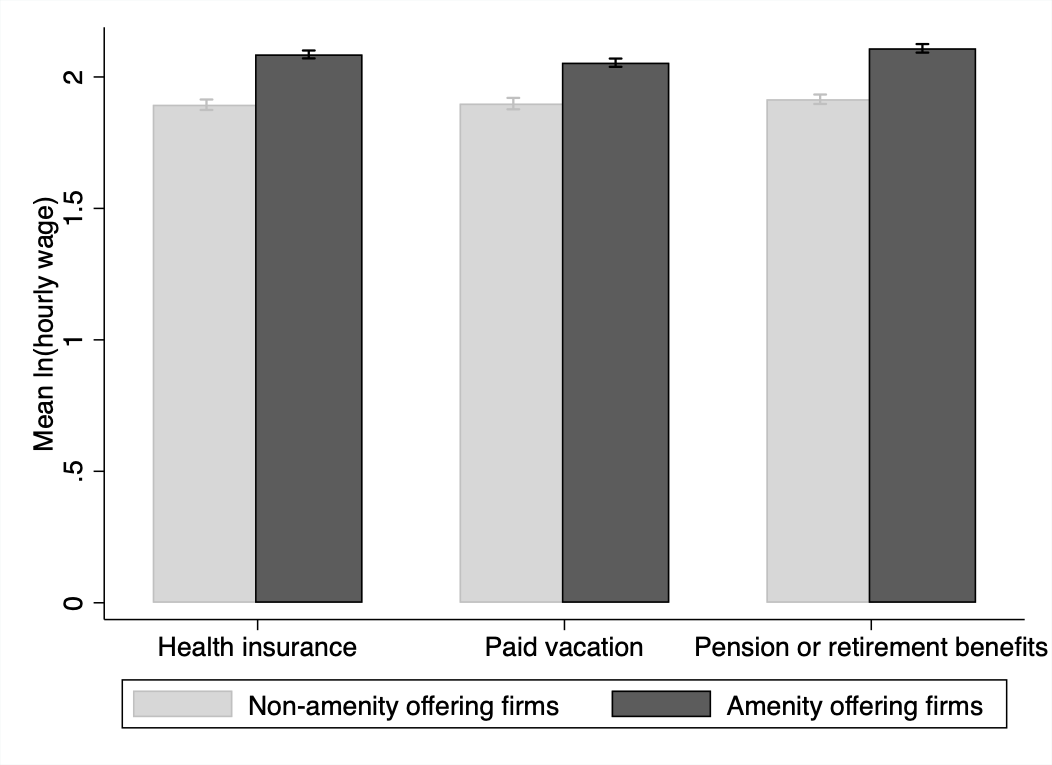}
    \begin{tabular}{p{.8\linewidth}}
    \footnotesize{Notes: Hourly wage calculated as weekly earnings divided by weekly hours for the employed.} 
    \end{tabular}
    \caption{Mean ln(hourly Wage) by firm amenity provision at week 208}
    \label{fig:hist_wk208}
\end{figure}

\clearpage

\begin{figure}[h!]
    \centering
    \subfigure[control units]{
    \includegraphics[width=.49\linewidth]{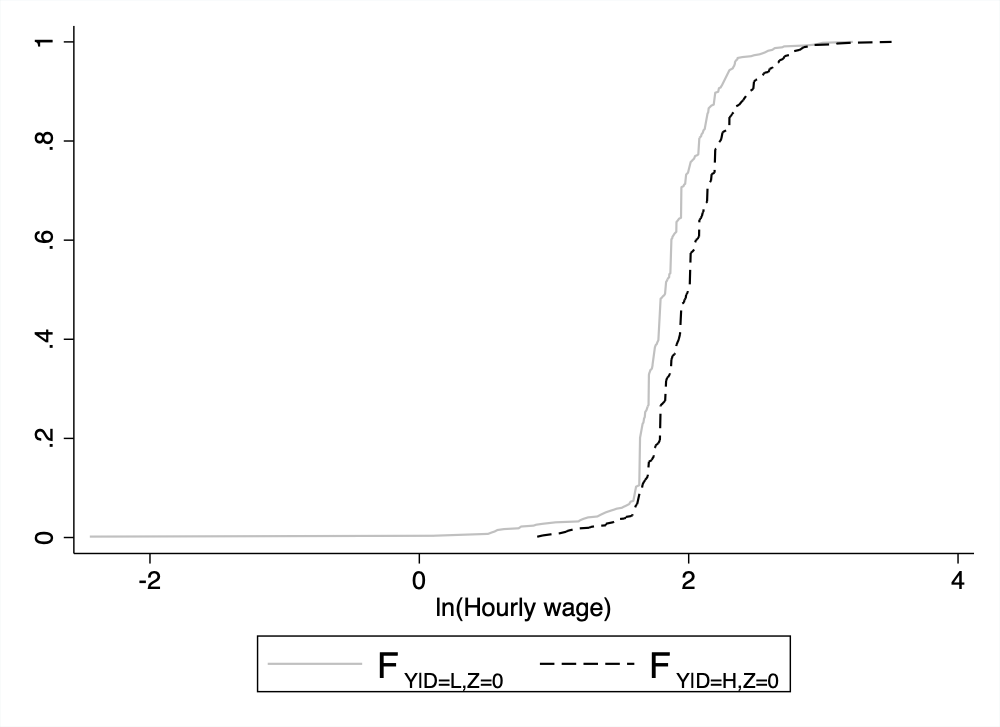}}
    \subfigure[treated units]{\includegraphics[width=.49\linewidth]{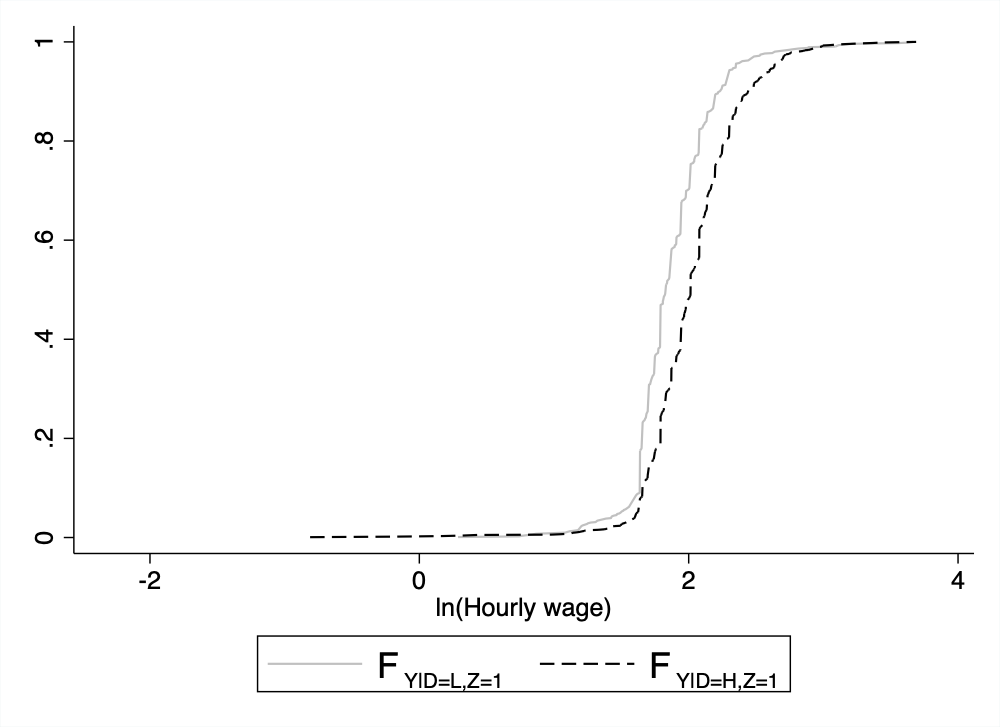}}  
    \begin{tabular}{p{.8\linewidth}}
    \footnotesize{Notes: Hourly wage calculated as weekly earnings divided by weekly hours for the employed.} 
    \end{tabular}
    \caption{Cumulative distribution function by firm type at week 135, amenity=health}
    \label{fig:}
\end{figure}

\begin{figure}[h!]
    \centering
    \subfigure[control units]{
    \includegraphics[width=.49\linewidth]{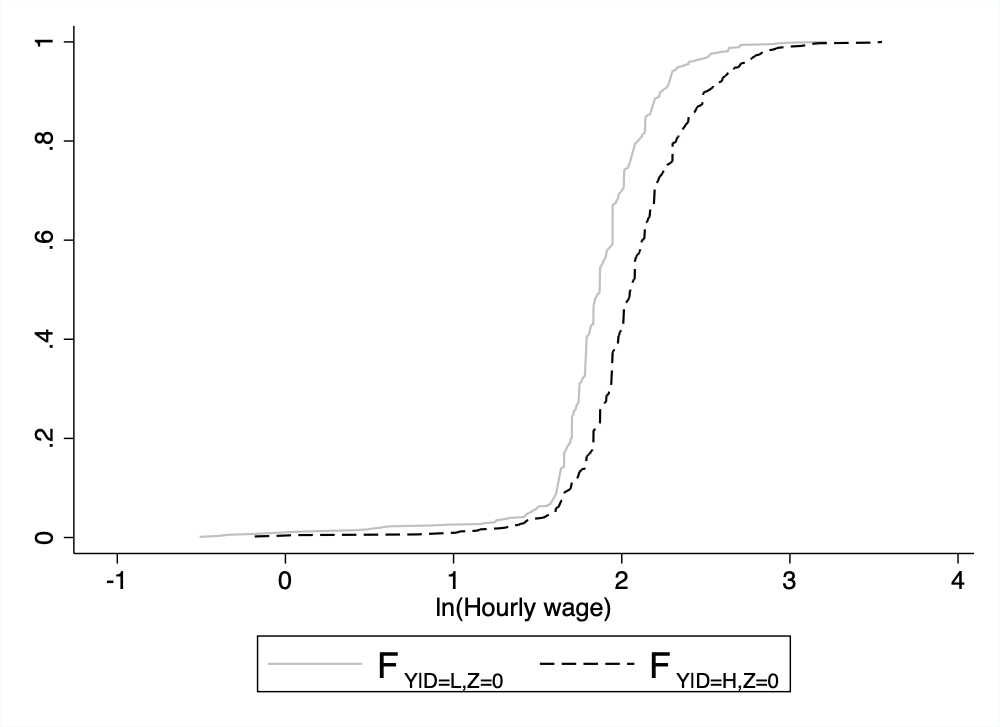}}
    \subfigure[treated units]{\includegraphics[width=.49\linewidth]{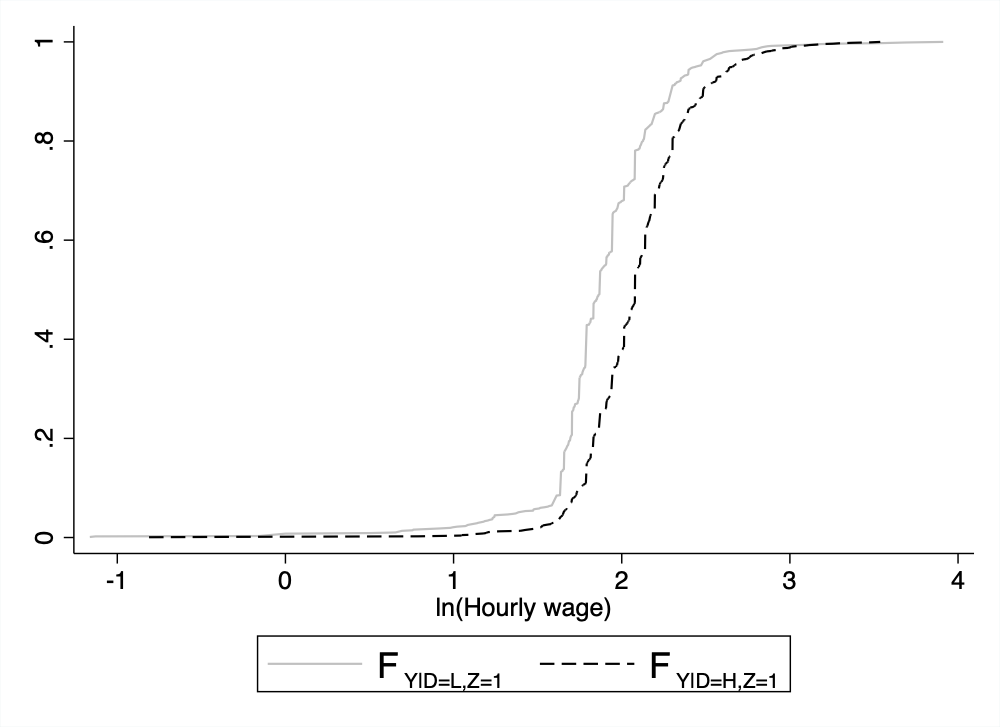}}
    \begin{tabular}{p{.8\linewidth}}
    \footnotesize{Notes: Hourly wage calculated as weekly earnings divided by weekly hours for the employed.} 
    \end{tabular}
    \caption{Cumulative distribution function by firm type at week 180, amenity=health}
    \label{fig:}
\end{figure}

\begin{figure}[h!]
    \centering
    \subfigure[control units]{
    \includegraphics[width=.49\linewidth]{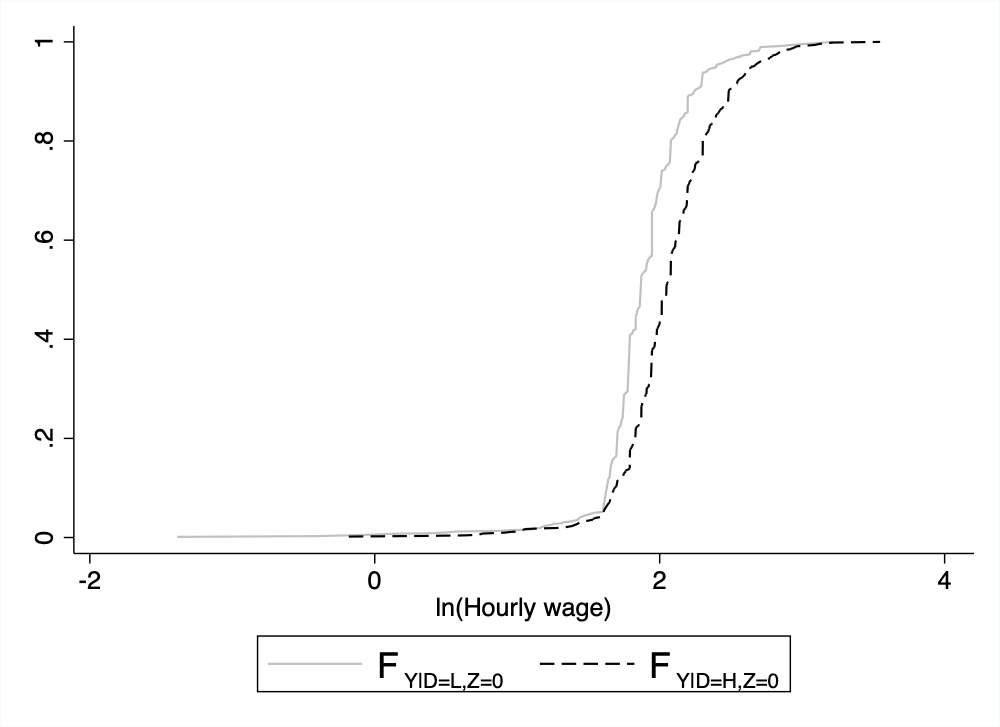}}
    \subfigure[treated units]{\includegraphics[width=.49\linewidth]{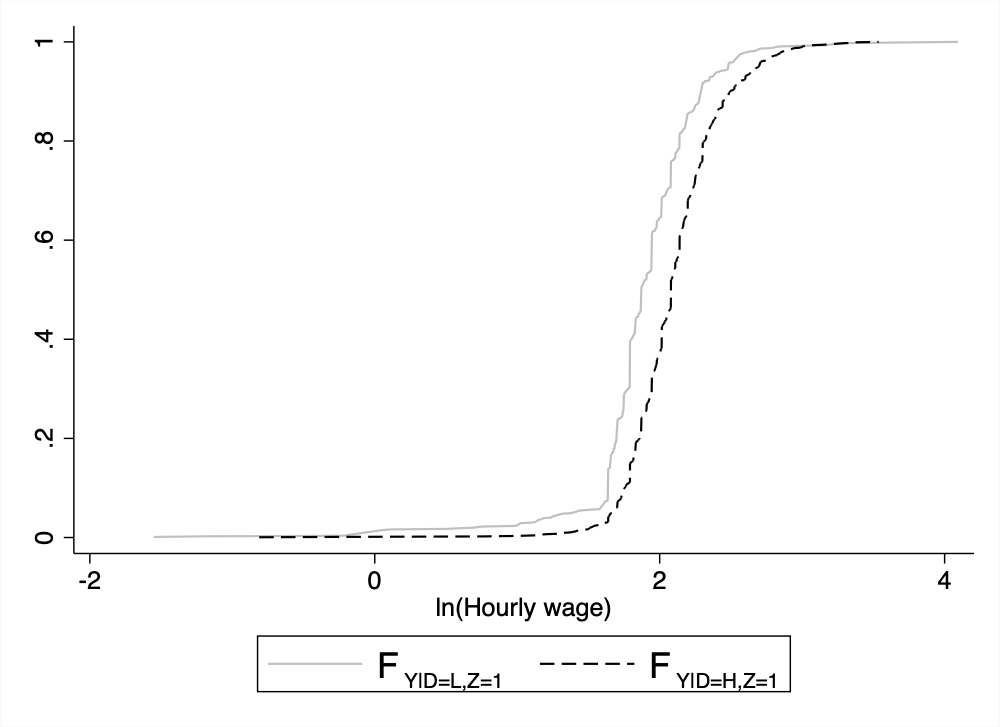}}
    \begin{tabular}{p{.8\linewidth}}
    \footnotesize{Notes: Hourly wage calculated as weekly earnings divided by weekly hours for the employed.} 
    \end{tabular}
    \caption{Cumulative distribution function by firm type at week 208, amenity=health}
    \label{fig:}
\end{figure}

\clearpage

\begin{table}[htbp]
    \centering
    \begin{tabular}{l*{3}{cc}}
        \toprule &
        \multicolumn{2}{c}{Control} & \multicolumn{2}{c}{Treated} & \multicolumn{2}{c}{Difference} \\
        \cmidrule(lr){2-3}
        \cmidrule(lr){4-5}
        \cmidrule(lr){6-7}
        \input empiric_tbls_figs/amen_dist/scho_t3_wk135_weighted_A_slides.tex 
        \midrule
        \input empiric_tbls_figs/amen_dist/scho_t3_wk135_weighted_B_slides.tex
        \bottomrule
    \end{tabular}
    \caption{Probability of working at amenity-providing firm at week 135 (conditional on employment)}
\end{table}

\begin{table}[htbp]
    \centering
    \begin{tabular}{l*{3}{cc}}
        \toprule &
        \multicolumn{2}{c}{Control} & \multicolumn{2}{c}{Treated} & \multicolumn{2}{c}{Difference} \\
        \cmidrule(lr){2-3}
        \cmidrule(lr){4-5}
        \cmidrule(lr){6-7}
        \input empiric_tbls_figs/amen_dist/scho_t3_wk180_weighted_A_slides.tex 
        \midrule
        \input empiric_tbls_figs/amen_dist/scho_t3_wk180_weighted_B_slides.tex
        \bottomrule
    \end{tabular}
    \caption{Probability of working at amenity-providing firm at week 180 (conditional on employment)}
\end{table}

\begin{table}[htbp]
    \centering
    \begin{tabular}{l*{3}{cc}}
        \toprule &
        \multicolumn{2}{c}{Control} & \multicolumn{2}{c}{Treated} & \multicolumn{2}{c}{Difference} \\
        \cmidrule(lr){2-3}
        \cmidrule(lr){4-5}
        \cmidrule(lr){6-7}
        \input empiric_tbls_figs/amen_dist/scho_t3_wk208_weighted_A_slides.tex 
        \midrule
        \input empiric_tbls_figs/amen_dist/scho_t3_wk208_weighted_B_slides.tex
        \bottomrule
    \end{tabular}
    \caption{Probability of working at amenity-providing firm at week 208 (conditional on employment)}
\end{table}

\clearpage

\clearpage

\subsection{Distribution of Job Types}\label{app:type_dist_appendix} \Cref{app:type_dist_appendix} presents the distribution of workers by the amenity category their job falls into, as discussed in \Cref{fn:type_dist_discussion}, for all weeks of interest (90, 135, 180 and 208). The sample size in these distributions decreases to 6,232 individuals (=2,454 control units + 3,778 treated units). Constructing these distributions requires restricting the sample to workers who have non-missing amenity status for all three amenities simultaneously across weeks of interest. This is a stronger restriction than only requiring non-missing amenity status across weeks on a per amenity basis, as is the case for our primary sample of analysis. \\

\begin{table}[htbp]
    \centering
    \begin{tabular}{l*{8}{c}}
        \toprule &
        \multicolumn{2}{c}{Control} & \multicolumn{2}{c}{Treated} & \multicolumn{2}{c}{Difference} & \multicolumn{2}{c}{Pooled} \\
        \cmidrule(lr){2-3}
        \cmidrule(lr){4-5}
        \cmidrule(lr){6-7}
        \cmidrule(lr){8-9}
        \input empiric_tbls_figs/types_dist/typesdist_wk90_weighted_A.tex 
        \midrule
        \input empiric_tbls_figs/types_dist/typesdist_wk90_weighted_B.tex
        \bottomrule
        \multicolumn{8}{l}{\footnotesize Notes: Hourly wage calculated as weekly earnings divided by weekly hours for the employed.}\\
    \end{tabular}
    \caption{Distribution of job types at week 90 (conditional on employment) \label{typedist90}}
\end{table}

\begin{table}[htbp]
    \centering
    \begin{tabular}{l*{8}{c}}
        \toprule &
        \multicolumn{2}{c}{Control} & \multicolumn{2}{c}{Treated} & \multicolumn{2}{c}{Difference} & \multicolumn{2}{c}{Pooled} \\
        \cmidrule(lr){2-3}
        \cmidrule(lr){4-5}
        \cmidrule(lr){6-7}
        \cmidrule(lr){8-9}
        \input empiric_tbls_figs/types_dist/typesdist_wk135_weighted_A.tex 
        \midrule
        \input empiric_tbls_figs/types_dist/typesdist_wk135_weighted_B.tex
        \bottomrule
        \multicolumn{8}{l}{\footnotesize Notes: Hourly wage calculated as weekly earnings divided by weekly hours for the employed.}\\
    \end{tabular}
    \caption{Distribution of job types at week 135 (conditional on employment)}
\end{table}

\begin{table}[htbp]
    \centering
    \begin{tabular}{l*{8}{c}}
        \toprule &
        \multicolumn{2}{c}{Control} & \multicolumn{2}{c}{Treated} & \multicolumn{2}{c}{Difference} & \multicolumn{2}{c}{Pooled} \\
        \cmidrule(lr){2-3}
        \cmidrule(lr){4-5}
        \cmidrule(lr){6-7}
        \cmidrule(lr){8-9}
        \input empiric_tbls_figs/types_dist/typesdist_wk180_weighted_A.tex 
        \midrule
        \input empiric_tbls_figs/types_dist/typesdist_wk180_weighted_B.tex
        \bottomrule
        \multicolumn{8}{l}{\footnotesize Notes: Hourly wage calculated as weekly earnings divided by weekly hours for the employed.}\\
    \end{tabular}
    \caption{Distribution of job types at week 180 (conditional on employment)}
\end{table}

\begin{table}[htbp]
    \centering
    \begin{tabular}{l*{8}{c}}
        \toprule &
        \multicolumn{2}{c}{Control} & \multicolumn{2}{c}{Treated} & \multicolumn{2}{c}{Difference} & \multicolumn{2}{c}{Pooled} \\
        \cmidrule(lr){2-3}
        \cmidrule(lr){4-5}
        \cmidrule(lr){6-7}
        \cmidrule(lr){8-9}
        \input empiric_tbls_figs/types_dist/typesdist_wk208_weighted_A.tex 
        \midrule
        \input empiric_tbls_figs/types_dist/typesdist_wk208_weighted_B.tex
        \bottomrule
        \multicolumn{8}{l}{\footnotesize Notes: Hourly wage calculated as weekly earnings divided by weekly hours for the employed.}\\
    \end{tabular}
    \caption{Distribution of job types at week 208 (conditional on employment)}
\end{table}

\clearpage

\subsection{Replication of \cite{lee_training_2009} Bounds}\label{app:leebounds_rep_app} As discussed in Section 5, Appendix \Cref{leeboundsjc} reports our replication of the bounds reported in \cite{lee_training_2009} for weeks 90, 135, 180 and 208 along with the trimming proportion $p\equiv\mathbb P(AE)$, e.g., the share of the always-employed among individuals receiving job training. We do not report bounds for week 45 since we discovered that the monotonicity assumption is violated.

\Cref{leeboundsjc} reports Lee's bounds when treating $ln(\text{hourly wage})$ as a continuous variable (as we do throughout the paper). All quantities are very close to the estimates in \cite{lee_training_2009}. There is a small difference that arises in the bounds due to Lee's use of vingtiles of $ln(\text{hourly wage})$. \Cref{leeboundsjc_vingtiles} shows that when we use vingtiles of $ln(\text{hourly wage})$, the bounds are identical to the ones reported in \cite{lee_training_2009}. \\

\begin{table}[htbp]
    \centering
    \begin{tabular}{l*{3}{cc}}
        \toprule
        & & & & \multicolumn{2}{c}{$\mathbb E[Y_{1,D_1}-Y_{0,D_0}|D_0>0,D_1>0]$} \\
        \cmidrule(lr){5-6}
        &  $\mathbb P[D>0|Z=0]$ &  $\mathbb P[D>0|Z=1]$ &$p$&      lower&       upper\\
        \midrule
        \input empiric_tbls_figs/bounds/lee_bounds_weighted
        \bottomrule
        \multicolumn{6}{l}{\footnotesize Notes: Treatment bounds are for ln(hourly wage); hourly wage calculated as weekly earnings divided by}\\
        \multicolumn{6}{l}{\footnotesize weekly hours for the employed. Propensity scores and trimming proportion are numerically equivalent to} \\
        \multicolumn{6}{l}{\footnotesize Lee; slight numerical difference in bounds occurs as Lee uses vingtiles of ln(hourly wage) and we do not.} \\
        \multicolumn{6}{l}{\footnotesize See \Cref{leeboundsjc_vingtiles} for identical treatment bounds to Lee.}
    \end{tabular}
    \caption{Lee's bounds: continuous ln(hourly wage)\label{leeboundsjc}}
\end{table}

\begin{table}[htbp]
    \centering
    \begin{tabular}{l*{3}{cc}}
        \toprule
        & & & & \multicolumn{2}{c}{$\mathbb E[Y_{1,D_1}-Y_{0,D_0}|D_0>0,D_1>0]$} \\
        \cmidrule(lr){5-6}
        &  $\mathbb P[D>0|Z=0]$ &  $\mathbb P[D>0|Z=1]$ &$p$&      lower&       upper\\
        \midrule
        \input empiric_tbls_figs/bounds/lee_bounds_vingtiles
        \bottomrule
        \multicolumn{6}{l}{\footnotesize Notes: Treatment bounds are for vingtiles of ln(hourly wage); hourly wage calculated as weekly}\\
        \multicolumn{6}{l}{\footnotesize earnings divided by weekly hours for the employed.}
    \end{tabular}
    \caption{Lee's bounds: vingtiles of ln(hourly wage)\label{leeboundsjc_vingtiles}}
\end{table}

\clearpage

\subsection{Identified Sets for Response Types}\label{app:pscores}
\Cref{app:pscores} provides propensity scores for when classifying firm type based on the provision of paid vacation and retirement/pension benefits. \Cref{app:pscores} also provides the identified sets for $p_{L,L},p_{H,H}$ for all weeks and amenities, with the exception of week 90 for health insurance which is provided in the main text.

\begin{table}[htbp]
    \centering
    \begin{tabular}{l*{4}{c}}
        \toprule
        &  $\mathbb P[D=H|Z=0]$ & $\mathbb P[D=H|Z=1]$&  $\mathbb P[D=L|Z=0]$&  $\mathbb P[D=L|Z=1]$\\
        \midrule
        \input empiric_tbls_figs/pscores/vacat_pscores_weighted.tex
        \bottomrule
    \end{tabular}
    \caption{Propensity scores by week, amenity: paid vacation\label{vacatpscores}}
\end{table}

\begin{figure}
    \centering
    \includegraphics[width=\textwidth]{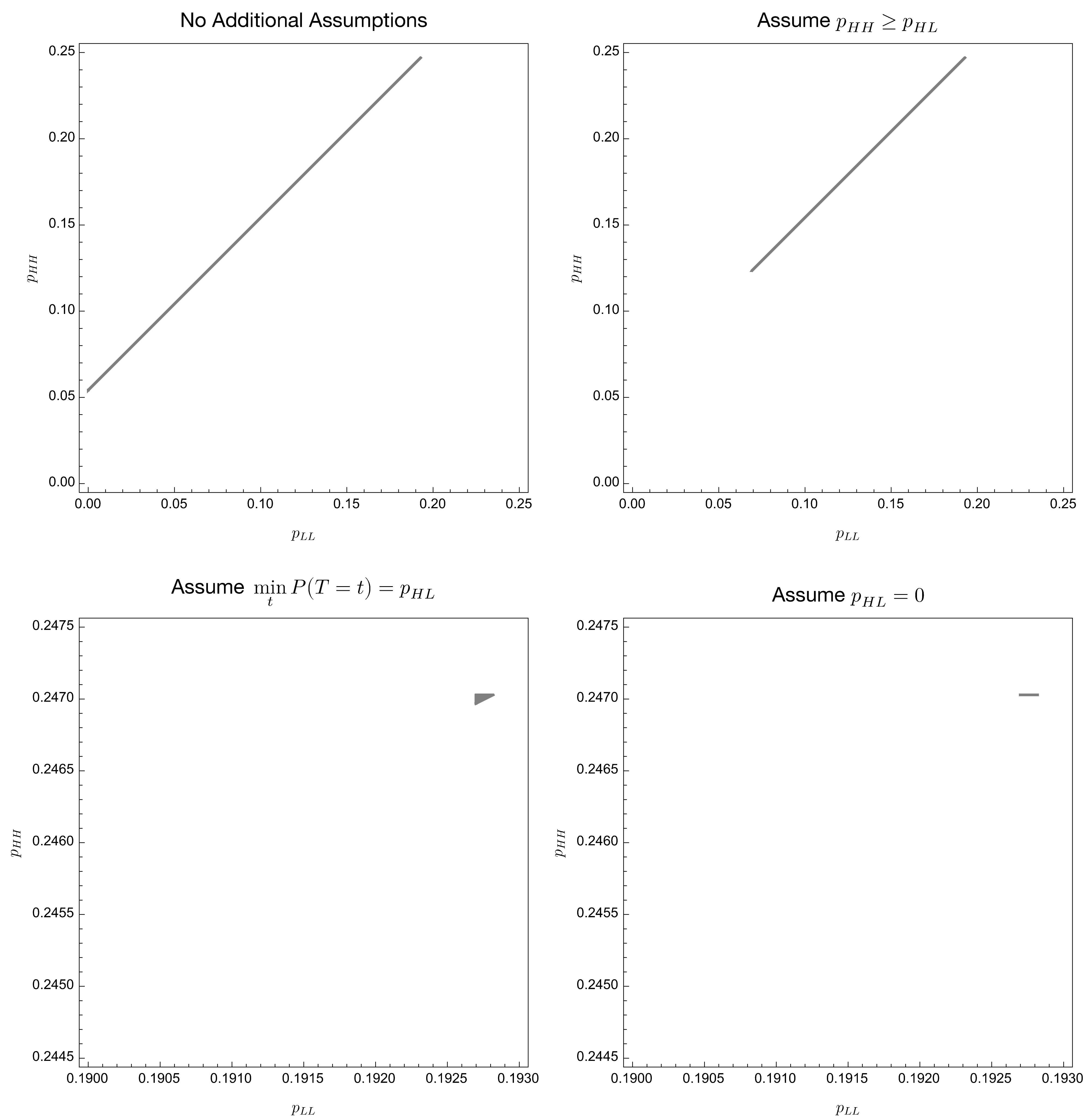}
    \caption{Identified set for $p_{L,L},p_{H,H}$ at week 90. Amenity=paid vacation. The first panel illustrates the identified set when $\mathcal R_T = \{\text{Assumption } \ref{Ass:LM}\}$. The remaining panels illustrate the identified set of additional assumptions imposed over  \text{Assumption } \ref{Ass:LM}. The scale of the axis in the bottom two panels is shrunk to see more clearly the identified set.}
    \label{fig:Vacat90idset}
\end{figure}

\begin{figure}
    \centering
    \includegraphics[width=\textwidth]{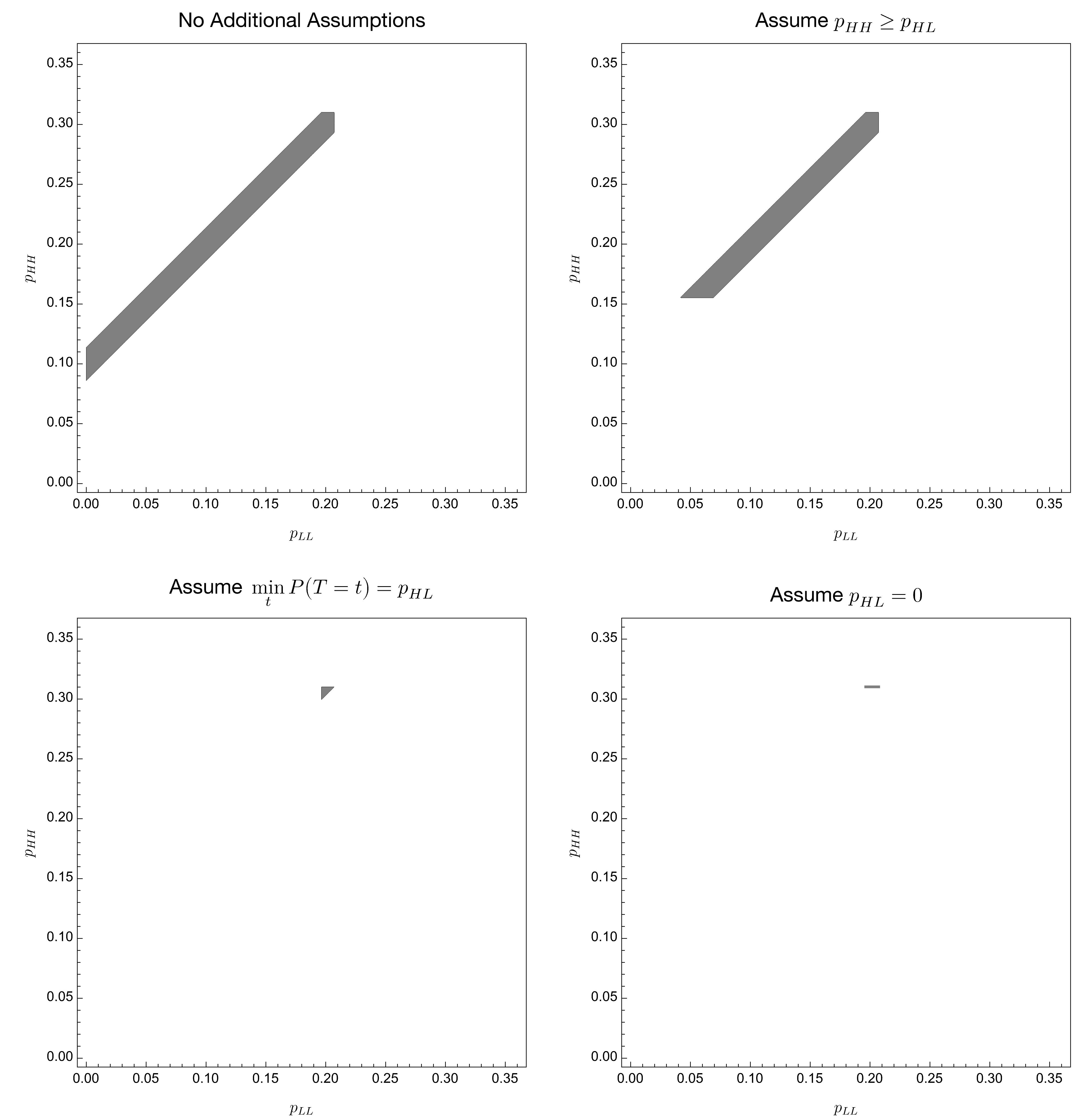}
    \caption{Identified set for $p_{L,L},p_{H,H}$ at week 135. Amenity=paid vacation. The first panel illustrates the identified set when $\mathcal R_T = \{\text{Assumption } \ref{Ass:LM}\}$. The remaining panels illustrate the identified set of additional assumptions imposed over  \text{Assumption } \ref{Ass:LM}. }
    \label{fig:Vacat135idset}
\end{figure}

\begin{figure}
    \centering
    \includegraphics[width=\textwidth]{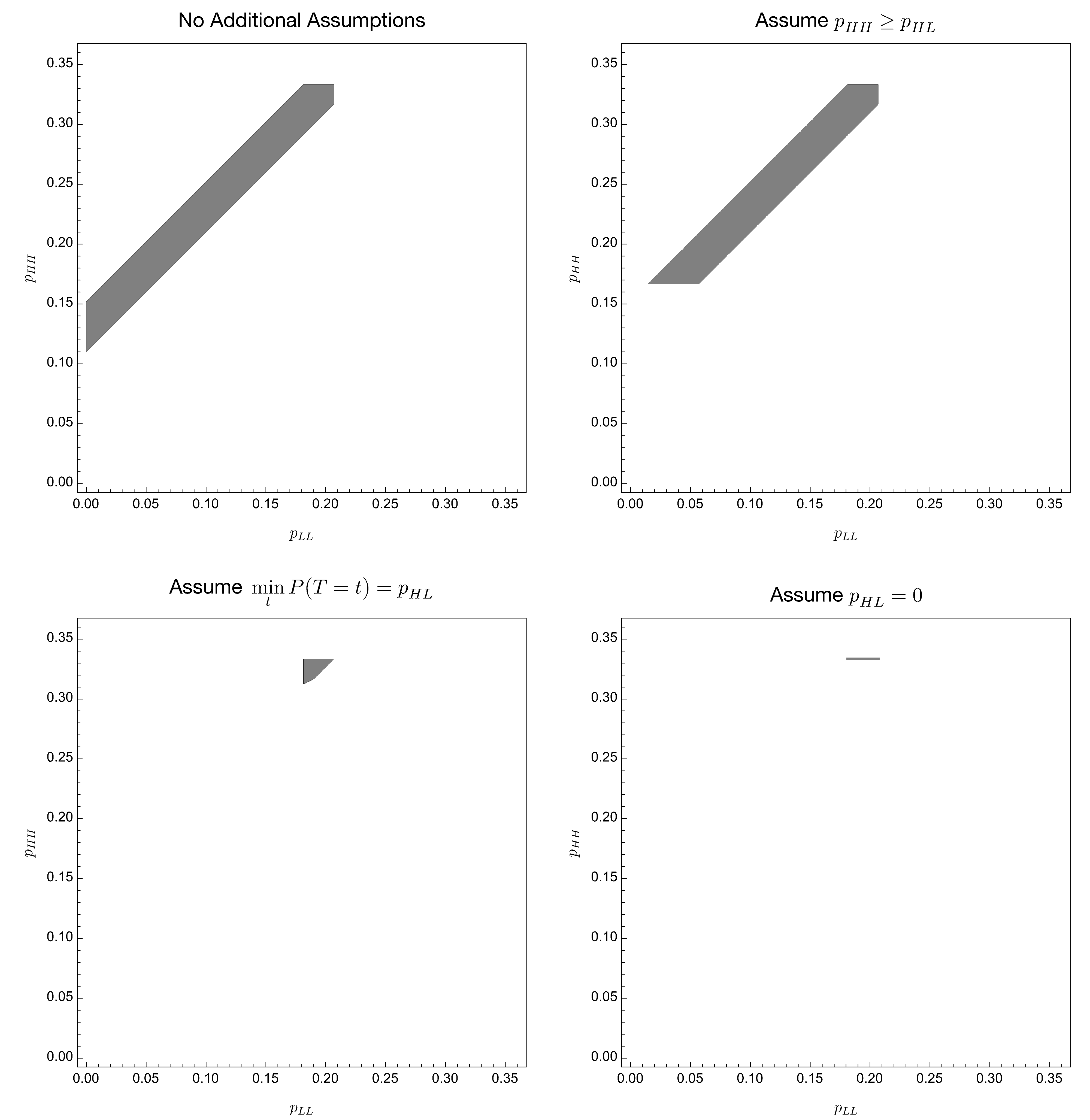}
    \caption{Identified set for $p_{L,L},p_{H,H}$ at week 180. Amenity=paid vacation. The first panel illustrates the identified set when $\mathcal R_T = \{\text{Assumption } \ref{Ass:LM}\}$. The remaining panels illustrate the identified set of additional assumptions imposed over  \text{Assumption } \ref{Ass:LM}. }
    \label{fig:Vacat180idset}
\end{figure}

\begin{figure}
    \centering
    \includegraphics[width=\textwidth]{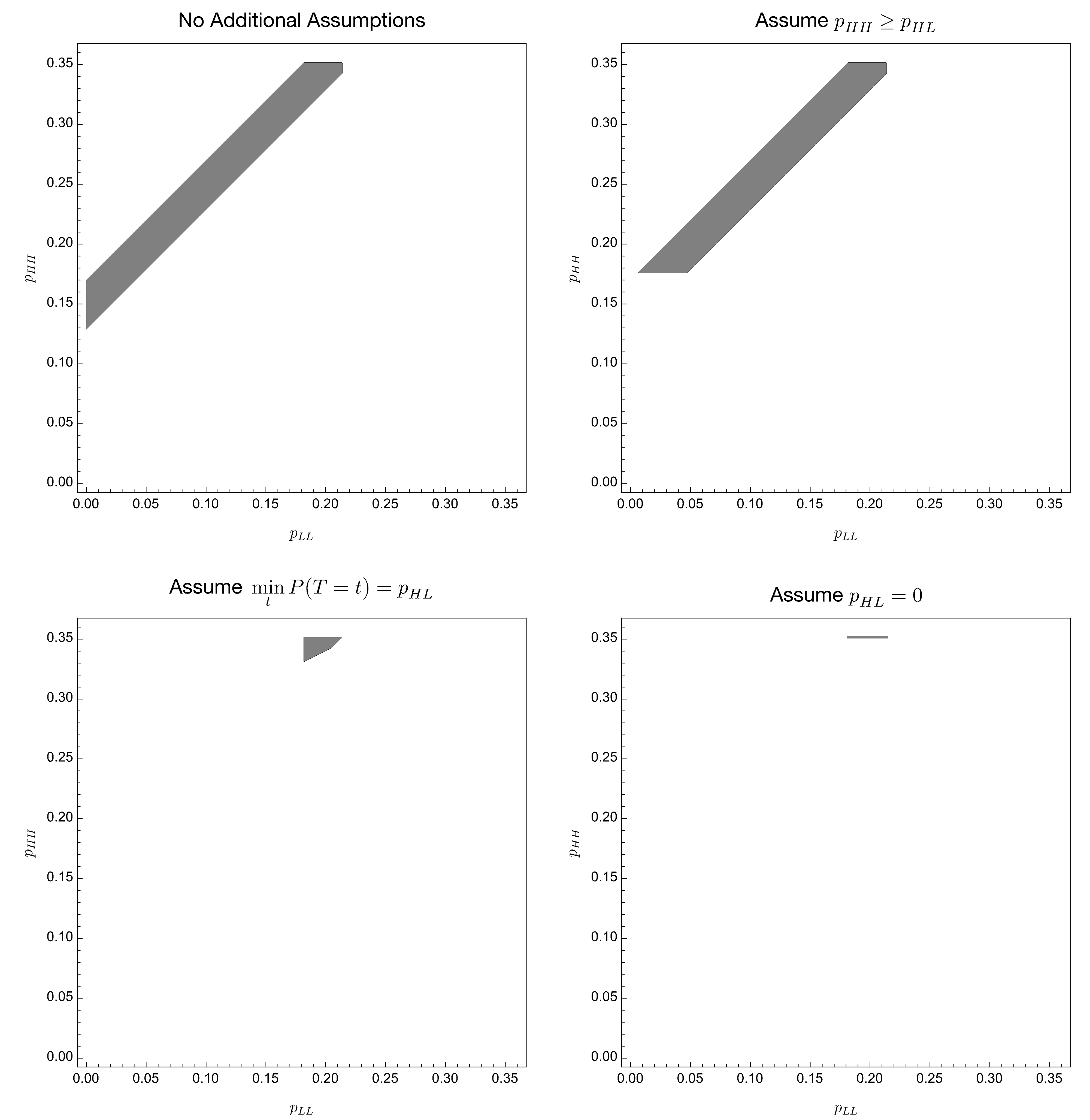}
    \caption{Identified set for $p_{L,L},p_{H,H}$ at week 208. Amenity=paid vacation. The first panel illustrates the identified set when $\mathcal R_T = \{\text{Assumption } \ref{Ass:LM}\}$. The remaining panels illustrate the identified set of additional assumptions imposed over  \text{Assumption } \ref{Ass:LM}. }
    \label{fig:Vacat208idset}
\end{figure}

\clearpage

\begin{table}[htbp]
    \centering
    \begin{tabular}{l*{4}{c}}
        \toprule
        &  $\mathbb P[D=H|Z=0]$ & $\mathbb P[D=H|Z=1]$&  $\mathbb P[D=L|Z=0]$&  $\mathbb P[D=L|Z=1]$\\
        \midrule
        \input empiric_tbls_figs/pscores/pension_pscores_weighted.tex
        \bottomrule
    \end{tabular}
    \caption{Propensity scores by week, amenity: retirement/pension benefits\label{retpscores}}
\end{table}

\begin{figure}
    \centering
    \includegraphics[width=\textwidth]{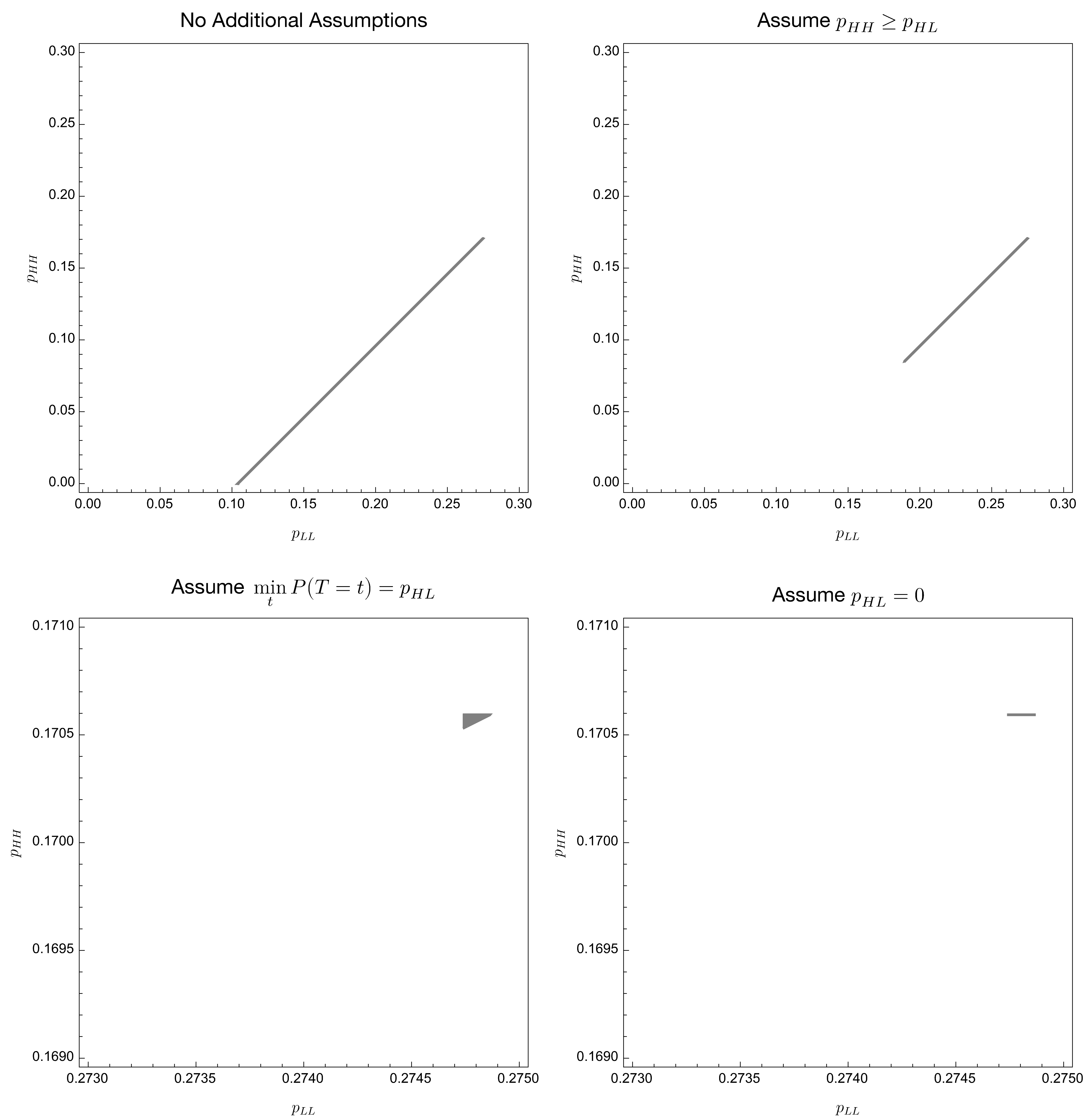}
    \caption{Identified set for $p_{L,L},p_{H,H}$ at week 90. Amenity=retirement/pension benefits. The first panel illustrates the identified set when $\mathcal R_T = \{\text{Assumption } \ref{Ass:LM}\}$. The remaining panels illustrate the identified set of additional assumptions imposed over  \text{Assumption } \ref{Ass:LM}. The scale of the axis in the bottom two panels is shrunk to see more clearly the identified set.}
    \label{fig:Pension90idset}
\end{figure}

\begin{figure}
    \centering
    \includegraphics[width=\textwidth]{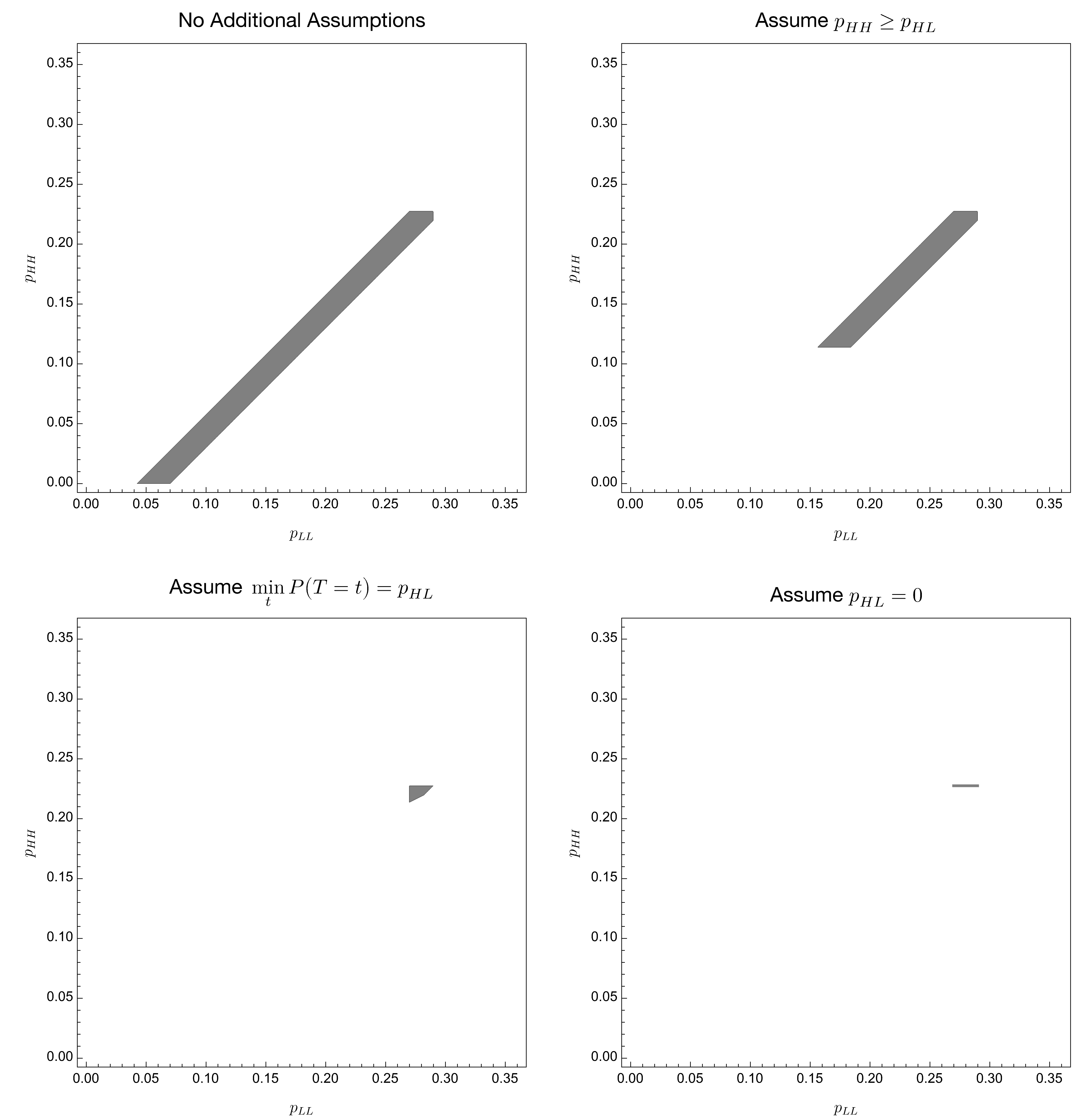}
    \caption{Identified set for $p_{L,L},p_{H,H}$ at week 135. Amenity=retirement/pension benefits. The first panel illustrates the identified set when $\mathcal R_T = \{\text{Assumption } \ref{Ass:LM}\}$. The remaining panels illustrate the identified set of additional assumptions imposed over  \text{Assumption } \ref{Ass:LM}. }
    \label{fig:Pension135idset}
\end{figure}

\begin{figure}
    \centering
    \includegraphics[width=\textwidth]{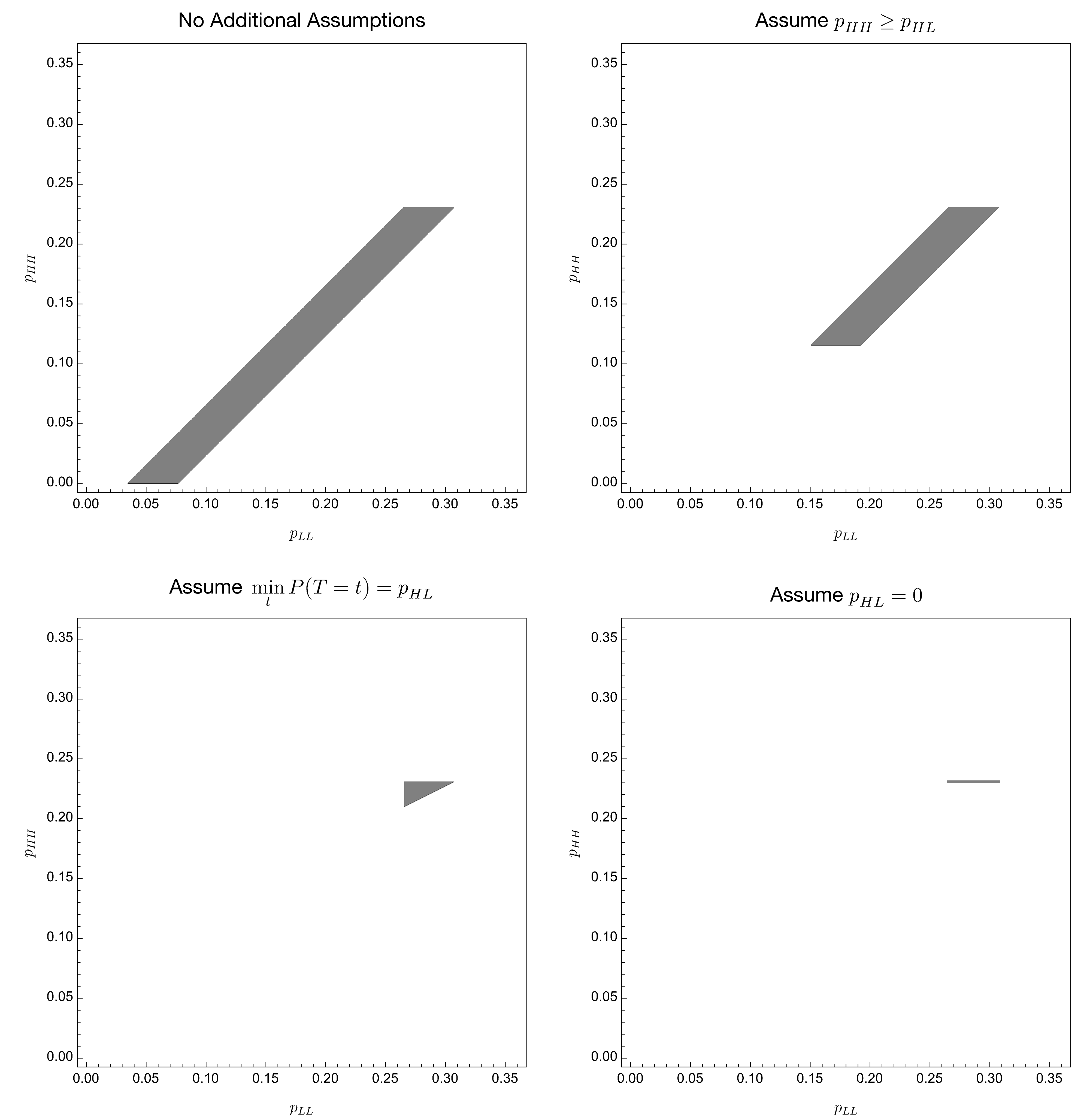}
    \caption{Identified set for $p_{L,L},p_{H,H}$ at week 180. Amenity=retirement/pension benefits. The first panel illustrates the identified set when $\mathcal R_T = \{\text{Assumption } \ref{Ass:LM}\}$. The remaining panels illustrate the identified set of additional assumptions imposed over  \text{Assumption } \ref{Ass:LM}. }
    \label{fig:Pension180idset}
\end{figure}

\begin{figure}
    \centering
    \includegraphics[width=\textwidth]{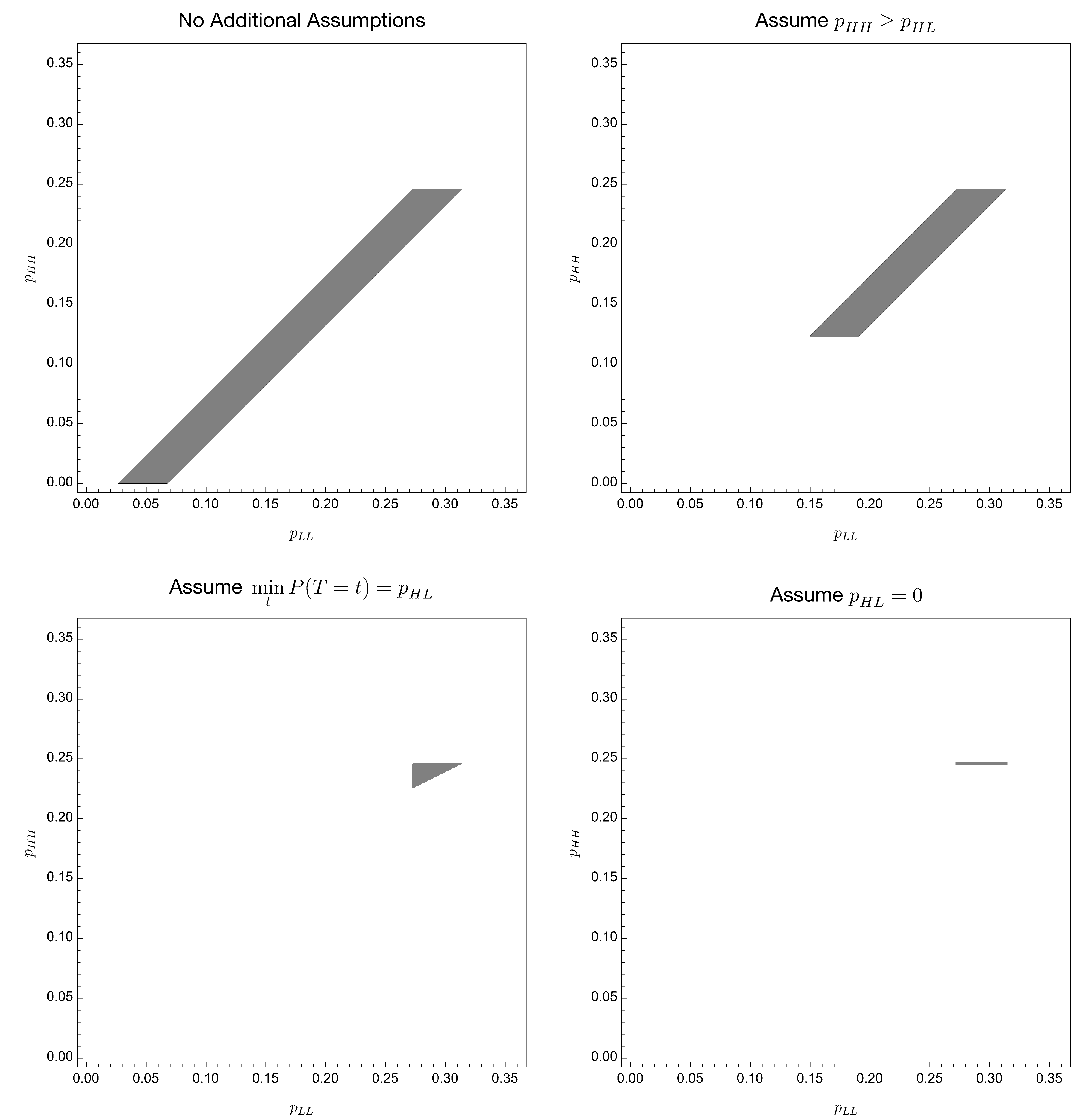}
    \caption{Identified set for $p_{L,L},p_{H,H}$ at week 208. Amenity=retirement/pension benefits. The first panel illustrates the identified set when $\mathcal R_T = \{\text{Assumption } \ref{Ass:LM}\}$. The remaining panels illustrate the identified set of additional assumptions imposed over  \text{Assumption } \ref{Ass:LM}. }
    \label{fig:Pension208idset}
\end{figure}

\clearpage

\begin{figure}
    \centering
    \includegraphics[width=\textwidth]{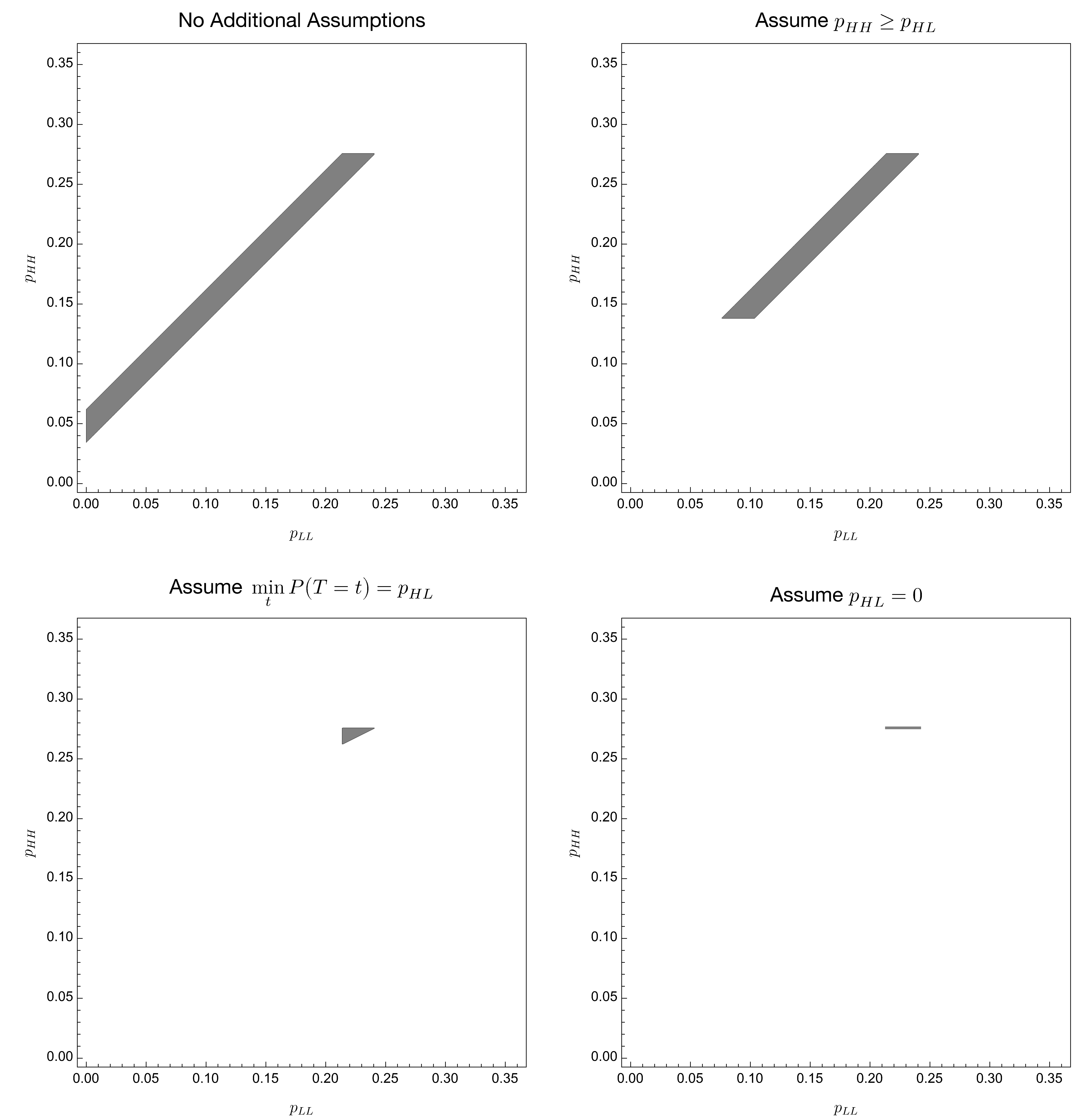}
    \caption{Identified set for $p_{L,L},p_{H,H}$ at week 135. Amenity=health insurance. The first panel illustrates the identified set when $\mathcal R_T = \{\text{Assumption } \ref{Ass:LM}\}$. The remaining panels illustrate the identified set of additional assumptions imposed over  \text{Assumption } \ref{Ass:LM}. }
    \label{fig:health135idset}
\end{figure}

\begin{figure}
    \centering
    \includegraphics[width=\textwidth]{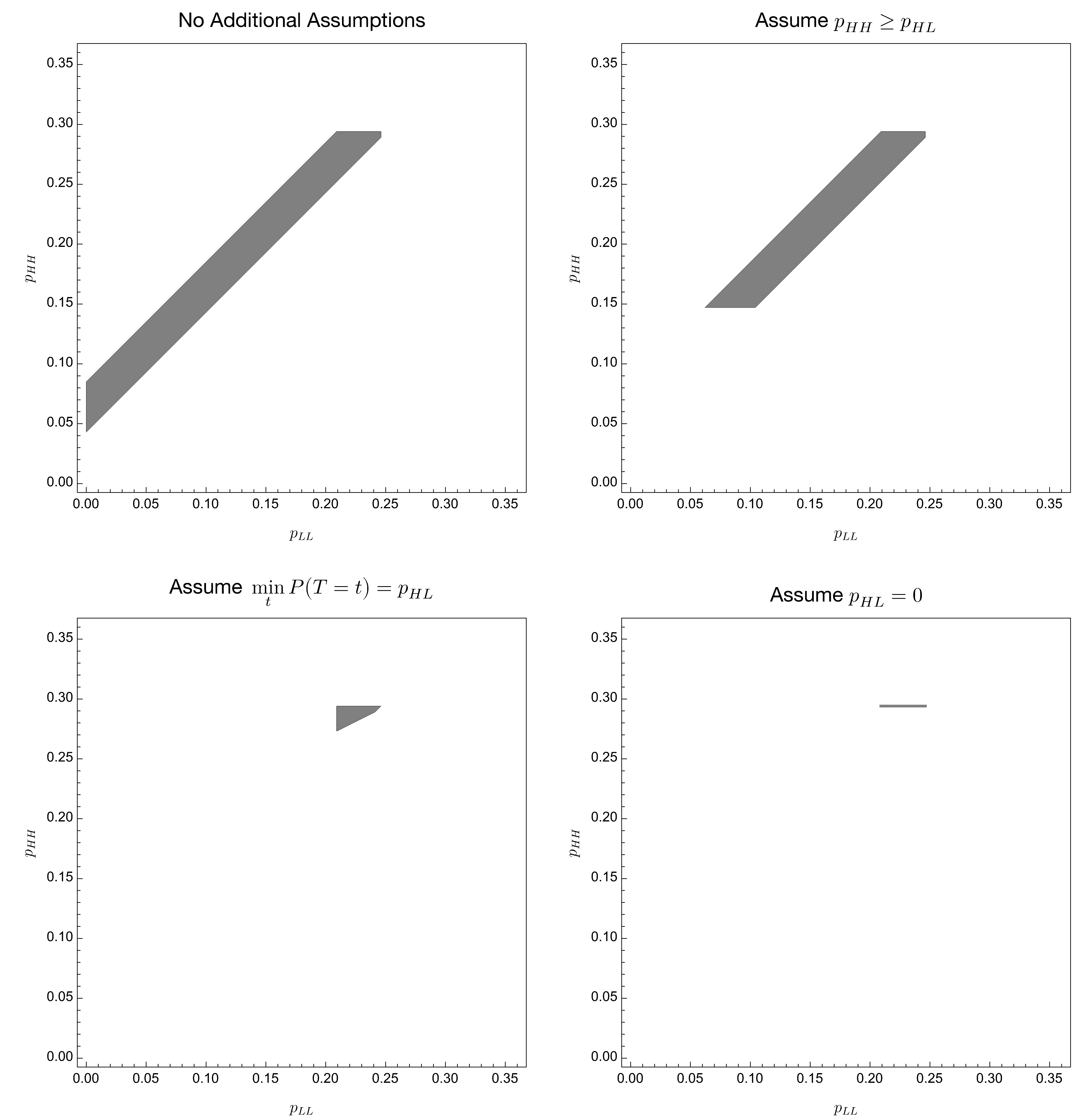}
    \caption{Identified set for $p_{L,L},p_{H,H}$ at week 180. Amenity=health insurance. The first panel illustrates the identified set when $\mathcal R_T = \{\text{Assumption } \ref{Ass:LM}\}$. The remaining panels illustrate the identified set of additional assumptions imposed over  \text{Assumption } \ref{Ass:LM}. }
    \label{fig:health180idset}
\end{figure}

\begin{figure}
    \centering
    \includegraphics[width=\textwidth]{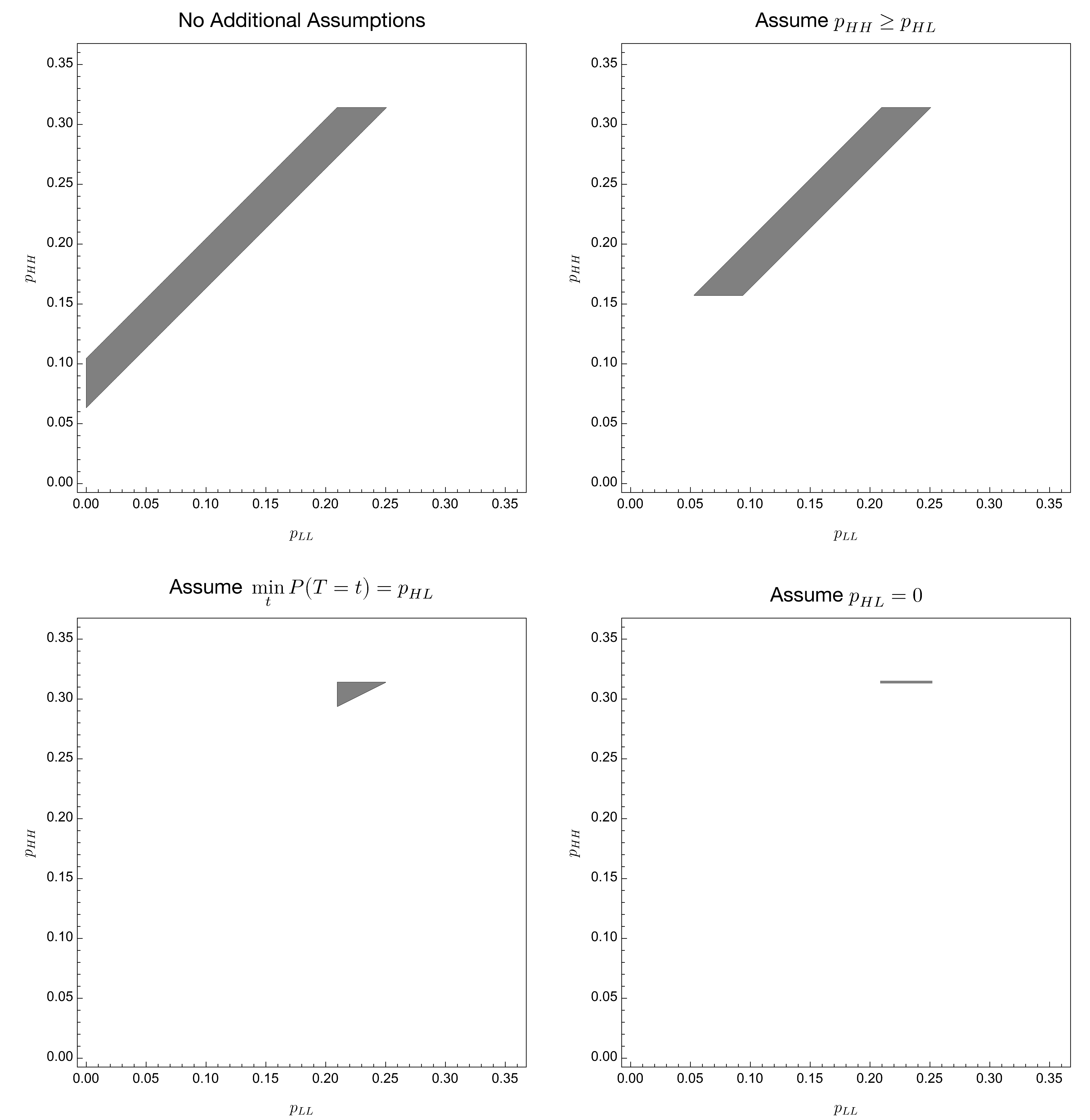}
    \caption{Identified set for $p_{L,L},p_{H,H}$ at week 208. Amenity=health insurance. The first panel illustrates the identified set when $\mathcal R_T = \{\text{Assumption } \ref{Ass:LM}\}$. The remaining panels illustrate the identified set of additional assumptions imposed over  \text{Assumption } \ref{Ass:LM}. }
    \label{fig:health208idset}
\end{figure}

\clearpage

\subsection{Multilayered Bounds}\label{app:mlbounds}
\Cref{app:mlbounds} provides aggregate multilayered bounds for all classifications of firm type, and within-firm-type multilayered bounds for when classifying firm type based on the provision of paid vacation and retirement/pension benefits.

\begin{table}[htbp]
    \centering
       \begin{tabular}{l*{6}c}
        
        \toprule
        & 
        \multirow{1}{*}{$ p^{*}_{HH} $} &  \multirow{1}{*}{$ p^{*}_{LL} $}
       & \multirow{1}{*}{$ p^{*}_{HH} $} &  \multirow{1}{*}{$ p^{*}_{LL} $} & \multicolumn{2}{c}{\tiny $\displaystyle\sum_{d \in \{L,H\}} \frac{p_{d,d}}{p_{HH}+p_{LL}} \mathbb E[Y_{1,d}-Y_{0,d}|T=(d,d)]$} \\
        \cmidrule(lr){6-7}
        \textbf{Week 90} &   \multicolumn{2}{c}{\tiny(for lower bound)}& \multicolumn{2}{c}{\tiny(for upper bound)} &  lower & upper
        \input empiric_tbls_figs/agg_bounds/health_wk90_agg_bounds_weighted
        \midrule
        \textbf{Week 135} & & & & &
        \input empiric_tbls_figs/agg_bounds/health_wk135_agg_bounds_weighted
        \midrule
        \textbf{Week 180} & & & & &
        \input empiric_tbls_figs/agg_bounds/health_wk180_agg_bounds_weighted
        \midrule
        \textbf{Week 208} & & & & &
        \input empiric_tbls_figs/agg_bounds/health_wk208_agg_bounds_weighted
        \bottomrule 
         & & & & &\\
        \multicolumn{7}{l}{\footnotesize Notes: Treatment bounds are for ln(hourly wage); hourly wage calculated as weekly earnings divided by weekly}\\
        \multicolumn{7}{l}{\footnotesize hours for the employed. $p^{*}_{t}$ is the optimal value of $p_t$ over the joint identified set for response-types under the} \\
        \multicolumn{7}{l}{\footnotesize given assumption; $p^{*}_{t}$ may differ for lower and upper bounds and is therefore presented separately.}
    \end{tabular}
    \caption{\small
    Aggregate multilayered bounds for Job Corps; health insurance amenity. \label{tab:agg_healthbnds}}
\end{table}
\clearpage

\begin{table}[htbp]
    \centering
    \begin{tabular}{l*{7}{c}}
        \toprule
        & & & \multicolumn{2}{c}{\footnotesize $\mathbb E(Y_{1,H}-Y_{0,H}|T = (H,H))$} & \multicolumn{2}{c}{\footnotesize $\mathbb E(Y_{1,L}-Y_{0,L}|T = (L,L))$} \\
        \cmidrule(lr){4-5}
        \cmidrule(lr){6-7}
        \textbf{Week 90} &  $ p^{*}_{HH} $&  $ p^{*}_{LL} $&       lower &       upper &       lower&       upper 
        \input empiric_tbls_figs/bounds/vacat_wk90_bounds_weighted.tex
        \midrule
        \textbf{Week 135} & & & & & & 
        \input empiric_tbls_figs/bounds/vacat_wk135_bounds_weighted.tex
        \midrule
        \textbf{Week 180} & & & & & & 
        \input empiric_tbls_figs/bounds/vacat_wk180_bounds_weighted.tex
        \midrule
        \textbf{Week 208} & & & & & & 
        \input empiric_tbls_figs/bounds/vacat_wk208_bounds_weighted.tex
        \bottomrule
        \multicolumn{8}{l}{\footnotesize Notes: Treatment bounds are for ln(hourly wage); hourly wage calculated as weekly earnings divided by}\\
        \multicolumn{8}{l}{\footnotesize weekly hours for the employed. $p^{*}_{t}$ is the minimum value of $p_t$ over the identified set for response-types} \\
        \multicolumn{8}{l}{\footnotesize under the given assumption.}
    \end{tabular}
    \caption{Multilayered bounds by week, amenity: paid vacation.\label{tab:vacatbnds}}
\end{table}


\begin{table}[htbp]
    \centering
    \begin{tabular}{l*{6}{c}}
        \toprule
        & \multicolumn{2}{c}{Lower}& \multicolumn{2}{c}{Upper} & \multicolumn{2}{c}{\tiny $\sum_{d \in \{L,H\}} \frac{p_{d,d}}{\sum_{d'\in \{L,H\}}p_{d',d'}} \mathbb E[Y_{1,d}-Y_{0,d}|T=(d,d)]$} \\
        \cmidrule(lr){2-3}
        \cmidrule(lr){4-5}
        \cmidrule(lr){6-7}
        \textbf{Week 90} & $ p^{*}_{HH} $ &  $ p^{*}_{LL} $ & $ p^{*}_{HH} $ &  $ p^{*}_{LL} $ & lower & upper
        \input empiric_tbls_figs/agg_bounds/vacat_wk90_agg_bounds_weighted
        \midrule
        \textbf{Week 135} & & & & &
        \input empiric_tbls_figs/agg_bounds/vacat_wk135_agg_bounds_weighted
        \midrule
        \textbf{Week 180} & & & & &
        \input empiric_tbls_figs/agg_bounds/vacat_wk180_agg_bounds_weighted
        \midrule
        \textbf{Week 208} & & & & &
        \input empiric_tbls_figs/agg_bounds/vacat_wk208_agg_bounds_weighted
        \bottomrule
        \multicolumn{7}{l}{\footnotesize Notes: Treatment bounds are for ln(hourly wage); hourly wage calculated as weekly earnings divided by weekly}\\
        \multicolumn{7}{l}{\footnotesize hours for the employed. $p^{*}_{t}$ is the optimal value of $p_t$ over the joint identified set for response-types under the} \\
        \multicolumn{7}{l}{\footnotesize given assumption; $p^{*}_{t}$ may differ for lower and upper bounds and is therefore presented separately.}
    \end{tabular}
    \caption{Aggregate multilayered bounds by week. Amenity=paid vacation. \label{tab:agg_vacatbnds}}
\end{table}

\begin{landscape}
\begin{figure}
    \centering
    \includegraphics[width=1.2\textwidth]{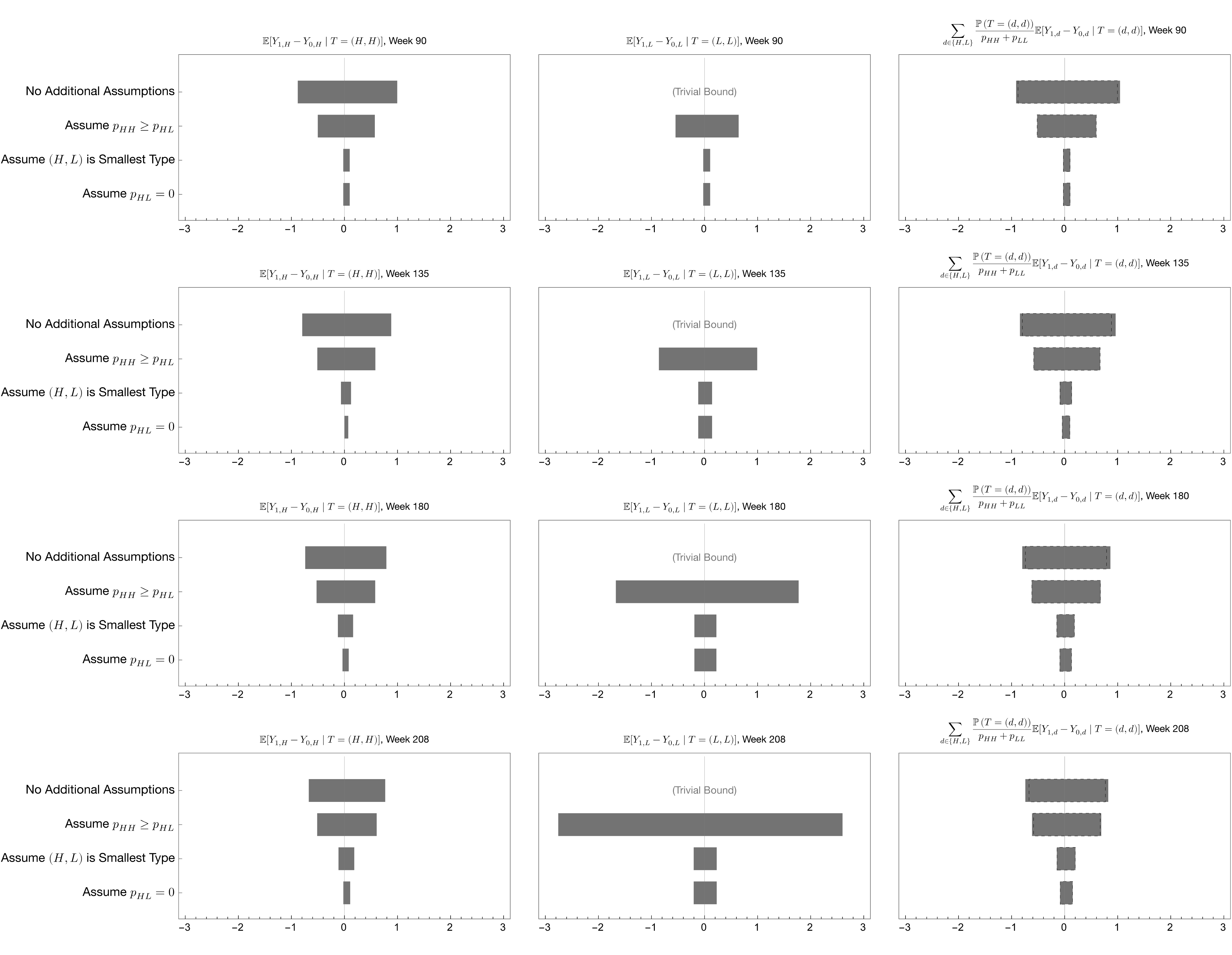}
    \begin{tabular}{p{\textwidth}}
    \footnotesize{Notes: Treatment bounds are for ln(hourly wage); hourly wage calculated as weekly earnings divided by weekly hours for the employed.} 
    \end{tabular}
    \caption{Multilayered bounds by week. Amenity=paid vacation.}
    \label{fig:vacatbnds}
\end{figure}
\end{landscape}


\begin{table}[htbp]
    \centering
    \begin{tabular}{l*{7}{c}}
        \toprule
        & & & \multicolumn{2}{c}{\footnotesize $\mathbb E(Y_{1,H}-Y_{0,H}|T = (H,H))$} & \multicolumn{2}{c}{\footnotesize $\mathbb E(Y_{1,L}-Y_{0,L}|T = (L,L))$} \\
        \cmidrule(lr){4-5}
        \cmidrule(lr){6-7}
        \textbf{Week 90} &  $ p^{*}_{HH} $&  $ p^{*}_{LL} $&       lower &       upper &       lower&       upper 
        \input empiric_tbls_figs/bounds/pension_wk90_bounds_weighted.tex
        \midrule
        \textbf{Week 135} & & & & & & 
        \input empiric_tbls_figs/bounds/pension_wk135_bounds_weighted.tex
        \midrule
        \textbf{Week 180} & & & & & & 
        \input empiric_tbls_figs/bounds/pension_wk180_bounds_weighted.tex
        \midrule
        \textbf{Week 208} & & & & & & 
        \input empiric_tbls_figs/bounds/pension_wk208_bounds_weighted.tex
        \bottomrule
        \multicolumn{8}{l}{\footnotesize Notes: Treatment bounds are for ln(hourly wage); hourly wage calculated as weekly earnings divided by}\\
        \multicolumn{8}{l}{\footnotesize weekly hours for the employed. $p^{*}_{t}$ is the minimum value of $p_t$ over the identified set for response-types} \\
        \multicolumn{8}{l}{\footnotesize under the given assumption.}
    \end{tabular}
    \caption{Multilayered bounds by week, amenity: retirement/pension benefits. \label{tab:pensionbnds}}
\end{table}

\begin{table}[htbp]
    \centering
    \begin{tabular}{l*{6}{c}}
        \toprule
        & \multicolumn{2}{c}{Lower}& \multicolumn{2}{c}{Upper} & \multicolumn{2}{c}{\tiny $\sum_{d \in \{L,H\}} \frac{p_{d,d}}{\sum_{d'\in \{L,H\}}p_{d',d'}} \mathbb E[Y_{1,d}-Y_{0,d}|T=(d,d)]$} \\
        \cmidrule(lr){2-3}
        \cmidrule(lr){4-5}
        \cmidrule(lr){6-7}
        \textbf{Week 90} & $ p^{*}_{HH} $ &  $ p^{*}_{LL} $ & $ p^{*}_{HH} $ &  $ p^{*}_{LL} $ & lower & upper
        \input empiric_tbls_figs/agg_bounds/pension_wk90_agg_bounds_weighted
        \midrule
        \textbf{Week 135} & & & & &
        \input empiric_tbls_figs/agg_bounds/pension_wk135_agg_bounds_weighted
        \midrule
        \textbf{Week 180} & & & & &
        \input empiric_tbls_figs/agg_bounds/pension_wk180_agg_bounds_weighted
        \midrule
        \textbf{Week 208} & & & & &
        \input empiric_tbls_figs/agg_bounds/pension_wk208_agg_bounds_weighted
        \bottomrule
        \multicolumn{7}{l}{\footnotesize Notes: Treatment bounds are for ln(hourly wage); hourly wage calculated as weekly earnings divided by weekly}\\
        \multicolumn{7}{l}{\footnotesize hours for the employed. $p^{*}_{t}$ is the optimal value of $p_t$ over the joint identified set for response-types under the} \\
        \multicolumn{7}{l}{\footnotesize given assumption; $p^{*}_{t}$ may differ for lower and upper bounds and is therefore presented separately.}
    \end{tabular}
    \caption{Aggregate multilayered bounds by week. Amenity=retirement/pension benefits. \label{tab:agg_pensionbnds}}
\end{table}

\begin{landscape}
\begin{figure}
    \centering
    \includegraphics[width=1.2\textwidth]{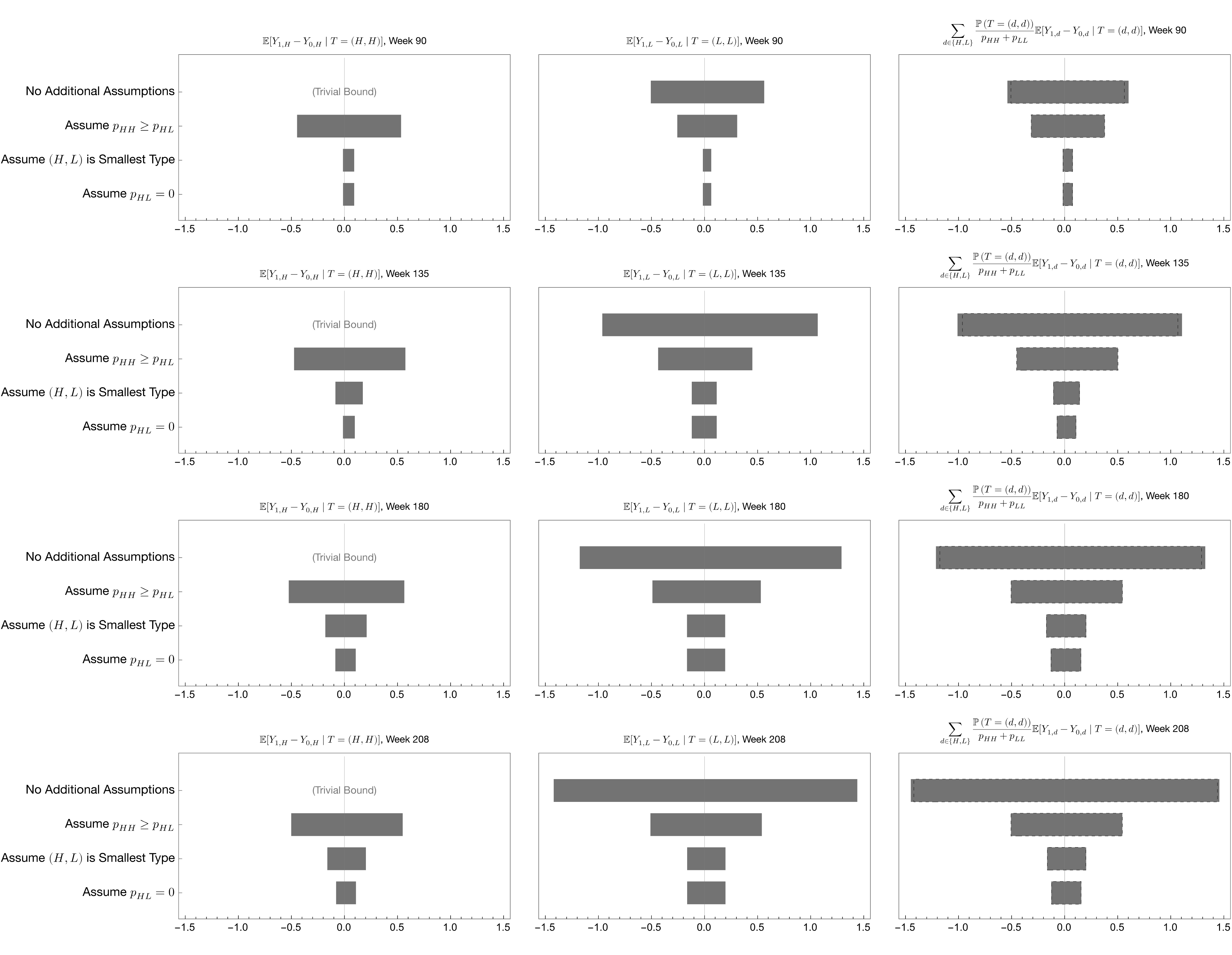}
    \begin{tabular}{p{\textwidth}}
    \footnotesize{Notes: Treatment bounds are for ln(hourly wage); hourly wage calculated as weekly earnings divided by weekly hours for the employed.} 
    \end{tabular}
    \caption{Multilayered bounds by week. Amenity=pension benefits.}
    \label{fig:pensionbnds}
\end{figure}
\end{landscape}

\clearpage
\section{Additional Empirical Tables and Figures for WorkAdvance RCTs}
\subsection{Summary Statistics and Differential Sorting of Treatment and Control Workers}

This section provides descriptive statistics on the WorkAdvance RCT.

\begin{table}[htbp]
    \centering
    \begin{tabular}{l*{3}{cc}}
        \toprule &
        \multicolumn{2}{c}{Control} & \multicolumn{2}{c}{Treated} & \multicolumn{2}{c}{Difference} \\
        \cmidrule(lr){2-3}
        \cmidrule(lr){4-5}
        \cmidrule(lr){6-7}
        \input empiric_tbls_figs/workadv_bounds/sumstats_okla_1.tex
        \input empiric_tbls_figs/workadv_bounds/sumstats_okla_3.tex
    \midrule
    Sample size & 344 & & 353 & & 697 \\

        \bottomrule
        \multicolumn{7}{l}{\footnotesize Notes: Quarterly earnings are not conditional on employment (i.e., includes 0s for the unemployed).} \\
    \end{tabular}

    \caption{Summary statistics by treatment status, RCT $=$ Madison Strategies}
    \label{tab:ms_summ_stats}
\end{table}

\begin{table}[htbp]
    \centering
    \begin{tabular}{l*{3}{cc}}
        \toprule &
        \multicolumn{2}{c}{Control} & \multicolumn{2}{c}{Treated} & \multicolumn{2}{c}{Difference} \\
        \cmidrule(lr){2-3}
        \cmidrule(lr){4-5}
        \cmidrule(lr){6-7}
        \input empiric_tbls_figs/workadv_bounds/sumstats_ohio_1.tex
        \input empiric_tbls_figs/workadv_bounds/sumstats_ohio_3.tex
    \midrule
    Sample size & 349 & & 349 & & 698 \\

        \bottomrule
        \multicolumn{7}{l}{\footnotesize Notes: Quarterly earnings are not conditional on employment (i.e., includes 0s for the unemployed).} \\
    \end{tabular}

    \caption{Summary statistics by treatment status, RCT $=$ Towards Employment}
     \label{tab:te_summ_stats}
\end{table}

\begin{figure}[h!]
    \centering
    \includegraphics[width=.8\linewidth]{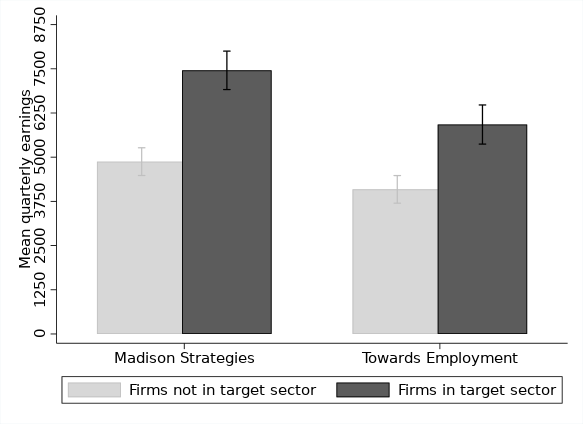}
    \begin{tabular}{p{.8\linewidth}}
        \footnotesize{Notes: Quarterly earnings means are among the employed (i.e., do not include 0s for the unemployed).} 
    \end{tabular}
    \caption{Mean quarterly earnings by firm sector}
    \label{fig:hist_wa}
\end{figure}

\begin{figure}[h!]
    \centering
    \subfigure[control units]{
    \includegraphics[width=.49\linewidth]{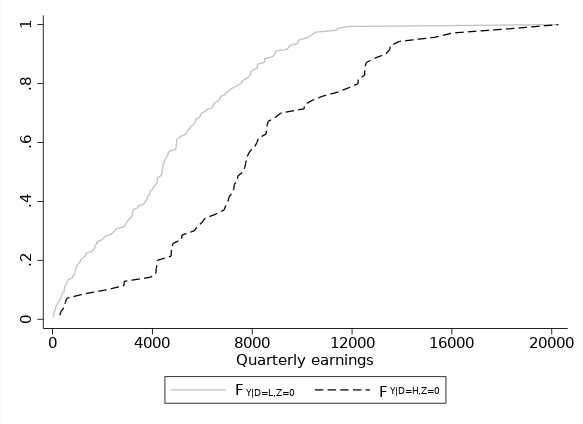}}
    \subfigure[treated units]{\includegraphics[width=.49\linewidth]{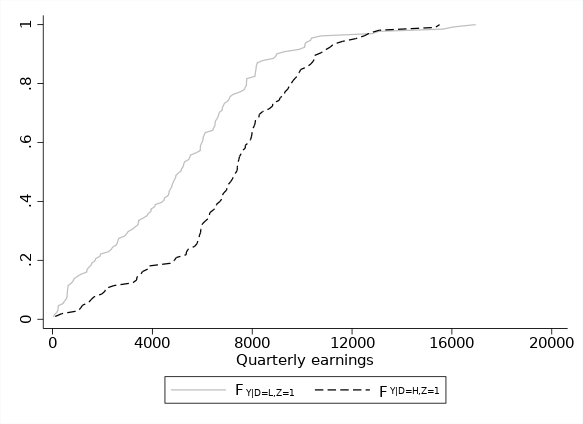}}  
    \begin{tabular}{p{.8\linewidth}}
        \footnotesize{Notes: Quarterly earnings are for the employed (i.e., does not include 0s for the unemployed).} 
    \end{tabular}
    \caption{Cumulative distribution function by firm sector, RCT= Madison Strategies}
    \label{fig:ms_cdf}
\end{figure}

\begin{figure}[h!]
    \centering
    \subfigure[control units]{
    \includegraphics[width=.49\linewidth]{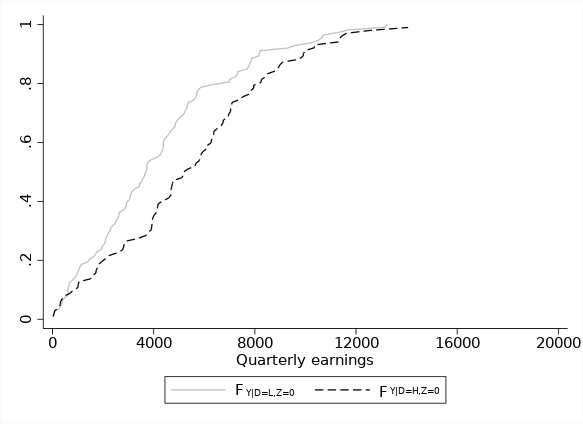}}
    \subfigure[treated units]{\includegraphics[width=.49\linewidth]{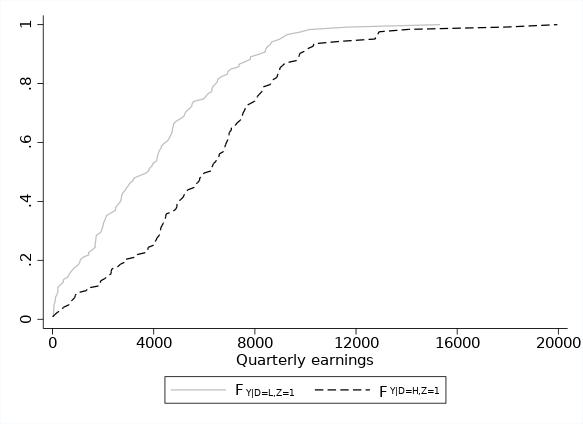}}  
    \begin{tabular}{p{.8\linewidth}}
        \footnotesize{Notes: Quarterly earnings are for the employed (i.e., does not include 0s for the unemployed).} 
    \end{tabular}
    \caption{Cumulative distribution function by firm sector, RCT= Towards Employment}
    \label{fig:te_cdf}
\end{figure}

\clearpage

\subsection{Aggregate multilayered bounds for WorkAdvance RCT} This section presents aggregate multilayered bounds for the WorkAdvance RCT.
\begin{table}[htbp]
    \centering
    \caption{Aggregate multilayered bounds for WorkAdvance RCT. \label{tab:agg_workadvance}}
    \begin{tabular}{l*{6}{c}}
        \toprule
        & 
        \multirow{1}{*}{$ p^{*}_{HH} $} &  \multirow{1}{*}{$ p^{*}_{LL} $}
       & \multirow{1}{*}{$ p^{*}_{HH} $} &  \multirow{1}{*}{$ p^{*}_{LL} $} & \multicolumn{2}{c}{\tiny $\displaystyle\sum_{d \in \{L,H\}} \frac{p_{d,d}}{p_{HH}+p_{LL}} \mathbb E[Y_{1,d}-Y_{0,d}|T=(d,d)]$} \\
        \cmidrule(lr){6-7}
        \textbf{Madison Strategies} &   \multicolumn{2}{c}{\tiny(for lower bound)}& \multicolumn{2}{c}{\tiny(for upper bound)} &  lower & upper
        \input empiric_tbls_figs/workadv_bounds/agg_bounds_okla_c
        \midrule
        \textbf{Towards Employment} & & & & &
        \input empiric_tbls_figs/workadv_bounds/agg_bounds_ohio_c
        \bottomrule
    \end{tabular}
        \vspace{0.1em}
        
        \begin{minipage}{\textwidth}
        \footnotesize
            Notes: 
            Outcome is quarterly wages. $p^{*}_{t}$ is the minimum value of $p_t$ over the identified set for response-types under the given assumption; $p^{*}_{t}$ may differ for lower and upper bounds and is therefore presented separately.
        \end{minipage}
\end{table}
\clearpage

\subsection{\cite{lee_training_2009} Bounds}
This section generates Lee's bounds for the WorkAdvance RCT.
\begin{table}[htbp]
    \centering
    \begin{tabular}{l*{3}{cc}}
        \toprule
        & & & & \multicolumn{2}{c}{$\mathbb E[Y_{1,D_1}-Y_{0,D_0}|D_0>0,D_1>0]$} \\
        \cmidrule(lr){5-6}
        &  $\mathbb P[D>0|Z=0]$ &  $\mathbb P[D>0|Z=1]$ &$p$&      lower&       upper\\
        \midrule
        \input empiric_tbls_figs/workadv_bounds/lee_bounds.tex
        \bottomrule
        \multicolumn{6}{l}{\footnotesize Notes: Treatment bounds are for quarterly wages.}
    \end{tabular}
    \caption{Lee's bounds for WorkAdvance RCT.\label{wa_leebounds}}
    \label{tab:wa_leebounds}
\end{table}

\clearpage

\clearpage

\section{`Top 5' papers (potentially) collapsing multilayered selection to single layered selection}\label{sec:app_d_litreview}
In a literature survey which we detail in \Cref{sec:intro}, we counted 56 papers published in `top 5' general interest economic journals that cited \cite{lee_training_2009} and 42 that empirically implemented Lee bounds. \Cref{tab:app_d_tab} details 7 of these papers that feature multilayered selection, where researchers simplified the sample selection problem by collapsing it to a single dimension.\footnote{Only outcomes for which Lee's bounds are estimated are noted in column 3.} \\

\begin{longtable}[c]{|p{.22\textwidth}|p{.22\textwidth}|p{.15\textwidth}|p{.22\textwidth}|p{.22\textwidth}|}
    \hline
    \multicolumn{3}{|c|}{\textbf{Paper}} & \multicolumn{2}{|c|}{\textbf{Sample Selection}} \\
    \cline{1-3}
    \cline{4-5}
    Application & Treatment & Outcomes & Layer addressed & Additional layer \\
    \endfirsthead
    \multicolumn{5}{@{}l}{\ldots continued} \\
    \hline
    Application & Treatment & Outcomes & Layer addressed & Additional layer \\
    \hline
    \endhead
    \hline
    \multicolumn{5}{|l|}{\textit{\cite{daruich_effects_2023}}} \\*
    \hline
    Study 2001 Italian reform lifting constraints on the employment of temporary contract workers but maintaining employment protection laws for permanent contract employees; exploit staggered implementation across collective bargaining agreements. &
    See ``Application". &
    Individuals' earnings. &
    Only observe employment outcomes after labor market entry; utilize Lee bounds to address that the reform affects the entry margin into the labor market. &
    Conditional on labor market entry, altering temporary contract worker regulations likely also affects worker sorting across industries and/or firms; indeed this paper finds meaningful changes in the shares of temporary contract workers in certain industries. \\
    \hline
    \pagebreak
    \multicolumn{5}{|l|}{\textit{\cite{cullen_equilibrium_2023}}} \\
    \hline
    Study state-level laws in the U.S. protecting the right of private sector workers to discuss salary information with co-workers. &
    See ``Application". &
    Worker wages. &
    Treatment effects could be driven by compositional changes of private sector workers if high-paid (low-paid) workers disproportionately leave (join) the private sector; estimate Lee bounds to address this challenge. &
    Conditional on private sector entry, treatment may also affect the sorting of workers across industries or firms within the private sector; for example, knowledge of co-workers' salaries could cause workers to sort to firms with flatter pay hierarchies (i.e., fairness concerns). \\
    \hline
    \pagebreak
    \multicolumn{5}{|l|}{\textit{\cite{bianchi_dynamics_2022}}} \\
    \hline
    Study long-term and spillover effects of management interventions on firm performance; estimate effect of the Training Within Industry program, a U.S. government training program intended to be provided to all firms involved in war production between 1940 and 1945. &
    Exploit that constraints resulted in only 7\% of applicant firms receiving full training, 48\% receiving no training and the remainder receiving partial training; this paper compares applicant firms who received training to applicant firms who did not. &
    Firm total factor productivity. &
    Estimate Lee bounds to address the higher attrition rate of untrained firms; treated firms had 90\% survival rate at least 10 years following treatment whereas control firms only had 64\% survival rate. &
    Conditional on firm survival, training may also affect the sorting of firms across industries or other important dimensions. This is particularly plausible in this paper's setting where: (i) trained firms undertook structural changes transforming them into larger and more complex organizations; (ii) this paper estimates treatment bounds for outcomes observed post-Second World War, when many firms would have plausibly switched industries (i.e., left war production). \\
    \hline
    \pagebreak
    \multicolumn{5}{|l|}{\textit{\cite{fink_seasonal_2020}}} \\
    \hline
    Complete experiment offering subsidized loans to randomly selected villages in rural Zambia where farmers suffer from liquidity constraints in the months prior to harvest (lean season). &
    Offered cash and maize loans at the start of the lean season (January) with repayment due at harvest (July) in either cash, maize or both. &
    Individual- and village-level earnings. &
    Estimate Lee bounds to address that the likelihood of entering the labor market decreases with the loan treatment. &
    Conditional on entry to the labor market, treatment has potential to affect the types of jobs individuals accept; this is particularly plausible in this paper's setting as labor sales occur within villages between better- and worse-off farmers at individually negotiated rates. \\
    \hline
    \pagebreak
    \multicolumn{5}{|l|}{\textit{\cite{giorcelli_long-term_2019}}} \\
    \hline
    Study long-run effects of management on firm performance; estimate effects of US Technical Assistance and Productivity Program (USTAPP) which provided management training and technologically advanced machines to Italian firms from 1952 to 1958. &
    Exploit unexpected USTAPP budget cut which occurred after the firm application submission and review period which resulted in many firms not receiving training; compare applicant firms who received training to those who did not. &
    Firm-level: sales, number of employees, total factor productivity revenue. &
    Estimate Lee bounds to address the treatment-control difference in firm survival probability. &
    As in \cite{bianchi_dynamics_2022}, conditional on firm survival, management training likely affects the sorting of firms across industries and/or other important dimensions. \\
    \hline
    \pagebreak
    \multicolumn{5}{|l|}{\textit{\cite{fisman_cultural_2017}}} \\
    \hline
    Estimate effect of cultural proximity (e.g., shared codes, language, religion) on loan outcomes for lenders and borrowers using dyadic data on religion and caste for lending officers and borrowers from a state-owned Indian bank. &
    Exploit lending officer rotation policy providing variation in officer-borrower matching. &
    Cultural group-level: amount of debt received, total number of borrowers and average loan size. &
    Estimate Lee bounds as column 3 outcomes are only observed conditional on a group receiving credit. &
    Conditional on a group receiving credit, ``same group matches" also plausibly affect the type of loans a group receive. For example, ``same group match" borrowers may receive favorable loan terms; this paper indeed notes the potential for these effects but is constrained by data limitations. \\
    \hline
    \multicolumn{5}{|l|}{\textit{\cite{blanco_effects_2013}}} \\
    \hline
    Estimate bounds on average and quantile treatment effects of Job Corps. &
    Job Corps program; see description in \Cref{sec:jc_prog_desc}.
    &
    Worker wages. &
    Estimate Lee bounds to address that Job Corps training may affect labor supply along the extensive margin. &
    See \Cref{sec:diff_sort_treat_ctl}. \\
    \hline
    \\
    \caption{`Top 5' papers potentially collapsing multilayered sample selection to a single dimension \label{tab:app_d_tab}}
\end{longtable}

\clearpage

\end{document}

%% file: empiric_tbls_figs/pscores/health_pscores_weighted.tex
 
Week 90     &      0.2239&      0.2372&      0.2361&      0.2229\\
Week 135    &      0.2758&      0.3037&      0.2415&      0.2414\\
Week 180    &      0.2941&      0.3313&      0.2462&      0.2512\\
Week 208    &      0.3142&      0.3559&      0.2513&      0.2509\\

%% file: empiric_tbls_figs/bounds/health_wk90_bounds_weighted.tex
\\
\midrule
Baseline    &      0.0010&      0.0000&     -2.1415&      2.3907&           &           \\
$ p_{H,H} \geq p_{H,L} $&      0.1120&      0.1108&     -0.4214&      0.5020&     -0.4002&      0.4542\\
(H,L) is smallest type&      0.2239&      0.2228&     -0.0023&      0.0754&     -0.0191&      0.0673\\
$ p_{H,L} = 0 $&      0.2239&      0.2228&     -0.0018&      0.0750&     -0.0191&      0.0673\\

%% file: empiric_tbls_figs/bounds/health_wk135_bounds_weighted.tex
\\
\midrule
Baseline    &      0.0344&      0.0000&     -1.1228&      1.1454&           &           \\
$ p_{H,H} \geq p_{H,L} $&      0.1379&      0.0758&     -0.5075&      0.5433&     -0.6064&      0.7090\\
(H,L) is smallest type&      0.2619&      0.2137&     -0.1135&      0.1369&     -0.1180&      0.1672\\
$ p_{H,L} = 0 $&      0.2758&      0.2137&     -0.0529&      0.0732&     -0.1180&      0.1672\\

%% file: empiric_tbls_figs/bounds/health_wk180_bounds_weighted.tex
\\
\midrule
Baseline    &      0.0429&      0.0000&     -1.0454&      1.1410&           &           \\
$ p_{H,H} \geq p_{H,L} $&      0.1471&      0.0620&     -0.5063&      0.5525&     -0.7710&      0.8503\\
(H,L) is smallest type&      0.2730&      0.2090&     -0.1421&      0.1704&     -0.1806&      0.2110\\
$ p_{H,L} = 0 $&      0.2941&      0.2090&     -0.0552&      0.0854&     -0.1806&      0.2110\\

%% file: empiric_tbls_figs/bounds/health_wk208_bounds_weighted.tex
\\
\midrule
Baseline    &      0.0633&      0.0000&     -0.8821&      0.9895&           &           \\
$ p_{H,H} \geq p_{H,L} $&      0.1571&      0.0525&     -0.4888&      0.5670&     -0.9059&      0.8730\\
(H,L) is smallest type&      0.2935&      0.2096&     -0.1217&      0.1797&     -0.1914&      0.2082\\
$ p_{H,L} = 0 $&      0.3142&      0.2096&     -0.0430&      0.1016&     -0.1914&      0.2082\\

%% file: empiric_tbls_figs/workadv_bounds/pscores.tex
 
Madison Strategies&      0.2035&      0.2975&      0.4535&      0.3711\\
Towards Employment&      0.2923&      0.3524&      0.3238&      0.3410\\

%% file: empiric_tbls_figs/workadv_bounds/bounds_okla_weighted.tex
\\
\midrule
Baseline    &      0.0000&      0.1560&           &           &  -6172.4634&   7139.9806\\
$ p_{H,H} \geq p_{H,L} $&      0.1017&      0.2578&  -7380.1460&   5628.8654&  -3353.0004&   4562.1007\\
(H,L) is smallest type&      0.1977&      0.3595&  -2652.6847&   1216.5151&   -798.7005&   2016.2079\\
$ p_{H,L} = 0 $&      0.2035&      0.3595&  -2353.4366&    806.8709&   -798.7005&   2016.2079\\

%% file: empiric_tbls_figs/workadv_bounds/bounds_ohio_weighted.tex
\\
\midrule
Baseline    &      0.0000&      0.0000&           &           &           &           \\
$ p_{H,H} \geq p_{H,L} $&      0.1461&      0.1175&  -5784.7451&   6488.1176&  -6510.9268&   5975.5610\\
(H,L) is smallest type&      0.2536&      0.2636&  -2044.9605&   3064.8644&  -2335.5870&   1932.3913\\
$ p_{H,L} = 0 $&      0.2923&      0.2636&   -740.4216&   1356.9608&  -2335.5870&   1932.3913\\

%% file: empiric_tbls_figs/sum_stats/sumstats_weighted_1.tex
            &        Mean&          S.D.&        Mean&          S.D.&  Difference&          S.E.\\
\midrule
Female      &        0.46&        0.50&        0.45&        0.50&       -0.01&        0.01\\
Age at baseline&       18.35&        2.10&       18.44&        2.16&        0.09&        0.05\\
White, non-Hispanic&        0.26&        0.44&        0.27&        0.44&        0.00&        0.01\\
Black, non-Hispanic&        0.49&        0.50&        0.49&        0.50&        0.00&        0.01\\
Hispanic    &        0.17&        0.38&        0.17&        0.37&       -0.00&        0.01\\
Other race/ethnicity&        0.07&        0.26&        0.07&        0.26&       -0.00&        0.01\\
Never married&        0.92&        0.28&        0.92&        0.28&        0.00&        0.01\\
Married     &        0.02&        0.15&        0.02&        0.14&       -0.00&        0.00\\
Living together&        0.04&        0.20&        0.04&        0.19&       -0.00&        0.00\\
Separated   &        0.02&        0.14&        0.02&        0.15&        0.00&        0.00\\
Has child   &        0.19&        0.39&        0.19&        0.39&       -0.00&        0.01\\
Number of children&        0.27&        0.64&        0.27&        0.65&        0.00&        0.01\\
Education   &       10.11&        1.54&       10.11&        1.56&        0.01&        0.03\\
Mother's education&       11.46&        2.59&       11.48&        2.56&        0.02&        0.06\\
Father's education&       11.54&        2.79&       11.39&        2.85&       -0.15&        0.08\\
Ever arrested&        0.25&        0.43&        0.25&        0.43&       -0.00&        0.01\\

%% file: empiric_tbls_figs/sum_stats/sumstats_weighted_2.tex
 
$ < $3,000  &        0.25&        0.43&        0.25&        0.44&        0.00&        0.01\\
3,000-6,000 &        0.21&        0.41&        0.21&        0.40&       -0.00&        0.01\\
6,000-9,000 &        0.11&        0.32&        0.12&        0.32&        0.00&        0.01\\
9,000-18,000&        0.24&        0.43&        0.24&        0.43&       -0.00&        0.01\\
$ > $18,000 &        0.18&        0.39&        0.18&        0.38&       -0.00&        0.01\\

%% file: empiric_tbls_figs/sum_stats/sumstats_weighted_3.tex
 
$ < $3,000  &        0.79&        0.41&        0.79&        0.41&       -0.00&        0.01\\
3,000-6,000 &        0.13&        0.34&        0.13&        0.33&       -0.00&        0.01\\
6,000-9,000 &        0.05&        0.21&        0.05&        0.22&        0.01&        0.00\\
$ > $9,000  &        0.03&        0.18&        0.03&        0.17&       -0.00&        0.00\\

%% file: empiric_tbls_figs/sum_stats/sumstats_weighted_4.tex
 
Have job    &        0.19&        0.39&        0.20&        0.40&        0.01&        0.01\\
Mths. empl. prev. yr.&        3.53&        4.24&        3.60&        4.25&        0.07&        0.09\\
Had job, prev. yr.&        0.63&        0.48&        0.63&        0.48&        0.01&        0.01\\
Earnings, prev. yr.&     2810.48&     4435.62&     2906.45&     6401.33&       95.97&      117.10\\
Usual hours/week&       20.91&       20.70&       21.82&       21.05&        0.91&        0.45\\
Usual weekly earn.&      102.89&      116.46&      110.99&      350.61&        8.10&        5.09\\

%% file: empiric_tbls_figs/sum_stats/sumstats_weighted_5.tex
 
Week 52 hours&       17.78&       23.39&       15.30&       22.68&       -2.49&        0.49\\
Week 104 hours&       21.98&       26.08&       22.64&       26.25&        0.67&        0.56\\
Week 156 hours&       23.88&       26.15&       25.88&       26.57&        2.00&        0.56\\
Week 208 hours&       25.83&       26.25&       27.79&       25.74&        1.95&        0.56\\
Week 52 earn.&      103.80&      159.89&       91.55&      149.28&      -12.25&        3.33\\
Week 104 earn.&      150.41&      210.24&      157.42&      200.27&        7.02&        4.42\\
Week 156 earn.&      180.88&      224.43&      203.71&      239.80&       22.84&        4.94\\
Week 208 earn.&      200.50&      230.66&      227.91&      250.22&       27.41&        5.11\\

%% file: empiric_tbls_figs/sum_stats/sumstats_weighted_6.tex
Total 4 yr. earn.&    30006.69&    26893.60&    30800.41&    26437.39&      793.72&      571.83\\

%% file: empiric_tbls_figs/sum_stats/sumstats_weighted_7.tex
 
Sample size &        3599&           &        5546&           &        9145&           \\

%% file: empiric_tbls_figs/amen_dist/scho_t3_wk90_weighted_A_slides.tex
            &        Mean&          S.D.&        Mean&          S.D.&  Difference&          S.E.\\
\midrule
Health insurance&      0.4860&      0.5000&      0.5072&      0.5001&      0.0211&      0.0186\\
Paid sick leave&      0.3922&      0.4884&      0.4251&      0.4945&      0.0329&      0.0183\\
Paid vacation&      0.5407&      0.4985&      0.5800&      0.4937&      0.0393&      0.0180\\
Childcare assistance&      0.1311&      0.3376&      0.1422&      0.3494&      0.0111&      0.0127\\
Flexible hours&      0.5330&      0.4991&      0.5568&      0.4969&      0.0237&      0.0183\\
Employer-provided transportation&      0.1973&      0.3981&      0.1876&      0.3905&     -0.0097&      0.0147\\
Pension or retirement benefits&      0.3670&      0.4822&      0.3938&      0.4887&      0.0268&      0.0180\\
Dental plan &      0.3863&      0.4871&      0.4288&      0.4950&      0.0425&      0.0182\\
Tuition reimbursement&      0.2212&      0.4152&      0.2602&      0.4389&      0.0390&      0.0158\\

%% file: empiric_tbls_figs/amen_dist/scho_t3_wk90_weighted_B_slides.tex
 
Employment  &      0.4368&      0.4961&      0.4391&      0.4963&      0.0022&      0.0111\\
Sample size &   3288&           &   5091&           &   8379&           \\

%% file: empiric_tbls_figs/amen_dist/scho_t3_wk135_weighted_A_slides.tex
            &        Mean&          S.D.&        Mean&          S.D.&  Difference&          S.E.\\
\midrule
Health insurance&      0.5375&      0.4988&      0.5514&      0.4975&      0.0139&      0.0177\\
Paid sick leave&      0.4614&      0.4987&      0.4552&      0.4981&     -0.0061&      0.0177\\
Paid vacation&      0.6067&      0.4887&      0.5954&      0.4909&     -0.0114&      0.0171\\
Childcare assistance&      0.1254&      0.3313&      0.1411&      0.3482&      0.0157&      0.0120\\
Flexible hours&      0.5600&      0.4966&      0.5794&      0.4938&      0.0194&      0.0174\\
Employer-provided transportation&      0.1899&      0.3924&      0.1867&      0.3898&     -0.0032&      0.0139\\
Pension or retirement benefits&      0.4355&      0.4960&      0.4428&      0.4968&      0.0073&      0.0176\\
Dental plan &      0.4745&      0.4995&      0.4733&      0.4994&     -0.0012&      0.0178\\
Tuition reimbursement&      0.2574&      0.4374&      0.2883&      0.4531&      0.0309&      0.0158\\

%% file: empiric_tbls_figs/amen_dist/scho_t3_wk135_weighted_B_slides.tex
 
Employment  &      0.4923&      0.5000&      0.5242&      0.4995&      0.0319&      0.0112\\
Sample size &   3288&           &   5091&           &   8379&           \\

%% file: empiric_tbls_figs/amen_dist/scho_t3_wk180_weighted_A_slides.tex
            &        Mean&          S.D.&        Mean&          S.D.&  Difference&          S.E.\\
\midrule
Health insurance&      0.5503&      0.4976&      0.5754&      0.4944&      0.0251&      0.0155\\
Paid sick leave&      0.4639&      0.4988&      0.4952&      0.5001&      0.0313&      0.0156\\
Paid vacation&      0.6286&      0.4833&      0.6341&      0.4818&      0.0055&      0.0147\\
Childcare assistance&      0.1423&      0.3495&      0.1596&      0.3663&      0.0173&      0.0111\\
Flexible hours&      0.5622&      0.4963&      0.5839&      0.4930&      0.0217&      0.0153\\
Employer-provided transportation&      0.1841&      0.3877&      0.1974&      0.3981&      0.0133&      0.0123\\
Pension or retirement benefits&      0.4393&      0.4965&      0.4754&      0.4995&      0.0361&      0.0156\\
Dental plan &      0.4726&      0.4994&      0.5001&      0.5001&      0.0274&      0.0156\\
Tuition reimbursement&      0.2676&      0.4428&      0.3047&      0.4604&      0.0371&      0.0141\\

%% file: empiric_tbls_figs/amen_dist/scho_t3_wk180_weighted_B_slides.tex
 
Employment  &      0.5216&      0.4996&      0.5651&      0.4958&      0.0435&      0.0111\\
Sample size &   3288&           &   5091&           &   8379&           \\

%% file: empiric_tbls_figs/amen_dist/scho_t3_wk208_weighted_A_slides.tex
            &        Mean&          S.D.&        Mean&          S.D.&  Difference&          S.E.\\
\midrule
Health insurance&      0.5619&      0.4963&      0.5945&      0.4911&      0.0326&      0.0151\\
Paid sick leave&      0.4752&      0.4995&      0.5187&      0.4997&      0.0435&      0.0153\\
Paid vacation&      0.6363&      0.4812&      0.6488&      0.4774&      0.0126&      0.0143\\
Childcare assistance&      0.1441&      0.3513&      0.1680&      0.3739&      0.0239&      0.0110\\
Flexible hours&      0.6038&      0.4892&      0.6088&      0.4881&      0.0049&      0.0147\\
Employer-provided transportation&      0.1877&      0.3906&      0.2038&      0.4029&      0.0160&      0.0121\\
Pension or retirement benefits&      0.4500&      0.4976&      0.4883&      0.5000&      0.0382&      0.0152\\
Dental plan &      0.4882&      0.5000&      0.5175&      0.4998&      0.0293&      0.0153\\
Tuition reimbursement&      0.2951&      0.4562&      0.3150&      0.4646&      0.0199&      0.0140\\

%% file: empiric_tbls_figs/amen_dist/scho_t3_wk208_weighted_B_slides.tex
 
Employment  &      0.5510&      0.4975&      0.5902&      0.4918&      0.0393&      0.0111\\
Sample size &   3288&           &   5091&           &   8379&           \\

%% file: empiric_tbls_figs/types_dist/typesdist_wk90_weighted_A.tex
            &        Mean&          S.D.&        Mean&          S.D.&  Difference&          S.E.&ln(Hourly wage)&          S.D.\\
\midrule
H=1, R=1, V=1&      0.2914&      0.4547&      0.3352&      0.4722&      0.0438&      0.0198&      1.9757&      0.3192\\
H=1, R=1, V=0&      0.0303&      0.1715&      0.0192&      0.1375&     -0.0110&      0.0068&      1.9308&      0.3114\\
H=1, R=0, V=1&      0.1123&      0.3160&      0.1073&      0.3096&     -0.0050&      0.0135&      1.8983&      0.2501\\
H=1, R=0, V=0&      0.0500&      0.2180&      0.0485&      0.2149&     -0.0015&      0.0093&      1.8232&      0.2830\\
H=0, R=1, V=1&      0.0288&      0.1672&      0.0341&      0.1817&      0.0054&      0.0074&      1.8834&      0.2968\\
H=0, R=1, V=0&      0.0195&      0.1383&      0.0129&      0.1130&     -0.0066&      0.0055&      1.8118&      0.2250\\
H=0, R=0, V=1&      0.0758&      0.2648&      0.0913&      0.2881&      0.0155&      0.0118&      1.7606&      0.3116\\
H=0, R=0, V=0&      0.3920&      0.4885&      0.3514&      0.4776&     -0.0406&      0.0208&      1.7558&      0.3098\\

%% file: empiric_tbls_figs/types_dist/typesdist_wk90_weighted_B.tex
 
Sample size &        2454&           &        3778&           &           &           &        6232&           \\

%% file: empiric_tbls_figs/types_dist/typesdist_wk135_weighted_A.tex
            &        Mean&          S.D.&        Mean&          S.D.&  Difference&          S.E.&ln(Hourly wage)&          S.D.\\
\midrule
H=1, R=1, V=1&      0.3598&      0.4802&      0.3806&      0.4857&      0.0208&      0.0192&      2.0418&      0.3401\\
H=1, R=1, V=0&      0.0262&      0.1597&      0.0270&      0.1620&      0.0008&      0.0064&      2.0489&      0.3276\\
H=1, R=0, V=1&      0.1118&      0.3152&      0.1007&      0.3011&     -0.0110&      0.0123&      1.9947&      0.3217\\
H=1, R=0, V=0&      0.0345&      0.1825&      0.0445&      0.2063&      0.0101&      0.0076&      1.8928&      0.3839\\
H=0, R=1, V=1&      0.0302&      0.1713&      0.0284&      0.1661&     -0.0019&      0.0067&      1.9083&      0.2917\\
H=0, R=1, V=0&      0.0222&      0.1475&      0.0188&      0.1357&     -0.0035&      0.0057&      1.8744&      0.3267\\
H=0, R=0, V=1&      0.0843&      0.2779&      0.0625&      0.2421&     -0.0218&      0.0106&      1.8479&      0.3182\\
H=0, R=0, V=0&      0.3311&      0.4709&      0.3376&      0.4730&      0.0064&      0.0188&      1.8545&      0.3716\\

%% file: empiric_tbls_figs/types_dist/typesdist_wk135_weighted_B.tex
 
Sample size &        2454&           &        3778&           &           &           &        6232&           \\

%% file: empiric_tbls_figs/types_dist/typesdist_wk180_weighted_A.tex
            &        Mean&          S.D.&        Mean&          S.D.&  Difference&          S.E.&ln(Hourly wage)&          S.D.\\
\midrule
H=1, R=1, V=1&      0.3648&      0.4816&      0.3899&      0.4878&      0.0251&      0.0177&      2.1175&      0.3441\\
H=1, R=1, V=0&      0.0202&      0.1409&      0.0324&      0.1772&      0.0122&      0.0057&      2.1131&      0.3140\\
H=1, R=0, V=1&      0.1208&      0.3260&      0.1045&      0.3059&     -0.0163&      0.0117&      2.0218&      0.3153\\
H=1, R=0, V=0&      0.0386&      0.1928&      0.0381&      0.1915&     -0.0005&      0.0070&      1.8936&      0.4835\\
H=0, R=1, V=1&      0.0289&      0.1677&      0.0306&      0.1721&      0.0016&      0.0062&      1.9896&      0.3512\\
H=0, R=1, V=0&      0.0114&      0.1060&      0.0183&      0.1340&      0.0069&      0.0043&      1.9150&      0.3633\\
H=0, R=0, V=1&      0.0851&      0.2792&      0.0711&      0.2571&     -0.0140&      0.0099&      1.8927&      0.3296\\
H=0, R=0, V=0&      0.3301&      0.4704&      0.3151&      0.4647&     -0.0149&      0.0171&      1.8606&      0.4033\\

%% file: empiric_tbls_figs/types_dist/typesdist_wk180_weighted_B.tex
 
Sample size &        2454&           &        3778&           &           &           &        6232&           \\

%% file: empiric_tbls_figs/types_dist/typesdist_wk208_weighted_A.tex
            &        Mean&          S.D.&        Mean&          S.D.&  Difference&          S.E.&ln(Hourly wage)&          S.D.\\
\midrule
H=1, R=1, V=1&      0.3663&      0.4820&      0.4083&      0.4916&      0.0420&      0.0172&      2.1271&      0.3271\\
H=1, R=1, V=0&      0.0256&      0.1579&      0.0298&      0.1702&      0.0043&      0.0057&      2.0602&      0.3268\\
H=1, R=0, V=1&      0.1198&      0.3248&      0.1073&      0.3096&     -0.0125&      0.0113&      2.0046&      0.3436\\
H=1, R=0, V=0&      0.0418&      0.2003&      0.0387&      0.1929&     -0.0031&      0.0070&      1.9273&      0.4623\\
H=0, R=1, V=1&      0.0291&      0.1681&      0.0276&      0.1639&     -0.0015&      0.0059&      1.9958&      0.3719\\
H=0, R=1, V=0&      0.0133&      0.1148&      0.0170&      0.1293&      0.0037&      0.0043&      1.9189&      0.3324\\
H=0, R=0, V=1&      0.0875&      0.2827&      0.0717&      0.2580&     -0.0158&      0.0097&      1.9253&      0.2689\\
H=0, R=0, V=0&      0.3166&      0.4653&      0.2996&      0.4582&     -0.0170&      0.0163&      1.8768&      0.4223\\

%% file: empiric_tbls_figs/types_dist/typesdist_wk208_weighted_B.tex
 
Sample size &        2454&           &        3778&           &           &           &        6232&           \\

%% file: empiric_tbls_figs/bounds/lee_bounds_weighted.tex
 
Week 90     &      0.4600&      0.4601&      0.0003&      0.0468&      0.0484\\
Week 135    &      0.5173&      0.5451&      0.0509&     -0.0072&      0.0842\\
Week 180    &      0.5403&      0.5825&      0.0724&     -0.0325&      0.0901\\
Week 208    &      0.5655&      0.6068&      0.0680&     -0.0217&      0.0989\\

%% file: empiric_tbls_figs/bounds/lee_bounds_vingtiles.tex
 
Week 90     &      0.4600&      0.4601&      0.0003&      0.0423&      0.0428\\
Week 135    &      0.5173&      0.5451&      0.0509&     -0.0159&      0.0757\\
Week 180    &      0.5403&      0.5825&      0.0724&     -0.0325&      0.0868\\
Week 208    &      0.5655&      0.6068&      0.0680&     -0.0194&      0.0933\\

%% file: empiric_tbls_figs/pscores/vacat_pscores_weighted.tex
 
Week 90     &      0.2470&      0.2673&      0.2129&      0.1928\\
Week 135    &      0.3102&      0.3209&      0.2071&      0.2241\\
Week 180    &      0.3334&      0.3589&      0.2069&      0.2236\\
Week 208    &      0.3516&      0.3839&      0.2139&      0.2229\\

%% file: empiric_tbls_figs/pscores/pension_pscores_weighted.tex
 
Week 90     &      0.1706&      0.1852&      0.2894&      0.2749\\
Week 135    &      0.2275&      0.2475&      0.2898&      0.2976\\
Week 180    &      0.2309&      0.2748&      0.3094&      0.3076\\
Week 208    &      0.2460&      0.2930&      0.3195&      0.3138\\

%% file: empiric_tbls_figs/agg_bounds/health_wk90_agg_bounds_weighted.tex
\\
\midrule
Baseline    &      0.0014&      0.0004&      0.0014&      0.0004&     -2.4634&      2.4588\\
$ p_{H,H} \geq p_{H,L} $&      0.1120&      0.1109&      0.1120&      0.1109&     -0.4106&      0.4780\\
(H,L) is smallest type&      0.2235&      0.2224&      0.2235&      0.2224&     -0.0136&      0.0749\\
$ p_{H,L} = 0 $&      0.2235&      0.2224&      0.2235&      0.2224&     -0.0136&      0.0749\\

%% file: empiric_tbls_figs/agg_bounds/health_wk135_agg_bounds_weighted.tex
\\
\midrule
Baseline    &      0.0366&      0.0022&      0.0392&      0.0048&     -1.1602&      1.2762\\
$ p_{H,H} \geq p_{H,L} $&      0.1380&      0.0758&      0.1380&      0.0758&     -0.5423&      0.6017\\
(H,L) is smallest type&      0.2614&      0.2127&      0.2614&      0.2127&     -0.1179&      0.1532\\
$ p_{H,L} = 0 $&      0.2754&      0.2132&      0.2754&      0.2132&     -0.0833&      0.1165\\

%% file: empiric_tbls_figs/agg_bounds/health_wk180_agg_bounds_weighted.tex
\\
\midrule
Baseline    &      0.0475&      0.0046&      0.0485&      0.0056&     -1.1369&      1.2631\\
$ p_{H,H} \geq p_{H,L} $&      0.1471&      0.0620&      0.1471&      0.0620&     -0.5846&      0.6407\\
(H,L) is smallest type&      0.2727&      0.2081&      0.2727&      0.2081&     -0.1607&      0.1900\\
$ p_{H,L} = 0 $&      0.2937&      0.2086&      0.2937&      0.2086&     -0.1101&      0.1397\\

%% file: empiric_tbls_figs/agg_bounds/health_wk208_agg_bounds_weighted.tex
\\
\midrule
Baseline    &      0.0705&      0.0072&      0.0671&      0.0038&     -0.9987&      1.0704\\
$ p_{H,H} \geq p_{H,L} $&      0.1571&      0.0525&      0.1571&      0.0525&     -0.5932&      0.6436\\
(H,L) is smallest type&      0.2928&      0.2087&      0.2928&      0.2087&     -0.1532&      0.1941\\
$ p_{H,L} = 0 $&      0.3138&      0.2092&      0.3138&      0.2092&     -0.1049&      0.1461\\

%% file: empiric_tbls_figs/bounds/vacat_wk90_bounds_weighted.tex
\\
\midrule
Baseline    &      0.0542&      0.0000&     -0.8794&      0.9952&           &           \\
$ p_{H,H} \geq p_{H,L} $&      0.1235&      0.0692&     -0.5033&      0.5722&     -0.5478&      0.6430\\
(H,L) is smallest type&      0.2470&      0.1927&     -0.0228&      0.1002&     -0.0242&      0.1061\\
$ p_{H,L} = 0 $&      0.2470&      0.1927&     -0.0218&      0.0995&     -0.0242&      0.1061\\

%% file: empiric_tbls_figs/bounds/vacat_wk135_bounds_weighted.tex
\\
\midrule
Baseline    &      0.0860&      0.0000&     -0.7956&      0.8818&           &           \\
$ p_{H,H} \geq p_{H,L} $&      0.1551&      0.0413&     -0.5108&      0.5835&     -0.8611&      0.9932\\
(H,L) is smallest type&      0.2994&      0.1964&     -0.0655&      0.1239&     -0.1186&      0.1420\\
$ p_{H,L} = 0 $&      0.3102&      0.1964&      0.0012&      0.0725&     -0.1186&      0.1420\\

%% file: empiric_tbls_figs/bounds/vacat_wk180_bounds_weighted.tex
\\
\midrule
Baseline    &      0.1099&      0.0000&     -0.7394&      0.7904&           &           \\
$ p_{H,H} \geq p_{H,L} $&      0.1667&      0.0147&     -0.5265&      0.5794&     -1.6717&      1.7737\\
(H,L) is smallest type&      0.3123&      0.1814&     -0.1244&      0.1629&     -0.1909&      0.2229\\
$ p_{H,L} = 0 $&      0.3334&      0.1814&     -0.0361&      0.0806&     -0.1909&      0.2229\\

%% file: empiric_tbls_figs/bounds/vacat_wk208_bounds_weighted.tex
\\
\midrule
Baseline    &      0.1287&      0.0000&     -0.6719&      0.7683&           &           \\
$ p_{H,H} \geq p_{H,L} $&      0.1758&      0.0058&     -0.5136&      0.6070&     -2.7579&      2.6008\\
(H,L) is smallest type&      0.3310&      0.1816&     -0.1124&      0.1847&     -0.2046&      0.2294\\
$ p_{H,L} = 0 $&      0.3516&      0.1816&     -0.0206&      0.1083&     -0.2046&      0.2294\\

%% file: empiric_tbls_figs/agg_bounds/vacat_wk90_agg_bounds_weighted.tex
\\
\midrule
Baseline    &      0.0577&      0.0035&      0.0585&      0.0043&     -0.9155&      1.0441\\
$ p_{H,H} \geq p_{H,L} $&      0.1235&      0.0692&      0.1235&      0.0692&     -0.5192&      0.5976\\
(H,L) is smallest type&      0.2466&      0.1922&      0.2466&      0.1922&     -0.0284&      0.1068\\
$ p_{H,L} = 0 $&      0.2466&      0.1922&      0.2466&      0.1922&     -0.0284&      0.1068\\

%% file: empiric_tbls_figs/agg_bounds/vacat_wk135_agg_bounds_weighted.tex
\\
\midrule
Baseline    &      0.0921&      0.0060&      0.0933&      0.0072&     -0.8424&      0.9616\\
$ p_{H,H} \geq p_{H,L} $&      0.1551&      0.0413&      0.1551&      0.0413&     -0.5844&      0.6696\\
(H,L) is smallest type&      0.2978&      0.1955&      0.2978&      0.1955&     -0.0919&      0.1362\\
$ p_{H,L} = 0 $&      0.3098&      0.1960&      0.3098&      0.1960&     -0.0484&      0.1022\\

%% file: empiric_tbls_figs/agg_bounds/vacat_wk180_agg_bounds_weighted.tex
\\
\midrule
Baseline    &      0.1173&      0.0075&      0.1162&      0.0064&     -0.7990&      0.8620\\
$ p_{H,H} \geq p_{H,L} $&      0.1670&      0.0228&      0.1670&      0.0204&     -0.6207&      0.6776\\
(H,L) is smallest type&      0.3120&      0.1805&      0.3120&      0.1805&     -0.1506&      0.1868\\
$ p_{H,L} = 0 $&      0.3330&      0.1810&      0.3330&      0.1810&     -0.0935&      0.1328\\

%% file: empiric_tbls_figs/agg_bounds/vacat_wk208_agg_bounds_weighted.tex
\\
\midrule
Baseline    &      0.1374&      0.0087&      0.1352&      0.0065&     -0.7404&      0.8219\\
$ p_{H,H} \geq p_{H,L} $&      0.1761&      0.0249&      0.1761&      0.0186&     -0.6148&      0.6902\\
(H,L) is smallest type&      0.3302&      0.1807&      0.3302&      0.1807&     -0.1477&      0.2034\\
$ p_{H,L} = 0 $&      0.3512&      0.1812&      0.3512&      0.1812&     -0.0860&      0.1517\\

%% file: empiric_tbls_figs/bounds/pension_wk90_bounds_weighted.tex
\\
\midrule
Baseline    &      0.0000&      0.1042&           &           &     -0.5064&      0.5623\\
$ p_{H,H} \geq p_{H,L} $&      0.0853&      0.1894&     -0.4465&      0.5337&     -0.2568&      0.3074\\
(H,L) is smallest type&      0.1705&      0.2747&     -0.0141&      0.0921&     -0.0161&      0.0628\\
$ p_{H,L} = 0 $&      0.1706&      0.2747&     -0.0135&      0.0916&     -0.0161&      0.0628\\

%% file: empiric_tbls_figs/bounds/pension_wk135_bounds_weighted.tex
\\
\midrule
Baseline    &      0.0000&      0.0423&           &           &     -0.9636&      1.0670\\
$ p_{H,H} \geq p_{H,L} $&      0.1138&      0.1561&     -0.4749&      0.5746&     -0.4374&      0.4505\\
(H,L) is smallest type&      0.2136&      0.2699&     -0.0854&      0.1729&     -0.1203&      0.1144\\
$ p_{H,L} = 0 $&      0.2275&      0.2699&     -0.0143&      0.0977&     -0.1203&      0.1144\\

%% file: empiric_tbls_figs/bounds/pension_wk180_bounds_weighted.tex
\\
\midrule
Baseline    &      0.0000&      0.0345&           &           &     -1.1762&      1.2907\\
$ p_{H,H} \geq p_{H,L} $&      0.1155&      0.1500&     -0.5247&      0.5641&     -0.4911&      0.5305\\
(H,L) is smallest type&      0.2098&      0.2655&     -0.1802&      0.2097&     -0.1655&      0.1947\\
$ p_{H,L} = 0 $&      0.2309&      0.2655&     -0.0856&      0.1062&     -0.1655&      0.1947\\

%% file: empiric_tbls_figs/bounds/pension_wk208_bounds_weighted.tex
\\
\midrule
Baseline    &      0.0000&      0.0265&           &           &     -1.4213&      1.4398\\
$ p_{H,H} \geq p_{H,L} $&      0.1230&      0.1495&     -0.5019&      0.5491&     -0.5103&      0.5400\\
(H,L) is smallest type&      0.2254&      0.2725&     -0.1615&      0.2023&     -0.1640&      0.1975\\
$ p_{H,L} = 0 $&      0.2460&      0.2725&     -0.0784&      0.1084&     -0.1640&      0.1975\\

%% file: empiric_tbls_figs/agg_bounds/pension_wk90_agg_bounds_weighted.tex
\\
\midrule
Baseline    &      0.0075&      0.1117&      0.0084&      0.1126&     -0.5397&      0.6025\\
$ p_{H,H} \geq p_{H,L} $&      0.0852&      0.1894&      0.0852&      0.1894&     -0.3159&      0.3779\\
(H,L) is smallest type&      0.1702&      0.2744&      0.1702&      0.2744&     -0.0181&      0.0773\\
$ p_{H,L} = 0 $&      0.1702&      0.2744&      0.1702&      0.2744&     -0.0181&      0.0773\\

%% file: empiric_tbls_figs/agg_bounds/pension_wk135_agg_bounds_weighted.tex
\\
\midrule
Baseline    &      0.0032&      0.0455&      0.0038&      0.0461&     -1.0101&      1.1059\\
$ p_{H,H} \geq p_{H,L} $&      0.1141&      0.1565&      0.1141&      0.1565&     -0.4519&      0.5015\\
(H,L) is smallest type&      0.2131&      0.2689&      0.2131&      0.2689&     -0.1074&      0.1430\\
$ p_{H,L} = 0 $&      0.2271&      0.2694&      0.2271&      0.2694&     -0.0739&      0.1091\\

%% file: empiric_tbls_figs/agg_bounds/pension_wk180_agg_bounds_weighted.tex
\\
\midrule
Baseline    &      0.0024&      0.0371&      0.0024&      0.0371&     -1.2130&      1.3253\\
$ p_{H,H} \geq p_{H,L} $&      0.1154&      0.1501&      0.1154&      0.1501&     -0.5056&      0.5450\\
(H,L) is smallest type&      0.2095&      0.2646&      0.2095&      0.2646&     -0.1739&      0.2032\\
$ p_{H,L} = 0 $&      0.2305&      0.2651&      0.2305&      0.2651&     -0.1309&      0.1555\\

%% file: empiric_tbls_figs/agg_bounds/pension_wk208_agg_bounds_weighted.tex
\\
\midrule
Baseline    &      0.0010&      0.0276&      0.0014&      0.0280&     -1.4492&      1.4578\\
$ p_{H,H} \geq p_{H,L} $&      0.1230&      0.1496&      0.1230&      0.1496&     -0.5063&      0.5439\\
(H,L) is smallest type&      0.2246&      0.2716&      0.2246&      0.2716&     -0.1655&      0.2024\\
$ p_{H,L} = 0 $&      0.2456&      0.2721&      0.2456&      0.2721&     -0.1257&      0.1572\\

%% file: empiric_tbls_figs/workadv_bounds/sumstats_okla_1.tex
            &        Mean&          S.D.&        Mean&          S.D.&  Difference&          S.E.\\
\midrule
Female      &        0.14&        0.34&        0.18&        0.39&        0.05&        0.03\\
Adults $ \leq $ 24&        0.22&        0.42&        0.23&        0.42&        0.01&        0.03\\
Black       &        0.35&        0.48&        0.39&        0.49&        0.03&        0.04\\
HS/GED or less&        0.40&        0.49&        0.43&        0.50&        0.03&        0.04\\
\textbf{At baseline:} \\
Have job    &        0.28&        0.45&        0.26&        0.44&       -0.02&        0.03\\
Quarterly earnings&     2257.07&     2820.54&     1944.51&     2495.57&     -312.56&      201.91\\

%% file: empiric_tbls_figs/workadv_bounds/sumstats_okla_3.tex
 \textbf{8 quarters post randomization:} \\
Employment  &        0.66&        0.48&        0.67&        0.47&        0.01&        0.04\\
Share employment in target sector &        0.31&        0.46&        0.44&        0.50&        0.14&        0.04\\

Quarterly earnings&     3703.87&     4169.10&     4049.91&     4003.27&      346.04&      309.72\\

%% file: empiric_tbls_figs/workadv_bounds/sumstats_ohio_1.tex
            &        Mean&          S.D.&        Mean&          S.D.&  Difference&          S.E.\\
\midrule
Female      &        0.59&        0.49&        0.58&        0.49&       -0.01&        0.04\\
Adults $ \leq $ 24&        0.23&        0.42&        0.22&        0.42&       -0.01&        0.03\\
Black       &        0.73&        0.45&        0.77&        0.42&        0.05&        0.03\\
HS/GED or less&        0.42&        0.49&        0.44&        0.50&        0.03&        0.04\\
\textbf{At baseline:} \\
Have job    &        0.27&        0.45&        0.26&        0.44&       -0.01&        0.03\\
Quarterly earnings&     1504.59&     2201.72&     1723.72&     2452.60&      219.12&      176.42\\

%% file: empiric_tbls_figs/workadv_bounds/sumstats_ohio_3.tex
 \textbf{8 quarters post randomization:} \\
Employment  &        0.62&        0.49&        0.69&        0.46&        0.08&        0.04\\
Share employment in target sector &        0.47&        0.50&        0.51&        0.50&        0.03&        0.05\\
Quarterly earnings&     3040.57&     4024.21&     3497.54&     3684.01&      456.97&      292.04\\

%% file: empiric_tbls_figs/workadv_bounds/agg_bounds_okla_c.tex
\\
\midrule
Baseline    &      0.0145&      0.1710&      0.0090&      0.1655&  -6498.0976&   7255.9812\\
$ p_{H,H} \geq p_{H,L} $&      0.1010&      0.2570&      0.1010&      0.2570&  -4514.3517&   4886.1844\\
(H,L) is smallest type&      0.1960&      0.3545&      0.1960&      0.3545&  -1593.0171&   1848.4113\\
$ p_{H,L} = 0 $&      0.2010&      0.3570&      0.2010&      0.3570&  -1459.5360&   1688.9594\\

%% file: empiric_tbls_figs/workadv_bounds/agg_bounds_ohio_c.tex
\\
\midrule
Baseline    &      0.0000&      0.0000&      0.0000&      0.0000&           &           \\
$ p_{H,H} \geq p_{H,L} $&      0.1450&      0.1170&      0.1450&      0.1170&  -6142.9166&   6288.9131\\
(H,L) is smallest type&      0.2500&      0.2590&      0.2500&      0.2590&  -2318.5475&   2613.7796\\
$ p_{H,L} = 0 $&      0.2900&      0.2620&      0.2900&      0.2620&  -1562.2141&   1840.9057\\

%% file: empiric_tbls_figs/workadv_bounds/lee_bounds.tex
 
Madison Strategies&      0.6570&      0.6686&      0.0173&    244.1928&    524.9793\\
Towards Employment&      0.6160&      0.6934&      0.1116&   -676.8651&    705.6605\\

%% file: Extended-Lee-Bounds.bbl
\begin{thebibliography}{61}
\newcommand{\enquote}[1]{``#1''}
\expandafter\ifx\csname natexlab\endcsname\relax\def\natexlab#1{#1}\fi

\bibitem[\protect\citeauthoryear{Abowd, Kramarz, and Margolis}{Abowd et~al.}{1999}]{abowd_high_1999}
\textsc{Abowd, John~M., Francis Kramarz, and David~N. Margolis} (1999): \enquote{High {Wage} {Workers} and {High} {Wage} {Firms},} \emph{Econometrica}, 67, 251--333.

\bibitem[\protect\citeauthoryear{Ahn and Powell}{Ahn and Powell}{1993}]{ahn_semiparametric_1993}
\textsc{Ahn, Hyungtaik and James~L. Powell} (1993): \enquote{Semiparametric estimation of censored selection models with a nonparametric selection mechanism,} \emph{Journal of Econometrics}, 58, 3--29.

\bibitem[\protect\citeauthoryear{Andersson, Holzer, Lane, Rosenblum, and Smith}{Andersson et~al.}{2022}]{andersson_does_2022}
\textsc{Andersson, Fredrik, Harry~J. Holzer, Julia~I. Lane, David Rosenblum, and Jeffrey Smith} (2022): \enquote{Does {Federally}-{Funded} {Job} {Training} {Work}? {Nonexperimental} {Estimates} of {WIA} {Training} {Impacts} {Using} {Longitudinal} {Data} on {Workers} and {Firms},} \emph{Journal of Human Resources}, 0816--8185R1.

\bibitem[\protect\citeauthoryear{Bettinger, Gurantz, Kawano, Sacerdote, and Stevens}{Bettinger et~al.}{2019}]{bettinger_long-run_2019}
\textsc{Bettinger, Eric, Oded Gurantz, Laura Kawano, Bruce Sacerdote, and Michael Stevens} (2019): \enquote{The {Long}-{Run} {Impacts} of {Financial} {Aid}: {Evidence} from {California}’s {Cal} {Grant},} \emph{American Economic Journal: Economic Policy}, 11, 64--94.

\bibitem[\protect\citeauthoryear{Bianchi and Giorcelli}{Bianchi and Giorcelli}{2022}]{bianchi_dynamics_2022}
\textsc{Bianchi, Nicola and Michela Giorcelli} (2022): \enquote{The {Dynamics} and {Spillovers} of {Management} {Interventions}: {Evidence} from the {Training} within {Industry} {Program},} \emph{Journal of Political Economy}, 130, 1630--1675.

\bibitem[\protect\citeauthoryear{Blanco, Flores, and Flores-Lagunes}{Blanco et~al.}{2013}]{blanco_effects_2013}
\textsc{Blanco, German, Carlos~A Flores, and Alfonso Flores-Lagunes} (2013): \enquote{The {Effects} of {Job} {Corps} {Training} on {Wages} of {Adolescents} and {Young} {Adults},} \emph{American Economic Review}, 103, 418--422.

\bibitem[\protect\citeauthoryear{Bonhomme, Holzheu, Lamadon, Manresa, Mogstad, and Setzler}{Bonhomme et~al.}{2023}]{bonhomme_how_2023}
\textsc{Bonhomme, Stéphane, Kerstin Holzheu, Thibaut Lamadon, Elena Manresa, Magne Mogstad, and Bradley Setzler} (2023): \enquote{How {Much} {Should} {We} {Trust} {Estimates} of {Firm} {Effects} and {Worker} {Sorting}?} \emph{Journal of Labor Economics}, 41, 291--322.

\bibitem[\protect\citeauthoryear{Bonhomme, Lamadon, and Manresa}{Bonhomme et~al.}{2019}]{bonhomme_distributional_2019}
\textsc{Bonhomme, Stéphane, Thibaut Lamadon, and Elena Manresa} (2019): \enquote{A {Distributional} {Framework} for {Matched} {Employer} {Employee} {Data},} \emph{Econometrica}, 87, 699--739.

\bibitem[\protect\citeauthoryear{Canay, Santos, and Shaikh}{Canay et~al.}{2013}]{canay_testability_2013}
\textsc{Canay, Ivan, Andres Santos, and Azeem~M Shaikh} (2013): \enquote{On the {Testability} of {Identification} in {Some} {Nonparametric} {Models} {With} {Endogeneity},} \emph{Econometrica}, 81, 2535--2559.

\bibitem[\protect\citeauthoryear{Card, Cardoso, and Kline}{Card et~al.}{2016}]{card_bargaining_2016}
\textsc{Card, David, Ana~Rute Cardoso, and Patrick Kline} (2016): \enquote{Bargaining, {Sorting}, and the {Gender} {Wage} {Gap}: {Quantifying} the {Impact} of {Firms} on the {Relative} {Pay} of {Women} *,} \emph{The Quarterly Journal of Economics}, 131, 633--686.

\bibitem[\protect\citeauthoryear{Card, Heining, and Kline}{Card et~al.}{2013}]{card_workplace_2013}
\textsc{Card, David, Jörg Heining, and Patrick Kline} (2013): \enquote{Workplace {Heterogeneity} and the {Rise} of {West} {German} {Wage} {Inequality}*,} \emph{The Quarterly Journal of Economics}, 128, 967--1015.

\bibitem[\protect\citeauthoryear{Card, Kluve, and Weber}{Card et~al.}{2010}]{card_active_2010}
\textsc{Card, David, Jochen Kluve, and Andrea Weber} (2010): \enquote{Active {Labour} {Market} {Policy} {Evaluations}: {A} {Meta}‐{Analysis},} \emph{The Economic Journal}, 120, F452--F477.

\bibitem[\protect\citeauthoryear{Card, Kluve, and Weber}{Card et~al.}{2018}]{card_what_2018}
---\hspace{-.1pt}---\hspace{-.1pt}--- (2018): \enquote{What {Works}? {A} {Meta} {Analysis} of {Recent} {Active} {Labor} {Market} {Program} {Evaluations},} \emph{Journal of the European Economic Association}, 16, 894--931.

\bibitem[\protect\citeauthoryear{Cross and Manski}{Cross and Manski}{2002}]{cross_regressions_2002}
\textsc{Cross, Philip~J. and Charles~F. Manski} (2002): \enquote{Regressions, {Short} and {Long},} \emph{Econometrica}, 70, 357--368.

\bibitem[\protect\citeauthoryear{Cullen and Pakzad-Hurson}{Cullen and Pakzad-Hurson}{2023}]{cullen_equilibrium_2023}
\textsc{Cullen, Zoë~B. and Bobak Pakzad-Hurson} (2023): \enquote{Equilibrium {Effects} of {Pay} {Transparency},} \emph{Econometrica}, 91, 765--802.

\bibitem[\protect\citeauthoryear{Daruich, Di~Addario, and Saggio}{Daruich et~al.}{2023}]{daruich_effects_2023}
\textsc{Daruich, Diego, Sabrina Di~Addario, and Raffaele Saggio} (2023): \enquote{The {Effects} of {Partial} {Employment} {Protection} {Reforms}: {Evidence} from {Italy},} \emph{Review of Economic Studies}, 90, 2880--2942.

\bibitem[\protect\citeauthoryear{Dey and Flinn}{Dey and Flinn}{2005}]{dey_flinn_2005}
\textsc{Dey, Matthew~S. and Christopher~J. Flinn} (2005): \enquote{An Equilibrium Model of Health Insurance Provision and Wage Determination,} \emph{Econometrica}, 73, 571--627.

\bibitem[\protect\citeauthoryear{D’Haultfoeuille}{D’Haultfoeuille}{2011}]{dhaultfoeuille_completeness_2011}
\textsc{D’Haultfoeuille, Xavier} (2011): \enquote{On the {Completeness} {Condition} in {Nonparametric} {Instrumental} {Problems},} \emph{Econometric Theory}, 27, 460--471.

\bibitem[\protect\citeauthoryear{Fang and Santos}{Fang and Santos}{2019}]{fangInferenceDirectionallyDifferentiable2019}
\textsc{Fang, Zheng and Andres Santos} (2019): \enquote{Inference on {{Directionally Differentiable Functions}},} \emph{The Review of Economic Studies}, 86, 377--412.

\bibitem[\protect\citeauthoryear{Fang, Santos, Shaikh, and Torgovitsky}{Fang et~al.}{2023}]{fang_inference_2023}
\textsc{Fang, Zheng, Andres Santos, Azeem~M. Shaikh, and Alexander Torgovitsky} (2023): \enquote{Inference for {Large}‐{Scale} {Linear} {Systems} {With} {Known} {Coefficients},} \emph{Econometrica}, 91, 299--327.

\bibitem[\protect\citeauthoryear{Fink, Jack, and Masiye}{Fink et~al.}{2020}]{fink_seasonal_2020}
\textsc{Fink, Günther, B.~Kelsey Jack, and Felix Masiye} (2020): \enquote{Seasonal {Liquidity}, {Rural} {Labor} {Markets}, and {Agricultural} {Production},} \emph{American Economic Review}, 110, 3351--3392.

\bibitem[\protect\citeauthoryear{Fisman, Paravisini, and Vig}{Fisman et~al.}{2017}]{fisman_cultural_2017}
\textsc{Fisman, Raymond, Daniel Paravisini, and Vikrant Vig} (2017): \enquote{Cultural {Proximity} and {Loan} {Outcomes},} \emph{American Economic Review}, 107, 457--492.

\bibitem[\protect\citeauthoryear{Gallant and Nychka}{Gallant and Nychka}{1987}]{gallant_semi-nonparametric_1987}
\textsc{Gallant, A.~Ronald and Douglas~W. Nychka} (1987): \enquote{Semi-{Nonparametric} {Maximum} {Likelihood} {Estimation},} \emph{Econometrica}, 55, 363.

\bibitem[\protect\citeauthoryear{Gerard, Lagos, Severnini, and Card}{Gerard et~al.}{2021}]{gerard_assortative_2021}
\textsc{Gerard, François, Lorenzo Lagos, Edson Severnini, and David Card} (2021): \enquote{Assortative {Matching} or {Exclusionary} {Hiring}? {The} {Impact} of {Employment} and {Pay} {Policies} on {Racial} {Wage} {Differences} in {Brazil},} \emph{American Economic Review}, 111, 3418--3457.

\bibitem[\protect\citeauthoryear{Giorcelli}{Giorcelli}{2019}]{giorcelli_long-term_2019}
\textsc{Giorcelli, Michela} (2019): \enquote{The {Long}-{Term} {Effects} of {Management} and {Technology} {Transfers},} \emph{American Economic Review}, 109, 121--152.

\bibitem[\protect\citeauthoryear{Gottschalk, Green, and Sand}{Gottschalk et~al.}{2014}]{gottschalk_taking_2014}
\textsc{Gottschalk, Peter, David Green, and Ben Sand} (2014): \enquote{Taking {Selection} to {Task}: {Bounds} on {Trends} in {Occupational} {Task} {Prices} for the {U}.{S}., 1984-2014,} \emph{Unpublished Manuscript}.

\bibitem[\protect\citeauthoryear{Heckman}{Heckman}{1979}]{heckman_sample_1979}
\textsc{Heckman, James~J.} (1979): \enquote{Sample {Selection} {Bias} as a {Specification} {Error},} \emph{Econometrica}, 47, 153.

\bibitem[\protect\citeauthoryear{Heckman, Lalonde, and Smith}{Heckman et~al.}{1999}]{heckman_economics_1999}
\textsc{Heckman, James~J., Robert~J. Lalonde, and Jeffrey~A. Smith} (1999): \enquote{The {Economics} and {Econometrics} of {Active} {Labor} {Market} {Programs},} in \emph{Handbook of {Labor} {Economics}}, Elsevier, vol.~3, 1865--2097.

\bibitem[\protect\citeauthoryear{Heckman and Pinto}{Heckman and Pinto}{2018}]{heckman_unordered_2018}
\textsc{Heckman, James~J. and Rodrigo Pinto} (2018): \enquote{Unordered {Monotonicity},} \emph{Econometrica}, 86, 1--35.

\bibitem[\protect\citeauthoryear{Hendren and Sprung-Keyser}{Hendren and Sprung-Keyser}{2020}]{hendren_unified_2020}
\textsc{Hendren, Nathaniel and Ben Sprung-Keyser} (2020): \enquote{A {Unified} {Welfare} {Analysis} of {Government} {Policies}*,} \emph{The Quarterly Journal of Economics}, 135, 1209--1318.

\bibitem[\protect\citeauthoryear{Honoré and Hu}{Honoré and Hu}{2020}]{honore_selection_2020}
\textsc{Honoré, Bo~E. and Luojia Hu} (2020): \enquote{Selection {Without} {Exclusion},} \emph{Econometrica}, 88, 1007--1029.

\bibitem[\protect\citeauthoryear{Horowitz and Manski}{Horowitz and Manski}{1995}]{horowitz_identification_1995}
\textsc{Horowitz, Joel~L. and Charles~F. Manski} (1995): \enquote{Identification and {Robustness} with {Contaminated} and {Corrupted} {Data},} \emph{Econometrica}, 63, 281.

\bibitem[\protect\citeauthoryear{Huber, Laffers, and Mellace}{Huber et~al.}{2017}]{huber_sharp_2017}
\textsc{Huber, Martin, Lukas Laffers, and Giovanni Mellace} (2017): \enquote{Sharp {IV} {Bounds} on {Average} {Treatment} {Effects} on the {Treated} and {Other} {Populations} {Under} {Endogeneity} and {Noncompliance},} \emph{Journal of Applied Econometrics}, 32, 56--79.

\bibitem[\protect\citeauthoryear{Imbens and Angrist}{Imbens and Angrist}{1994}]{imbens_identification_1994}
\textsc{Imbens, Guido~W. and Joshua~D. Angrist} (1994): \enquote{Identification and {Estimation} of {Local} {Average} {Treatment} {Effects},} \emph{Econometrica}, 62, 467.

\bibitem[\protect\citeauthoryear{Johnson, Gritz, Jackson, Burghardt, Boussy, Leonard, and Orians}{Johnson et~al.}{1999}]{johnson_national_1999}
\textsc{Johnson, Terry, Mark Gritz, Russell Jackson, John Burghardt, Carol Boussy, Jan Leonard, and Carlyn Orians} (1999): \enquote{National {Job} {Corps} {Study}: {Report} on the {Process} {Analysis},} Tech. rep., Mathematica Policy Research Inc., Princeton, NJ.

\bibitem[\protect\citeauthoryear{Katz, Roth, Hendra, and Schaberg}{Katz et~al.}{2022}]{katz_why_2022}
\textsc{Katz, Lawrence~F., Jonathan Roth, Richard Hendra, and Kelsey Schaberg} (2022): \enquote{Why {Do} {Sectoral} {Employment} {Programs} {Work}? {Lessons} from {WorkAdvance},} \emph{Journal of Labor Economics}, 40, S249--S291.

\bibitem[\protect\citeauthoryear{Kitagawa}{Kitagawa}{2015}]{kitagawaTestInstrumentValidity2015}
\textsc{Kitagawa, Toru} (2015): \enquote{A {{Test}} for {{Instrument Validity}},} \emph{Econometrica}, 83, 2043--2063.

\bibitem[\protect\citeauthoryear{Kitagawa}{Kitagawa}{2021}]{kitagawa_identification_2021}
---\hspace{-.1pt}---\hspace{-.1pt}--- (2021): \enquote{The identification region of the potential outcome distributions under instrument independence,} \emph{Journal of Econometrics}, 225, 231--253.

\bibitem[\protect\citeauthoryear{Kwon and Roth}{Kwon and Roth}{2024}]{kwon2024testing}
\textsc{Kwon, Soonwoo and Jonathan Roth} (2024): \enquote{Testing Mechanisms,} \emph{arXiv preprint arXiv:2404.11739}.

\bibitem[\protect\citeauthoryear{Lachowska, Mas, and Woodbury}{Lachowska et~al.}{2020}]{lachowska_sources_2020}
\textsc{Lachowska, Marta, Alexandre Mas, and Stephen~A. Woodbury} (2020): \enquote{Sources of {Displaced} {Workers}’ {Long}-{Term} {Earnings} {Losses},} \emph{American Economic Review}, 110, 3231--3266.

\bibitem[\protect\citeauthoryear{Lamadon, Mogstad, and Setzler}{Lamadon et~al.}{2022}]{lamadon_imperfect_2022}
\textsc{Lamadon, Thibaut, Magne Mogstad, and Bradley Setzler} (2022): \enquote{Imperfect {Competition}, {Compensating} {Differentials}, and {Rent} {Sharing} in the {US} {Labor} {Market},} \emph{American Economic Review}, 112, 169--212.

\bibitem[\protect\citeauthoryear{Lee}{Lee}{2009}]{lee_training_2009}
\textsc{Lee, David~S.} (2009): \enquote{Training, {Wages}, and {Sample} {Selection}: {Estimating} {Sharp} {Bounds} on {Treatment} {Effects},} \emph{Review of Economic Studies}, 76, 1071--1102.

\bibitem[\protect\citeauthoryear{Li and Racine}{Li and Racine}{2007}]{li_nonparametric_2007}
\textsc{Li, Qi and Jeffrey Racine} (2007): \emph{Nonparametric econometrics: theory and practice}, Princeton Oxford: Princeton University Press.

\bibitem[\protect\citeauthoryear{Liu, Perry, Skinner, and Cody}{Liu et~al.}{2020}]{liu_estimating_2020}
\textsc{Liu, Albert, Clayton Perry, Robbie Skinner, and Scott Cody} (2020): \enquote{Estimating {Job} {Corps} {Cost} per {Enrollee} and {Cost} per {Graduate},} Tech. rep., Insight Policy Research, Arlington, VA.

\bibitem[\protect\citeauthoryear{Mourifi{\'e} and Wan}{Mourifi{\'e} and Wan}{2017}]{mourifieTestingLocalAverage2017}
\textsc{Mourifi{\'e}, Ismael and Yuanyuan Wan} (2017): \enquote{Testing {{Local Average Treatment Effect Assumptions}},} \emph{The Review of Economics and Statistics}, 99, 305--313.

\bibitem[\protect\citeauthoryear{Newey, Powell, and Walker}{Newey et~al.}{1990}]{newey_semiparametric_1990}
\textsc{Newey, Whitney, James Powell, and Walker} (1990): \enquote{Semiparametric {Estimation} of {Selection} {Models}: {Some} {Empirical} {Results},} \emph{American Economic Review}, 80, 324--328.

\bibitem[\protect\citeauthoryear{Olma}{Olma}{2021}]{olma_nonparametric_2021}
\textsc{Olma, Tomasz} (2021): \enquote{Nonparametric {Estimation} of {Truncated} {Conditional} {Expectation} {Functions},} Version Number: 1.

\bibitem[\protect\citeauthoryear{Pearl}{Pearl}{2001}]{pearl_direct_2001}
\textsc{Pearl, Judea} (2001): \enquote{Direct and {Indirect} {Effects},} Tech. rep.

\bibitem[\protect\citeauthoryear{Pearl}{Pearl}{2009}]{pearl_causal_2009}
---\hspace{-.1pt}---\hspace{-.1pt}--- (2009): \enquote{Causal inference in statistics: {An} overview,} \emph{Statistics Surveys}, 3.

\bibitem[\protect\citeauthoryear{Robins and Greenland}{Robins and Greenland}{1992}]{robins_identifiability_1992}
\textsc{Robins, James~M. and Sander Greenland} (1992): \enquote{Identifiability and {Exchangeability} for {Direct} and {Indirect} {Effects}:,} \emph{Epidemiology}, 3, 143--155.

\bibitem[\protect\citeauthoryear{Schmieder, Von~Wachter, and Heining}{Schmieder et~al.}{2023}]{schmieder_costs_2023}
\textsc{Schmieder, Johannes~F., Till Von~Wachter, and Jörg Heining} (2023): \enquote{The {Costs} of {Job} {Displacement} over the {Business} {Cycle} and {Its} {Sources}: {Evidence} from {Germany},} \emph{American Economic Review}, 113, 1208--1254.

\bibitem[\protect\citeauthoryear{Schochet, Bellotti, Cao, Glazerman, Grady, Gritz, McConnell, Johnson, and Burghardt}{Schochet et~al.}{2003}]{schochet_national_2003}
\textsc{Schochet, Peter, Jeanne Bellotti, Ruo-Jiao Cao, Steven Glazerman, April Grady, Mark Gritz, Sheena McConnell, Terry Johnson, and John Burghardt} (2003): \enquote{National {Job} {Corps} {Study}: {Data} {Documentation} and {Public} {Use} {Files}: {Volume} {I},} .

\bibitem[\protect\citeauthoryear{Schochet, Burghardt, and Glazerman}{Schochet et~al.}{2001}]{schochet_national_2001}
\textsc{Schochet, Peter, John Burghardt, and Steven Glazerman} (2001): \enquote{National {Job} {Corps} {Study}: {The} {Impacts} of {Job} {Corps} on {Participants}' {Employment} and {Related} {Outcomes},} Tech. rep., Mathematica Policy Research Inc., Princeton, NJ.

\bibitem[\protect\citeauthoryear{Schochet, Burghardt, and McConnell}{Schochet et~al.}{2008}]{schochet_does_2008}
\textsc{Schochet, Peter~Z, John Burghardt, and Sheena McConnell} (2008): \enquote{Does {Job} {Corps} {Work}? {Impact} {Findings} from the {National} {Job} {Corps} {Study},} \emph{American Economic Review}, 98, 1864--1886.

\bibitem[\protect\citeauthoryear{Semenova}{Semenova}{2020}]{semenova_generalized_2020}
\textsc{Semenova, Vira} (2020): \enquote{Generalized {Lee} {Bounds},} Publisher: [object Object] Version Number: 3.

\bibitem[\protect\citeauthoryear{Song, Price, Guvenen, Bloom, and Von~Wachter}{Song et~al.}{2019}]{song_firming_2019}
\textsc{Song, Jae, David~J Price, Fatih Guvenen, Nicholas Bloom, and Till Von~Wachter} (2019): \enquote{Firming {Up} {Inequality}*,} \emph{The Quarterly Journal of Economics}, 134, 1--50.

\bibitem[\protect\citeauthoryear{Słoczyński}{Słoczyński}{2020}]{sloczynski_when_2020}
\textsc{Słoczyński, Tymon} (2020): \enquote{When {Should} {We} ({Not}) {Interpret} {Linear} {IV} {Estimands} as {LATE}?} Publisher: [object Object] Version Number: 6.

\bibitem[\protect\citeauthoryear{{The Council of Economic Advisers}}{{The Council of Economic Advisers}}{2019}]{the_council_of_economic_advisers_government_2019}
\textsc{{The Council of Economic Advisers}} (2019): \enquote{Government {Employment} and {Training} {Programs}: {Assessing} the {Evidence} on their {Performance},} Tech. rep.

\bibitem[\protect\citeauthoryear{{United States Equal Employment Opportunity Commission}}{{United States Equal Employment Opportunity Commission}}{2009}]{united_states_equal_employment_opportunity_commission_federal_2009}
\textsc{{United States Equal Employment Opportunity Commission}} (2009): \enquote{Federal {Laws} {Prohibiting} {Job} {Discrimination} {Questions} and {Answers},} Tech. rep., United States Equal Employment Opportunity Commission, Washington, DC.

\bibitem[\protect\citeauthoryear{Vayalinkal}{Vayalinkal}{2024}]{vayalinkal_sharp_2024}
\textsc{Vayalinkal, Atom} (2024): \enquote{Sharp {Identification} {Regions} in {General} {Selection} {Models} with ({Un})ordered {Treatments} and {Discrete} {Instruments},} \emph{Unpublished Manuscript, University of Toronto}.

\bibitem[\protect\citeauthoryear{Zuo, Ghosh, Ding, and Yang}{Zuo et~al.}{2022}]{zuo_mediation_2022}
\textsc{Zuo, Shuozhi, Debashis Ghosh, Peng Ding, and Fan Yang} (2022): \enquote{Mediation analysis with the mediator and outcome missing not at random,} Version Number: 3.

\end{thebibliography}
